\documentclass[prd,preprint,superscriptaddress,amsmath,amssymb]{revtex4}

\usepackage{graphicx,color,slashed}

\begin{document}

{\begin{flushright}{KIAS-P18018}
\end{flushright}}

\title{\bf \Large Charged Higgs boson contribution to  $B^-_{q} \to \ell \bar \nu$ and $\bar B\to (P, V) \ell \bar\nu$ in a generic two-Higgs doublet model}

\author{Chuan-Hung Chen}
\email{physchen@mail.ncku.edu.tw}
\affiliation{Department of Physics, National Cheng-Kung University, Tainan 70101, Taiwan}

\author{Takaaki Nomura}
\email{nomura@kias.re.kr}
\affiliation{School of Physics, KIAS, Seoul 02455, Korea}

\date{\today}

\begin{abstract}
We comprehensively study the charged-Higgs contributions to the leptonic $B^-_q \to \ell \bar \nu$ ($q=u,c$) and semileptonic $\bar B \to X_q \ell \bar\nu$ ($X_u=\pi, \rho; X_c=D,D^*$) decays in the type-III two-Higgs-doublet model (2HDM). We employ the Cheng-Sher ansatz to  suppress the tree-level flavor-changing neutral currents (FCNCs) in the quark sector. When the strict constraints from the $\Delta B=2$, $b\to s \gamma$, and $pp(b\bar b)\to H/A \to \tau^+ \tau^-$ processes are considered,  parameters $\chi^u_{tq}$ from the quark couplings and $\chi^\ell_\ell$  from the lepton couplings dictate the leptonic and semileptonic $B$ decays. It is found that when the measured $B^-_u\to \tau \bar \nu$ and indirect bound of $B^-_c \to \tau \bar \nu$ obtained by LEP1 data are taken into account, $R(D)$ and $R(\pi)$ can have broadly allowed ranges; however,  the values of $R(\rho)$ and $R(D^*)$ are limited to  approximately the standard model (SM) results.  We also find that the same behaviors also occur in the $\tau$-lepton polarizations and forward-backward asymmetries ($A^{X_q,\tau}_{FB}$) of the semileptonic decays, with the exception of $A^{D^*,\tau}_{FB}$, for which the deviation from the SM due to the charged-Higgs effect is still sizable.  In addition, the $q^2$-dependent $A^{\pi,\tau}_{FB}$ and $A^{D,\tau}_{FB}$ can be very sensitive to the charged-Higgs effects and have completely different shapes from the SM.

\end{abstract}

\maketitle

\section{Introduction}

In spite of the success of the standard model (SM) in particle physics, we are still uncertain as to the solutions for  baryongenesis, neutrino mass, and dark matter. It is believed that the SM is an effective theory at the electroweak scale, and thus there should be plenty of room to explore the new  physics effects in  theoretical and experimental high energy physics.

A known extension of the SM is the two-Higgs-doublet model (2HDM), where the model can be used to resolve weak and strong CP problems~\cite{Lee:1973iz,Peccei:1977hh}. Due to  the involvement of new scalars, such as  one CP-even, one CP-odd, and two charged Higgses, despite its original motivation, the 2HDM provides rich phenomena in particle physics~\cite{Gunion:1989we,Chen:2014xva,Benbrik:2015evd,Chen:2016xju}, especially, the charged-Higgs, which causes lots of interesting effects in flavor physics. According to the imposed symmetry (e.g., soft $Z_2$ symmetry) to the Lagrangian in the literature, the 2HDM is classified as type-I, type-II, lepton-specific, and flipped models, for which  detailed introduction can be found in~\cite{Branco:2011iw}. Among these 2HDM schemes,  only the type-II model corresponds to the tree-level minimal supersymmetric standard model (MSSM) case. 

Recently, lepton-flavor universality has suffered challenges from tree-level $B$-meson decays. For instance, BaBar~\cite{Lees:2012xj,Lees:2013uzd}, Belle~\cite{Huschle:2015rga,Abdesselam:2016cgx,Hirose:2016wfn}, and LHCb~\cite{Aaij:2015yra,Aaij:2017uff} observed unexpected large branching ratios (BRs)  in $\bar B \to D^{(*)} \tau \bar\nu_\tau$, and the averaged observables  were defined and measured as~\cite{Amhis:2016xyh}:
\begin{align}
R(D)  & =\frac{BR(\bar B \to D \tau \bar\nu)}{BR(\bar B \to D \ell \bar\nu)} =  0.407 \pm 0.039 \pm 0.024\,,  \nonumber \\
 R(D^*)  & = \frac{BR(\bar B \to D^* \tau \bar\nu)}{BR(\bar B \to D^* \ell \bar\nu)}  =  0.304 \pm 0.013 \pm 0.007\,,
\end{align}
 where $\ell$ denotes the light leptons, and  the SM predictions using different approaches are closed to each other and obtained as $R(D) \approx 0.30$ \cite{Lattice:2015rga,Na:2015kha,Bernlochner:2017jka,Jaiswal:2017rve} and $R(D^*) \approx 0.25$~\cite{Fajfer:2012vx,Bigi:2017jbd,Bernlochner:2017jka,Jaiswal:2017rve}.  Intriguingly, when the $R(D)-R(D^*)$ correlation is taken into account, the deviation with respect to the SM prediction is $4.1\sigma$.   Based on these observations, possible extensions of the SM  for explaining the excesses are studied in~\cite{Fajfer:2012jt,Crivellin:2012ye,Datta:2012qk,Bailey:2012jg,Celis:2012dk,Tanaka:2012nw,HernandezSanchez:2012eg,Biancofiore:2013ki,Crivellin:2013wna,Dorsner:2013tla,Dutta:2013qaa,Sakaki:2013bfa,Bhattacharya:2014wla,Alonso:2015sja,Calibbi:2015kma,Freytsis:2015qca,Crivellin:2015hha,Bhattacharya:2015ida,Alonso:2016gym,Das:2016vkr,Li:2016vvp,Boucenna:2016qad,Becirevic:2016yqi,Altmannshofer:2016jzy,Bhattacharya:2016mcc,Bardhan:2016uhr,Dutta:2016eml,Bhattacharya:2016zcw,Alonso:2016oyd,Dutta:2017xmj,Chen:2017hir,Chen:2017eby,Megias:2017ove,Crivellin:2017zlb,Altmannshofer:2017yso,Ciuchini:2017mik,Celis:2017doq,Kamenik:2017tnu,Altmannshofer:2017poe,Alok:2017jaf,Choudhury:2017qyt,Buttazzo:2017ixm,Chen:2017usq,Akeroyd:2017mhr}.

 Moreover, when $|V_{ub}|\approx 3.72\times 10^{-3}$ is taken from the results of  lattice QCD~\cite{Lattice:2015tia} and light-cone sum rules (LCSRs)~\cite{Ball:2004ye,Ball:2006jz},  the SM result of $BR(B^-_u \to \tau \bar\nu)^{\rm SM}\approx 0.89 \times 10^{-4}$ is slightly smaller than the current measurement of  $BR(B^-_u \to \tau \bar\nu)^{\rm exp}=(1.09 \pm 0.24)\times 10^{-4}$~\cite{PDG}. 
In addition to the uncertainties of $V_{ub}$ and $B$-meson decay constant $f_B$, the difference between the SM prediction and experimental data may raise from new charged current effects~\cite{Isidori:2006pk,Chen:2006nua,Akeroyd:2007eh,Akeroyd:2008ac}.  Since the $\bar B\to D^{(*)} \tau \bar\nu$ and $B^-_u \to \tau \bar\nu$ processes are associated with the $W^{\pm}$-mediated $b\to (u,c) \tau \bar \nu$ decays in the SM, in this work, we  study the charged-Higgs contributions to the decays in detail in the 2HDM  framework.

 The charged-Higgs can be naturally taken as the origin of a lepton-flavor universality violation because its Yukawa coupling to a lepton is usually proportional to the lepton mass.  Due to the suppression of $m_\ell/v$ ($v\approx 246$ GeV), we thus need an extra factor in the coupling to enhance the charged-Higgs effect. In the 2HDM  schemes  mentioned above, it can be easily found that only the type-II model can have a $\tan^2\beta$ enhancement in the Hamiltonian of $b\to (u,c) \tau\bar\nu$. However, the type-II 2HDM cannot resolve the excesses  for the  following reasons: (i) the sign of type-II contribution is always destructive to the SM contributions in $b\to (u, c) \tau \bar\nu$, and (ii) the lower bound of the charged-Higgs mass  limited by $b\to s \gamma$ is now $m_{H^\pm} > 580$ GeV~\cite{Misiak:2017bgg}, so that the change due to the charged-Higgs effect is only at a percentage level.  Inevitably, we have to consider other schemes in the 2HDM that can retain the $\tan\beta$ enhancement, can be a constructive contribution to the SM, and can have a smaller $m_{H^\pm}$.  
 
The desired scheme can be achieved when the imposed  symmetry  is removed; that is, the two Higgs doublets can simultaneously couple to the up- and down-type quarks. This  scheme is called the type-III 2HDM in the literature~\cite{Benbrik:2015evd,Crivellin:2012ye,HernandezSanchez:2012eg}. In such a scheme, unless an extra assumption is made~\cite{Ahn:2010zza}, the flavor changing neutral currents  (FCNCs) are generally induced at the tree level. In order to  naturally suppress the tree-induced $\Delta F=2$ ($F=K,B_{d(s)},D$) processes, we can adopt the Cheng-Sher ansatz~\cite{Cheng:1987rs}, where the FCNC effects are parametrized to be the square-root of the production involving flavor masses.  We find that the same quark FCNC effects  also appear in the charged-Higgs couplings to the quarks. Using the Cheng-Sher ansatz,   it is found that in addition to the achievements of the $\tan\beta$ enhancement factor and a smaller $m_{H^\pm}$,  new unsuppressed factors denoted by $\chi^u_{tc(tu)}$ occur at the vertices $c(u)bH^\pm$, which play an important role in  $B^-_u \to \tau \bar\nu$ and $R(D^{(*)})$. We note that the type-II 2HDM and MSSM can generate the similar Yukawa couplings of the type-III model through the $Z_2$ soft-breaking term, which is from the Higgs potential, when loop effects are considered. Due to loop suppression factor, the loop-induced effects from type-II 2HDM in our study are small. Although the loop effects in supersymmetric (SUSY) models could be sizable, since we focus on the non-SUSY  models, the implications of loop-induced FCNCs in MSSM can be found in~\cite{Babu:1999hn,Isidori:2001fv,Isidori:2002qe,Dedes:2002rh}. 
 
 With the full $\Upsilon(4S)$ data set,  Belle recently reported the measurement of $B^-_u \to \mu \bar\nu$ with a $2.4\sigma$ significance, where the corresponding BR is $BR(B^-_u \to \mu \bar \nu)^{\rm exp} = (6.46 \pm 2.22 \pm 1.60)\times 10^{-7}$, and the SM result is $BR(B^-_u \to \mu \bar \nu)^{\rm SM} = (3.8 \pm 0.31)\times 10^{-7}$~\cite{Sibidanov:2017vph}.  
  The experimental measurement approaches the SM prediction, and it is expected that the improved measurement soon will be obtained at Belle II~\cite{Abe:2010gxa}.  In other words, in addition to the $B^-_u \to \tau \bar \nu$ channel, we can investigate the new charged current effect through a  precise measurement on the $B^-_u \to \mu \bar\nu$ decay. 
 
 In order to comprehensively understand the charged-Higgs contributions to the $b\to (u, c) \ell \bar \nu$ ($\ell=e, \mu,\tau)$ in the type-III 2HDM, in addition to the chiral suppression channels $B^-_u \to (\tau, \mu) \bar\nu$, we study various possible observables for the semileptonic processes $\bar B \to (P, V) \ell \bar\nu$ ($P=\pi, D; V=\rho, D^*)$, which include BRs, $R(P)$, $R(V)$, lepton helicity asymmetry,  and lepton forward-backward asymmetry. To constrain the free parameters, we not only study the constraints from the tree- and loop-induced $\Delta B=2$ processes, but also the $b\to s \gamma$ decay, which has arisen from the  new neutral scalars and charged-Higgs. Although  the neutral current contributions to $b\to s \gamma$ are much smaller than those from the charged-Higgs, for completeness, we also formulate their contributions in the paper. In addition,   the upper bound of $BR(B^-_c \to \tau \bar\nu) < 10\%$ obtained in~\cite{Akeroyd:2017mhr} is also taken into account  when we investigate the $\bar B \to D^* \tau \bar\nu$  decay.  
 
LHCb reported more than a $2\sigma$ deviation from the SM in $R(K) = BR(B^+\to K^+ \mu^+ \mu^-)/BR(B^+\to K^+ e^+ e^-)=0.745^{+0.090}_{-0.074} \pm 0.036$~\cite{Aaij:2014ora} and $R(K^*) = BR(B^0\to K^{*0} \mu^+ \mu^-)/BR(B^0\to K^{*0} e^+ e^-)= 0.69^{+0.11}_{-0.07} \pm 0.05$~\cite{Aaij:2017vbb}. 
Since we concentrate on the tree-level leptonic and semileptonic $B$ decays, we do not address this issue in this work. The charged-Higgs contributions to the $b\to s \ell^+ \ell^-$ processes can be found in~\cite{Hussain:2017tdf,Arnan:2017lxi,Arbey:2017gmh,Arhrib:2017yby,Choudhury:2017ijp,Iguro:2018qzf}.

The paper is organized as follows: In Section II, we discuss and parametrize the charged-Higgs Yukawa couplings to the quarks and leptons in the type-III 2HDM. In Section III, we study the charged-Higgs contributions to the leptonic $B^-_{u(c)} \to \ell \bar \nu$ and $\bar B \to (P, V) \ell \bar \nu$ decays, where  the interesting potential observables include the decay rate,  the branching fraction ratio, lepton helicity asymmetry, and lepton forward-backward asymmetry.  We study the tree- and loop-induced $\Delta B=2$ and loop-induced $b\to s \gamma$ processes in Section IV, where  the contributions of neutral scalar $H$, neutral pseudoscalar $A$, and charged-Higgs are taken into account.  The detailed numerical analysis and the current experimental bounds are shown in Section V, and a conclusion is given in Section VI.

\section{ Yukawa couplings  in the generic 2HDM }

To study the charged-Higgs contributions to the $b \to q \ell \bar \nu$ ($q=u,c$) decays in the type-III 2HDM,  we analyze  the relevant Yukawa couplings in this section, especially,  the charged-Higgs couplings to $ub$ and $cb$, where they can make significant contributions to  the leptonic and semileptonic $B$ decays. The characteristics of new Yukawa couplings in the type-III model will be also discussed.

\subsection{Formulation of $H^\pm$ Yukawa couplings to the quarks and leptons}

Since the charged-Higgs couplings to the quarks and the leptons in type-III 2HDM were derived before~\cite{Benbrik:2015evd}, we briefly introduce the relevant pieces in this  section. 
 We begin to write the Yukawa couplings in the  type-III model as:
\begin{align}
-{\cal L}_Y &=  \bar Q_L Y^d_1 D_R H_1 + \bar Q_L Y^{d}_2 D_R H_2
+ \bar Q_L Y^u_1 U_R \tilde H_1 + \bar Q_L Y^{u}_2 U_R \tilde H_2
 \nonumber \\
&+  \bar L Y^\ell_1 \ell_R H_1 + \bar L Y^{\ell}_2 \ell_R H_2 + H.c.\,, 
\label{eq:Yu}
\end{align}
where the flavor indices are suppressed; $Q^T_L=(u, d)_L$ and $L^T = (\nu, \ell)_L$ are the $SU(2)_L$ quark and lepton doublets, respectively; $f_R$ ($f=U,D,\ell$) is the singlet  fermion;  $Y^f_{1,2}$ are the $3\times 3$ Yukawa matrices, and $\tilde H_i = i\tau_2 H^*_i$ with $\tau_2$ being the  Pauli matrix. The components of  the Higgs doublets are  taken as:
\begin{align}
H_i &= \left(
            \begin{array}{c}
              \phi^+_i \\
              (v_i+\phi_i +i \eta_i)/\sqrt{2} \\
            \end{array}
          \right)\,, \label{eq:doublet}
\end{align}
and $v_i$ is the vacuum expectation value (VEV) of $H_i$. We note that Eq.~(\ref{eq:Yu}) can  recover  the type II 2HDM  when $Y^u_1$, $Y^d_2$, and $Y^\ell_2$ vanish. The physical states for scalars can then be expressed as: 
\begin{align}
 h &= -s_\alpha \phi_1 + c_\alpha  \phi_2 \,, \quad H= c_\alpha \phi_1 + s_\alpha \phi_2 \,, \nonumber \\
H^\pm (A) &= -s_\beta \phi^\pm_1 (\eta_1) + c_\beta \phi^\pm_2 (\eta_2) \,, \label{eq:hHHA}
\end{align}
where the mixing angles are defined as $c_\alpha (s_\alpha)= \cos\alpha (\sin\alpha)$, $c_\beta  = \cos\beta = v_1/v$, and $s_\beta= \sin\beta = v_2/v$ with $v=\sqrt{v^2_1 +v^2_2}$.  In this work, $h$ is the SM-like Higgs while $H$, $A$, and $H^\pm$ are new scalar bosons.

The fermion mass matrix can be formulated as:
 \begin{equation}
  \bar f_L {\bf M}^f f_R + H.c. \equiv  \frac{v}{\sqrt{2}}\bar f_L \left( c_\beta Y^f_1 + s_\beta Y^f_2 \right) f_R + H.c. 
 \end{equation}
Without assuming the relation between $Y^f_1$ and $Y^f_2$, both Yukawa matrices cannot be simultaneously diagonalized~\cite{Ahn:2010zza}. Thus, the FCNCs mediated by scalar bosons are induced at the tree level. We introduce unitary matrices $U^f_L$ and $U^f_R$  to diagonalize the fermion mass matrices by following $f^p_L = U^f_L f^w_L$ and  $f^p_R = U^f_R f^w_R$, where $f^{p(w)}_{L,R}$ denote the physical (weak) eigenstates. Then, the Yukawa couplings of  $H^\pm$  can be written as~\cite{Benbrik:2015evd}:
\begin{align}
-{\cal L}^{H^\pm}_Y &=  \sqrt{2} \bar u_R  \left[ - \frac{ 1}{v t_\beta } {\bf m_u} + \frac{{\bf X}^{u\dagger}}{ s_\beta}  \right] {\bf V} d_L H^+ + \sqrt{2} \bar u_L {\bf V} \left[ - \frac{t_\beta }{v } {\bf m_d} + \frac{{\bf X}^d}{  c_\beta}  \right] d_R H^+  \nonumber \\
&+\sqrt{2} \bar \nu_L \left[ -\frac{\tan\beta}{v } {\bf m_\ell} + \frac{{\bf X}^\ell }{ c_\beta}  \right] \ell_R H^+ + H.c.\,, \label{eq:YuCH}
\end{align}
where $t_\beta=\tan\beta=v_2/v_1$; ${\bf V}\equiv U^u_L U^{d\dagger}_L$ denotes the Cabibbo-Kobayashi-Maskawa (CKM) matrix, and the ${\bf X}s$ are defined as: %
 \begin{equation}
  {\bf X}^u= U^u_L \frac{Y^u_1}{\sqrt{2}}  U^{u\dagger}_R \,, \     {\bf X}^d= U^d_L \frac{Y^d_2}{\sqrt{2}} U^{d\dagger}_R \,, \ 
{\bf X}^\ell= U^\ell_L \frac{Y^\ell_2}{\sqrt{2}} U^{\ell\dagger}_R\,. \label{eq:Xs}
 \end{equation}

${\bf X}^{u,d}$ are the sources of tree-level FCNCs in the type-III model. In order to accommodate the strict constraints from the $\Delta F=2$ processes, such as $\Delta m_{P}$ ($P=K,B_{d,s},D$),  we adopt the so-called Cheng-Sher  ansatz~\cite{Cheng:1987rs} in the quark and lepton sectors, where ${\bf X}^f$ is  parametrized as: 
 \begin{equation}
X^f_{ij} =\frac{ \sqrt{m_{f_i} m_{f_j}}}{v} \chi^f_{ij}\,, \label{eq:CSA}
 \end{equation}
 and  $\chi^f_{ij}$ are the new free parameters.  Using this ansatz, it can be seen that $\Delta m_{P}$ arisen from the tree level is suppressed by $m_d m_s/v^2$ for $K$-meson, $m_{d(s)}m_b/v^2$ for $B_{d(s)}$, and $m_u m_c/v^2$ for $D$-meson.
Since we do not study the origin of neutrino mass, the neutrinos are taken as massless particles in this work. Nevertheless, even with a massive neutrino case, the influence on hadronic processes is small and negligible. In addition, to simplify the numerical analysis, in this work we use  the scheme with ${\bf X}^\ell_{ij} = (m_{\ell_i} /v )  \chi^\ell_{\ell_i} \delta_{\ell_i \ell_j}$, i.e. $\chi^\ell_{\ell_i \ell_j} = \chi^\ell_{\ell_i}  \delta_{\ell_i \ell_j}$; as a result, the Yukawa couplings of $H^\pm$ to the leptons can be expressed as:
  \begin{equation}
  {\cal L}^{H^\pm}_{Y,\ell} = \sqrt{2} \frac{\tan\beta \, m_\ell}{v} \left(1 - \frac{\chi^{\ell}_{\ell} }{s_\beta} \right) \bar \nu_{\ell }  P_R \ell  H^+ + H.c.\,, \label{eq:Clepton}
  \end{equation}
 with $P_{R(L)}=(1 \pm \gamma_5)/2$. The suppression factor $m_\ell /v$ could be moderated using the scheme of large $\tan\beta$. 

\subsection{  $b$-quark Yukawa couplings to $H^\pm$}

From Eq.~(\ref{eq:YuCH}), it can be seen that the coupling $u_{iR} b_L H^\pm$ $(u_i=u,c)$ in the type-II 2HDM (i.e. ${\bf X}^{d,u}=0$) is suppressed by $m_{u_i}/ (v t_\beta) V_{u_i b}$, and  this effect  can be neglected. However, the situation is changed in the type-III model. In addition to the disappearance of suppression factor $1/t_\beta$,  the new effect ${\bf X}^u$ accompanied with the CKM matrix in form of ${\bf X}^u V/v$ could lead to $\sqrt{m_{u_i} m_t}/v \chi^u_{u_i t}V_{tb}$, where  $\sqrt{m_{u_i} m_t}/v$ numerically plays the role of $|V_{u_i b}|$, and the magnitude of the coupling is dictated by the free parameter $\chi^{u}_{u_i t}$, which in principle is not  suppressed. Additionally, the  $u_{iL} b_R H^\pm$ coupling is also  remarkably  modified.  In order to  more comprehend the influence of the new charged-Higgs couplings on the $B$ decays, in the rest of this subsection, we discuss the $u_i b H^\pm$ coupling in detail. For convenience, we rewrite the $H^\pm$ couplings to the $b$-quark and light up-type quarks as:
 \begin{align}
 {\cal L}^{H^\pm}_Y &  \supset  \frac{\sqrt{2}}{v} \bar u_{i R} C^L_{u_i b} b_{L} H^+ +  \frac{\sqrt{2}}{v} \bar u_{i L} C^R_{u_i b} b_{R} H^+ + H.c.\,, \nonumber \\
 C^L_{u_i b} & =  \left(  \frac{m_{u_i}}{ t_\beta }  \delta_{u_i u_j} - \frac{\sqrt{m_{u_i} m_{u_j}}}{ s_\beta}  \chi^{u*}_{u_j u_i} \right) V_{u_j b} \,, \nonumber \\
 C^R_{u_i b} &= V_{u_i d_j} \left( t_\beta  m_b  \delta_{d_j b} - \frac{\sqrt{m_{d_j} m_{b}}}{  c_\beta} \chi^d_{d_j b}\right)\,, \label{eq:CLR}
 \end{align}
 where $u_j(d_j)$  indicates the sum of all possible up(down)-type quarks. 

 In the following, we analyze the characteristics of the $C^{L}_{u(c) b}$ and $C^{R}_{u(c) b}$ couplings in the type-III 2HDM with the Cheng-Sher ansatz. Due to $m_u V_{ub} \ll \sqrt{m_u m_c} V_{cb} \ll \sqrt{m_u m_t} V_{tb}$,  we can simplify  the $C^{L}_{ub}$ coupling as: 
 \begin{equation}
 \frac{\sqrt{2}}{v} C^L_{ub} \approx  -\sqrt{2} \frac{ \sqrt{m_u m_t} }{ v s_\beta} \chi^{u*}_{tu} V_{tb}\,. \label{eq:CLub}
 \end{equation}
With $m_u\sim 5.4$ MeV, $m_t\sim 165$ GeV, and $v\approx 246$ GeV, it can be found that  $\sqrt{m_u m_t}/v\approx 3.84\times 10^{-3}$ is very close to the value of $|V_{ub}|$; therefore,  $C^L_{ub}$ can be read as $\sqrt{2} C^L_{ub}/v  \sim - \sqrt{2}\chi^{u*}_{tu}  |V_{ub}| $, where $s_\beta \approx 1$  is applied.   Clearly, unlike the case in the type-II 2HDM, which   is highly suppressed by $m_u/(v t_\beta)$, $C^L_{ub}$ in the type-III model  is still proportional to $|V_{ub}|$, can be sizable, and is controlled by   $\chi^{u*}_{tu}$.  For the $C^R_{ub}$ coupling,  the decomposition from Eq.~(\ref{eq:CLR}) can be written as:
  \begin{equation}
 C^R_{ub} =- V_{ud}  \frac{\sqrt{m_d m_b} \chi^d_{db} }{c_\beta} -  V_{us} \frac{\sqrt{m_s m_b} \chi^{d}_{sb}}{ c_\beta} + V_{ub} m_b \left(  t_\beta- \frac{ \chi^{d}_{bb}}{c_\beta} \right)  \,. 
 \end{equation}
The  numerical values of the first two terms can be obtained as: $V_{ud} \sqrt{m_d/ m_b}\approx 0.047$ and $V_{us} \sqrt{m_s /m_b} \approx 0.032 \gg |V_{ub}|$. Unless $\chi^d_{db,sb}$ are strictly constrained,  each term with  different CKM factors may  be important and cannot be arbitrarily dropped. For clarity, we rewrite   $C^R_{ub}$ to be:
 \begin{align}
 \frac{\sqrt{2}}{v}  C^R_{ub} &=  \sqrt{2}\frac{m_b t_\beta}{v}V_{ub}  \left( 1 - \frac{\chi^R_{ub}}{s_\beta}\right)\,,   
 \label{eq:CRub}\\
  \chi^R_{ub} &=  \chi^d_{bb} + \frac{V_{ud}}{V_{ub}} \sqrt{ \frac{m_d}{m_b}} \chi^d_{db}+  \frac{V_{us}}{V_{ub}} \sqrt{ \frac{m_s}{m_b}} \chi^d_{sb}\,. \label{eq:chiRub}
 \end{align} 
 Due to $|V_{ub}|\ll V_{us,ud}$,  the magnitude of $\chi^R_{ub}$ in principle can be of $O(10)$, and  the resulted $C^R_{ub}$ is much larger than that in the type-II 2HDM. In order to avoid obtaining an $C^R_{ub}$ that is too large, we can require a cancellation between $V_{ud} \sqrt{m_d/m_b} \chi^d_{db}$ and $V_{us} \sqrt{m_s/m_b} \chi^d_{sb}$ when $\chi^d_{db,sb}$ both are sizable. However, we will show that $\chi^d_{db,sb}$ indeed are constrained by the measured $B_{d,s}$ mixing parameters and that their magnitudes should be less than $O(10^{-2})$. 
 
 For the processes dictated by the $b\to c$ decays,  due to  $\sqrt{m_u m_c} V_{ub} \ll m_c V_{cb} \ll \sqrt{m_c m_t} V_{tb}$,  the $H^\pm$ Yukawa coupling of $C^L_{cb}$ can be simplified as:
   \begin{align}
   \frac{\sqrt{2}}{v}  C^L_{cb} & \approx - \frac{\sqrt{2}}{v} \frac{\sqrt{m_c m_t}}{s_\beta} \chi^{u*}_{tc} V_{tb}  \nonumber \\
    & = - \sqrt{2} \frac{m_t}{v} V_{cb}  \left(  \frac{\chi^{u*}_{tc}}{s_\beta} \sqrt{\frac{m_c}{m_t}} \frac{V_{tb}}{V_{cb}}\right)\,,\label{eq:CLcb}
    \end{align} 
    where $m_c/t_\beta$ term has been ignored due to the use of large $t_\beta$ scheme, and the factor in parentheses can be numerically estimated to be $2.19 \chi^{u*}_{tc}$. This behavior is similar to $C^L_{ub}$, but it is $\chi^{u*}_{tc}$ that controls the magnitude. Clearly, if $\chi^{u*}_{tc}$ is not suppressed, it can make a signifiant contribution to the $b\to c$ transition. Using the fact that  $|V_{cd}| \sqrt{m_d m_b}\ll  V_{sc} \sqrt{m_s m_b}$, $V_{cb} m_b$, we can formulate the $C^R_{cb}$ coupling as:
   \begin{align}
   \frac{\sqrt{2}}{v} C^R_{cb} &  \approx  \sqrt{2} \frac{m_b t_\beta}{v}  V_{cb} \left(1 - \frac{\chi^R_{cb}}{s_\beta} \right) \,, \label{eq:CRcb} \\
 \chi^R_{cb} &= \chi^d_{bb} + \sqrt{\frac{m_s}{m_b}} \frac{V_{cs}}{V_{cb}}  \chi^d_{sb} \approx   \chi^d_{bb} + 3.69 \chi^d_{sb}\,. \nonumber 
   \end{align}
Since $C^R_{cb}$ has the $t_\beta$ enhancement, its magnitude is comparable with the SM $W$-gauge coupling of $g V_{cb}/\sqrt{2}$. 
For comparison, we also show the $tbH^\pm$ couplings as:
 \begin{align}
 \frac{\sqrt{2}}{v}C^L_{tb} & \approx \sqrt{2} \frac{m_t}{v} V_{tb} \left( \frac{1}{t_\beta}  - \frac{\chi^{u*}_{tt}}{s_\beta}\right)\,, \\
 \frac{\sqrt{2}}{v}C^R_{tb} & \approx \sqrt{2} \frac{m_b t_\beta}{v} V_{tb} \left( 1 - \frac{\chi^d_{bb}}{s_\beta}\right)\,,
 \end{align}
 where the small effects related to $V_{ub,cb}$ and $V_{ts,td}$ have been dropped.  Although there is a $m_t$ enhancement in the first term of $C^L_{tb}$, $1/t_\beta$ will reduce its contribution when a large $\tan\beta$ value is taken; therefore comparing with $\chi^{u*}_{tt}/s_\beta$, this term  can be ignored, i.e., $C^L_{tb} \approx - m_t V_{tb} \chi^{u*}_{tt} / s_\beta$. From the above analysis, it can be seen that $C^{L,R}_{ub,cb,tb}$  are different from those in the type-II model not only in magnitude but also in sign. For completeness,  the other Yukawa couplings of $H^\pm$ to the quarks are shown in detail  in the Appendix.

\section{ Phenomenological analysis}

The charged current interactions  in this model arise from the SM $W$-gauge and the charged-Higgs bosons. Based on the Yukawa couplings in Eqs.~(\ref{eq:Clepton}) and (\ref{eq:CLR}), 
the effective Hamiltonian for $b\to q \ell \bar\nu$  can be written as:
 \begin{align}
 {\cal H} (b\to q \ell \bar \nu)  =\frac{G_F}{\sqrt 2} V_{qb}& \left[(\bar q b)_{V-A}
 (\bar \ell \nu)_{V-A}+ C^{L,\ell}_{qb} (\bar q b)_{S-P}
 (\bar \ell \nu)_{S-P}  \right. \nonumber \\ 
 & \left. + C^{R,\ell}_{qb} (\bar q b)_{S+P}
 (\bar \ell \nu)_{S-P}
 \right]\,, \label{eq:Heff}
\end{align}
where the fermionic  currents are defined as $(\bar f' f)_{V\pm A}=\bar f' \gamma^\mu (1\pm \gamma_5) f$ and $(\bar f' f)_{S\pm P}=\bar f' (1\pm \gamma_5) f$, and the dimensionless coefficients for the $b\to u$ and $b\to c$ decays are given as:
\begin{subequations}
 \begin{align}
 C^{L,\ell}_{ub} &= \frac{m_t m_\ell t_\beta}{m^2_{H^\pm} s_\beta}  
 \left( 1 - \frac{\chi^\ell_\ell}{s_\beta}\right) \left( \sqrt{\frac{m_u}{ m_t}} \frac{V_{tb}}{V_{ub}}\chi^{u*}_{tu} \right) \,, \label{eq:cLlub}\\
 C^{R,\ell}_{ub} & = -\frac{m_b m_\ell t^2_\beta}{m^2_{H^\pm}}  \left( 1 - \frac{\chi^\ell_\ell}{s_\beta}\right) \left( 1- \frac{\chi^{R}_{ub}}{s_\beta}\right)\,, \label{eq:cRlub}\\
 C^{L,\ell}_{cb} & =  \frac{m_t m_\ell t_\beta}{m^2_{H^\pm} s_\beta}  
 \left( 1 - \frac{\chi^\ell_\ell}{s_\beta}\right)  \left( \sqrt{\frac{m_c}{m_t}} \frac{V_{tb}}{V_{cb}}\chi^{u*}_{tc}\right)\,, \label{eq:cLlcb} \\
 C^{R,\ell}_{cb} & = -\frac{m_b m_\ell t^2_\beta}{m^2_{H^\pm}}  \left( 1 - \frac{\chi^\ell_\ell}{s_\beta}\right) \left( 1 - \frac{\chi^R_{cb} }{s_\beta}\right)\,.   \label{eq:cRlcb}
 \end{align}
\label{eq:Cqb}
 \end{subequations}
Based on the interactions shown in Eqs.~(\ref{eq:Heff}) and (\ref{eq:Cqb}), we investigate  the  charged-Higgs  influence on  the  leptonic and semileptonic $B$ decays in the type-III 2HDM. 

\subsection{ \bf  Leptonic $B^-_{q} \to \ell \bar \nu$ decays}

The hadronic effect in a leptonic $B$ decay is the $B$-meson decay constant. The decay constant associated with an axial-vector current  for the  $B_q$-meson  is defined as:
 \begin{equation}
 \langle 0 | \bar q \gamma^\mu \gamma_5 b|  B^-_q (p_{B_q}) \rangle = -i f_{B_q} p^\mu_{B_q} \,.
 \end{equation}
Using the equation of motion, the decay constant associated with pseudoscalar current is given by:
 \begin{equation}
 \langle 0| \bar q \gamma_5 b | B^-_q(p) \rangle = i f_{B_q} \frac{m^2_{B_q} }{m_b + m_q }\,. 
 \end{equation}
 From the effective interactions in Eq.~(\ref{eq:Heff}), the decay rate for $B^-_q \to \ell \bar\nu$ can be formed as:
 \begin{align}
 \Gamma(B^-_q \to \ell \bar\nu) &= \Gamma^{\rm SM}(B^-_q \to \ell \bar\nu)  \left| 1 + \frac{m^2_{B_q} \left( C^{R,\ell}_{qb} - C^{L,\ell}_{qb} \right)}{m_\ell (m_b + m_q)}\right|^2\,, \label{eq:GaBqellnu} \\
 \Gamma^{\rm SM}(B^-_q \to \ell \bar\nu) &= \frac{G^2_F}{8\pi} |V_{qb}|^2 f^2_{B_q} m_{B_q} m^2_\ell \left( 1- \frac{m^2_\ell}{m^2_{B_q}}\right)^2 \,. \label{eq:Ga_SM}
 \end{align}
 Since  a leptonic meson decay is a chirality-suppressed process,  the decay rate in Eq.~(\ref{eq:Ga_SM}) is proportional to $m^2_\ell$.  From Eq.~(\ref{eq:cLlub}) to Eq.~(\ref{eq:cRlcb}), it can be seen that in the type-II 2HDM, $C^L_{ub} \sim C^L_{cb} \sim 0$ and $C^R_{ub, cb}$ are negative in sign;  therefore, the $H^\pm$ contribution to the $B^-_q \to \ell \bar\nu$ decay is always destructive.  The magnitude and the sign of $C^{R,L}_{qb}$ in the type-III can be changed due to the new effects of  $\chi^{u,d}_{ij}$ and $\chi^\ell_{\ell}$,. 
 
 Before doing  a detailed numerical analysis,  we can numerically understand the impact of 2HDM on the $B^-_q \to \ell \bar\nu$ decay as follows: taking $t_\beta = 50$ and $m_{H^\pm} = 300$ GeV,  we can see that the  charged-Higgs contributions to the $b\to u$ and $b\to c$ decays are respectively given as:
  \begin{align}
 \delta^{H^\pm,\ell}_{q}\equiv  \frac{m^2_{B_q} \left( C^{R,\ell}_{qb} - C^{L,\ell}_{qb} \right)}{m_\ell (m_b+m_q)} \approx \left\{ 
\begin{array}{c}
 - \left[ 0.77\left( 1 -\chi^R_{ub}/s_\beta \right) + 0.39  \chi^{u*}_{tu} e^{i  \phi_3} \right]\left( 1 -\chi^\ell_\ell/s_\beta \right)\,, \\
  -\left[1.09 \left( 1 -\chi^R_{cb}/s_\beta + 1.77  \chi^{u*}_{tc}  \right)\right]\left( 1 -\chi^\ell_\ell/s_\beta \right)\,,  \\  
\end{array} \right. \label{eq:RH}
  \end{align}
  where the sign can be positive when the parameters of $\chi^{u*}_{tu,tc}$ and $\chi^\ell_\ell$ are properly taken, and $\phi_3$ is the phase in $V_{ub}$. 
We note that the Yukawa coupling of the charged-Higgs  to lepton is proportional to the lepton mass; therefore, the ratio in Eq.~(\ref{eq:RH}) does not depend on $m_\ell$.  The lepton-flavor dependent effect is dictated by the $\chi^\ell_\ell$ parameter. 
 
 \subsection{  $B^-_q \to (P, V) \ell \bar\nu$ decays}
 
 Since the semileptonic $B$ decays involve the hadronic  QCD effects, in order to formulate the decays, we parametrize the form factors for  a $B$ decay to a pseudoscalar (P) meson as:
  \begin{align}
  \langle P (p_2)|\overline {q} \gamma^{\mu} b | \bar B (p_1)\rangle
   &= f^{B P}_1(q^2)\left[ P^{\mu}-\frac{P\cdot q}{q^2}q^{\mu} \right]  + f^{B P}_0(q^2)\frac{P\cdot q }{q^2}q^{\mu}, \nonumber \\
\langle P(p_2)| \overline{ q}  \, b |\bar B (p_1) \rangle & = ({m_{B}+m_{P}}) f^{B P}_S(q^2)\,,\label{eq:ffBP}
 \end{align}
where $P=p_1 + p_2$ and $q=p_1 -p_2$. The form factors for a $B$ decay to a vector (V) meson is defined as:
 \begin{align}
  \langle V(p_2,\epsilon_V)|\overline {q} \gamma^{\mu}b|  \bar B(p_1)\rangle
   &= \frac{V^{B V} (q^2)}{m_{B}+m_{V}}\varepsilon^{\mu\nu\rho\sigma}
     \epsilon^*_{V \nu}P_{\rho}q_{\sigma}, \nonumber\\
  \langle V(p_2,\epsilon_V)| \overline {q} \gamma^{\mu}\gamma_5 b|
  \bar B (p_1)\rangle
   &=2im_{V} A^{B V}_0(q^2)\frac{\epsilon^*_V \cdot q}{q^2}q^{\mu} \nonumber \\
  &  +i(m_{B}+m_{V})A^{B V}_1(q^2)\left[\epsilon^{*\mu}_{V}
    -\frac{\epsilon^*_V \cdot q}{q^2}q^{\mu} \right] \nonumber\\
    &-iA^{B V}_2(q^2)\frac{\epsilon^*_V \cdot q}{m_{B}+m_{V}}
     \left[ P^{\mu}-\frac{P\cdot q }{q^2}q^{\mu} \right],
     \nonumber\\
\langle V (p_2, \epsilon_V )| \overline {q} \gamma_5 b | \bar B (p_1)
\rangle & = -i \epsilon^*_V \cdot q f^{BV}_P(q^2). \label{eq:ffBV}
 \end{align}
 With the equation of motion, the form factors of $f^{BP}_S$ and $f^{BV}_P$ can be obtained  as:
\begin{eqnarray}
f^{BP}_S(q^2)\approx  \frac{m_B-m_P}{ m_b-m_q } f_0(q^2)\,, \;\;\;
 f^{BV}_P(q^2)\approx   \frac{2 m_{V}}{m_b +m_q}  A_0(q^2)\,. 
\end{eqnarray}
 Using the interactions in Eq.~(\ref{eq:Heff}) and the form factors defined above,  we can obtain the transition matrix elements for $\bar B\to (P,V) \ell \bar\nu$  as:
 \begin{align}
 {\cal M}_P &= \frac{G_F}{\sqrt{2}} V_{qb} \left[f^{BP}_1 \left(P^\mu -\frac{P\cdot q}{q^2}q^\mu \right) (\bar\ell \nu)_{V-A} \right. \nonumber \\
 &+ \left. \left( m_\ell f^{BP}_0 \frac{P\cdot q}{q^2} + (C^{R,\ell}_{qb} + C^{L,\ell}_{qb})(m_B + m_P) f^{BP}_S  \right) (\bar\ell \nu)_{S-P}\right]\,, \label{eq:MP} \\
{\cal M}^L_{V}&= -i \frac{G_F}{\sqrt{2}} V_{qb} \left\{\epsilon^*_V\cdot q \left( (C^{R,\ell}_{qb} -C^{L,\ell}_{qb} ) f^{BP}_P + 2 A^{BV}_0 \frac{m_{V} m_\ell}{q^2}\right) (\bar\ell \nu)_{S-P} \right. \nonumber \\
& \left. + \left[ (m_B + m_{V}) A^{BV}_1 \left( \epsilon^*_{V\mu}(L) - \frac{\epsilon^*_V \cdot q}{q^2} q_\mu\right) - \frac{A^{BV}_2 \epsilon^*_V \cdot q}{m_B + m_V}  \left(P_\mu -\frac{P\cdot q}{q^2}q_\mu \right)\right] (\bar\ell \nu)_{V-A}\right\}\,, \nonumber \\
{\cal M}^{T}_{V} &=  \frac{G_F}{\sqrt{2}} V_{qb} \left[ \frac{V^{BV}}{m_B + m_{V}} \varepsilon_{\mu\nu\rho \sigma} \epsilon^{*\nu}_{V}(T) P^\rho q^\sigma - i (m_B + m_{V})A^{BV}_1 \epsilon^*_{V\mu}(T)
\right] (\bar \ell \nu)_{V-A}\,, \label{eq:amps}
 \end{align}
where  $q^2$-dependence in the  form factors are hidden, and  ${\cal M}^L_{V}$ and ${\cal M}^T_{V}$ are the longitudinal and transverse $V$-meson components, respectively. From the formulations, we see that  the charged Higgs only affects ${\cal M}_{P}$ and the longitudinal part of the $V$-meson. 

\subsubsection{ Decay amplitudes in helicity basis}

To derive the angular differential decay rate,  we take the coordinates of  the kinematic variables in the rest frame of the $\ell \bar\nu$ invariant mass as:
 \begin{align}
q &=(\sqrt{q^2}, 0 , 0, 0)\,,~ p_{M} = ( E_{M}, 0, 0, p_M)\,,~ p_{M} = \frac{\sqrt{\lambda_M}}{2\sqrt{q^2}} \,,  \nonumber \\  
\lambda_M & = m^4_B + m^4_M + q^4 -2 m^2_B m^2_M -2 m^2_B q^2 -2 m^2_M q^2\,, \nonumber \\
p_{\nu}  &=E_\nu(1,  \sin\theta_\ell \cos\phi,  \sin\theta_\ell \sin\phi,  \cos\theta_\ell)\,,~p_{\ell} = (E_\ell , - \vec{p}_{\nu})\,, \nonumber \\
\epsilon_V(L)&=\frac{1}{m_{V}}( p_{V},0,0, E_{V})\,,~\epsilon_V(\pm) = \frac{1}{\sqrt{2}} (0, \mp 1, - i,0)\,,
 \end{align}
where $M$ denotes $P$- and $V$-meson; $\theta_\ell$ is the polar angle of a neutrino with respect to the moving direction of $M$ meson in the $q^2$ rest frame, and the components of $\vec{p}_{\ell}$ can be obtained from $\vec{p}_{\nu}$ by using  $\pi-\theta_\ell$ and $\phi + \pi$ instead of $\theta_\ell$ and $\phi$.

The solutions of the Dirac equation for positive and negative energy can be expressed as:
 \begin{align}
 u_{\pm} (p) &= \frac{1}{\sqrt{E+m} }
\left(
\begin{array}{c}
 \sqrt{E+m} \chi_{\pm} ( \vec{p})    \\
    \vec{\sigma}\cdot \vec{p} \chi_{\pm}(\vec{p})  
\end{array}
\right)\,, \quad  %
v_{\pm} (p) &= \frac{1}{\sqrt{E+m} }
\left(
\begin{array}{c}
  \vec{\sigma}\cdot \vec{p} \chi_{\mp}(\vec{p})    \\
     \sqrt{E+m} \chi_{\mp} ( \vec{p}) 
\end{array}
\right)\,, \label{eq:u_v}
 \end{align}
 where the $\pm$ indices in $\chi$ are the eigenvalues of $\vec{\sigma}\cdot \vec{p} /|\vec{p}|$, and  $+(-)$ denotes the left(right)-handed state. If the spatial momentum of a particle is taken as $\vec{p}= p(\sin\theta \cos\phi, \sin\theta\sin\phi, \cos\theta)$, the eigenstates of $\vec{\sigma} \cdot \vec{p}$ can be found as:
 \begin{align}
 \chi_+ (\vec{p}) = \left(
\begin{array}{c}
\cos\frac{\theta}{2}    \\
   e^{i\phi}
 \sin\frac{\theta}{2}\end{array}
\right)\,, \quad   \chi_{-}(\vec{p})= \left(
\begin{array}{c}
\sin\frac{\theta}{2}    \\
  - e^{i\phi} \cos\frac{\theta}{2}\end{array}
\right)\,. \label{eq:chi_pn}
 \end{align}
 With the Pauli-Dirac representation of  $\gamma$-matrices, which are defined as:
 \begin{align}
 \gamma^0 = \left(
\begin{array}{cc}
{\bf 1} & 0    \\
   0 &  -{\bf 1} \end{array}
\right)\,, 
\quad \gamma^i =  \left(
\begin{array}{cc}
0 &  \sigma^i   \\
   -\sigma^i & 0\end{array}
\right)\,,  \quad
 \gamma_5=\gamma^5 = \left(
\begin{array}{cc}
0 & {\bf 1}    \\
   {\bf 1} &  0 \end{array}
\right)\,,
 \end{align}
   we get $\bar \ell_{u_\pm} [...] (1-\gamma_5) \nu_{v_+} = 2 \bar \ell_{u_\pm} [...] \nu_{v_{+}}$ , where $[...] =\{ 1, \gamma^\mu, \sigma^{\mu \nu} \}$, and $\ell_{u_\pm}$ denote the charged-lepton in $u_{\pm}$ states. Since we take neutrinos as massless particles,  the neutrino states are always left-handed, i.e., $\bar \ell_{u_\pm} [...] (1-\gamma_5) \nu_{v_{-}}=0$.

With the chosen coordinates and the spinors in Eqs.~(\ref{eq:u_v}) and (\ref{eq:chi_pn}),  the  leptonic current in lepton helicity basis for the $\bar B\to P \ell \bar\nu$ decay can be derived as: 
 \begin{align}
 \bar\ell_{h=+} \slashed{e}_X (1- \gamma_5)  \nu &= 2 m_\ell \beta_\ell \cos\theta_\ell\,, \nonumber \\
 \bar\ell_{h=+}  (1- \gamma_5)  \nu &= -2 \sqrt{q^2} \beta_\ell  \,, \nonumber  \\
 \bar\ell_{h=-} \slashed{e}_X (1- \gamma_5)  \nu &= -2 \sqrt{q^2} \beta_\ell \sin\theta_\ell\,, \nonumber \\
 \bar\ell_{h=-}  (1- \gamma_5)  \nu  &= 0\,,
 \end{align}
where $\beta_\ell = \sqrt{1 - m^2_\ell/q^2}$, and the auxiliary polarization vector $e_X$ is defined as:
\begin{equation}
|\vec{P}| e^\mu_X  \equiv  P^\mu - \frac{P\cdot q }{q^2} q^\mu\,, ~~\epsilon^\mu_X \epsilon_{X \mu} =-1 \,, ~~|\vec{P}|= \sqrt{\frac{\lambda_P}{q^2}} \,. \nonumber
 \end{equation} 
In order to include the $V$-meson polarizations in the $\bar B \to V \ell \bar\nu$ decay, we separate a lepton current  in the lepton helicity basis into  longitudinal and transverse parts, where the longitudinal part of the $V$-meson is given as: 
\begin{align}
  \bar\ell_{h=+} \slashed{e}_Z (1- \gamma_5)  \nu  &= 2m_\ell \beta_\ell  \cos\theta_\ell \,, \\
    \bar\ell_{h=-} \slashed{e}_Z (1- \gamma_5)  \nu &= - 2\sqrt{q^2} \beta_\ell \sin\theta_\ell\,, 
\end{align}
while the two transverse parts of the $V$-meson are respectively given as:
\begin{align}
\bar\ell_{h=+} \slashed{e}_{V}(T) (1- \gamma_5)  \nu  &=  -2 m_\ell \beta_\ell  \left\{
\begin{array}{c}
 \frac{i}{\sqrt{2}} \sin\theta_\ell  e^{-i \phi}  ~(T=+) \,,   \\
   \frac{i }{\sqrt{2}} \sin\theta_\ell  e^{i \phi}  ~ (T= -) \,,
  \end{array}
   \right.
\end{align}

\begin{align}
\bar\ell_{h=-} \slashed{e}_{V}(T) (1- \gamma_5)  \nu &=  -2\sqrt{q^2} \beta_\ell  \left\{
\begin{array}{c}
 \frac{- i }{\sqrt{2}} (1- \cos\theta_\ell)  e^{-i \phi}  ~(T=+) \,,   \\
   \frac{i }{\sqrt{2}} (1 + \cos\theta_\ell ) e^{i \phi}  ~ (T= -) \,.
  \end{array}
   \right.
\end{align}
The auxiliary polarizations $e_Z$ and $e_V(T)$ are defined as:
\begin{equation}
 \frac{E_{V}}{m_{V}} e^\mu_Z  \equiv \epsilon^\mu_V (L) - \frac{\epsilon \cdot q }{q^2} q^\mu \,, 
 \quad \sqrt{\frac{\lambda_{V}}{2}} e^\mu_{V}(T)  \equiv  \varepsilon^{\mu \nu \rho \sigma} \epsilon_{V\nu}(T) P_\rho q_\sigma\,. \nonumber
 \end{equation}

 Using the helicity basis and the lepton currents discussed before,  the $\bar B \to P \ell \bar\nu$ decay amplitudes with  the charged-lepton positive and negative helicity are respectively obtained as:
 \begin{align}
 {\cal M}^{ h=+}_{P} & = \frac{G_F V_{qb}}{\sqrt{2}} 
 \left( 2 m_\ell \beta_\ell \frac{\sqrt{\lambda_{P}}}{\sqrt{q^2}} f^{BP}_1 \cos\theta_\ell - 2\beta_\ell \sqrt{q^2} X^{0\ell}_{P}\right)\,, \\
 {\cal M}^{h=-}_{P} & = \frac{G_F V_{qb}}{\sqrt{2}} \left( -2  \beta_{\ell}  \sqrt{\lambda_{P}} f^{BP}_1 \sin\theta_\ell \right)
\,, \\
 X^{0\ell}_P & =  \frac{m^2_B - m^2_P}{q^2}  m_\ell f^{BP}_0 + (m_B + m_P) \left( C^{R,\ell}_{qb} + C^{L,\ell}_{qb}\right) f^{BP}_S \,. \label{eq:X0P}
 \end{align}
As mentioned earlier, since the $V$-meson  carries spin degrees of freedom, we separate each lepton helicity amplitude into longitudinal (L) and transverse (T) parts to show  the $V$-meson polarization effects. Therefore, we write the helicity amplitudes of $\bar B\to V \ell \bar\nu$ for the longitudinal polarization of the $V$-meson as:
\begin{align}
{\cal M}^{L,h=+}_{V} & = -i\frac{G_F V_{qb}}{\sqrt{2}} \left(2 m_\ell \beta_\ell h^0_{V} \cos\theta_\ell - 2 \beta_\ell \frac{\sqrt{\lambda_{V}}}{ \sqrt{q^2} } X^{0\ell}_{V}\right) \,, \label{eq:ALV}\\
{\cal M}^{L,h=-}_{V} & =-i\frac{G_F V_{qb}}{\sqrt{2}} \left( -2\sqrt{q^2} \beta_\ell  h^0_{V} \sin\theta_\ell \right) \,, \\
 h^0_{V}(q^2) &= \frac{ 1}{2 m_{V} \sqrt{q^2}}\left[(m_B^2-m_{V}^2-q^2)(m_B+m_{V})A^{BV}_1-
\frac{{\lambda_{V} }}{m_B+m_{V}}A^{BV}_2\right]\,, \nonumber \\
X^{0\ell}_{V} & = m_\ell A^{BV}_0 + \frac{q^2}{2m_V } (C^{R,\ell}_{qb} - C^{L,\ell}_{qb}) f^{BV}_P\,.  \label{eq:X0V}
\end{align}
It can be seen that the formulae for ${\cal M}^{L,h=\pm}_{V}$ are similar to those for ${\cal M}^{ h=\pm}_{P}$. The helicity  amplitudes for the transverse polarizations of $V$-meson can be derived as:
\begin{align}
{\cal M}^{T=\pm, h=+}_{V} &= i\frac{G_F V_{qb}}{\sqrt{2}}  \left[-\sqrt{2} m_{\ell} \beta_{\ell}  \sin\theta_\ell e^{\mp i \phi} \right]  h^{\pm}_{V} \,, \\
{\cal M}^{T=\pm, h=-}_{V} &= \mp i \frac{G_F V_{qb}}{\sqrt{2}}  \left[-\sqrt{2} \sqrt{q^2} \beta_\ell (1\mp \cos\theta_\ell)  e^{\mp i \phi} \right]h^{\pm}_{V} \,, \\
h^{\pm}_{V} & = \frac{\sqrt{\lambda_{V}} }{m_B + m_{V}} V^{BV} \mp (m_B + m_{V} ) A^{BV}_1\,. \nonumber
\end{align}
Since the charged-Higgs only affects the longitudinal part, ${\cal M}^{T=\pm, h=\pm}_{V}$  are dictated by   the SM.  From  these obtained helicity amplitudes, it can be seen that due to angular-momentum conservation, ${\cal M}^{ h=+}_{P}$ and ${\cal M}^{L(T), h=+}_{V}$, which come from $\bar\ell \gamma_\mu (1-\gamma_5)\nu$,  are chirality-suppressed and proportional to $m_\ell$.  However, the charged lepton in $\bar\ell (1-\gamma_5) \nu$, which arises from the charged-Higgs interaction, prefers the $h=+$ state, and the associated contribution in principle exhibits no  chiral suppression factor. Nevertheless,  the $m_\ell$ factor indeed exists in our case due to the Cheng-Sher ansatz. 

 \subsubsection{ Angular differential decay rate, lepton helicity asymmetry, and forward-backward asymmetry}

When  the three-body phase space is included, the differential decay rates with  lepton helicity and $V$ polarization as a function of $q^2$ and $\cos\theta_\ell$ can be obtained as:
\begin{align}
 \frac{d\Gamma^{h=\pm}_{P\ell }}{dq^2 d\cos\theta_\ell} & = \frac{\sqrt{\lambda_{P}}}{512 \pi^3 m^3_B} \beta^2_{\ell}\,  |{\cal M}^{h=\pm }_{P}|^2   \,, \nonumber \\ 
 \frac{d\Gamma^{L(T), h=\pm}_{V\ell }}{dq^2 d\cos\theta_\ell}  & = \frac{\sqrt{\lambda_{V}}}{512 \pi^3 m^3_B} \beta^2_{\ell}\,  |{\cal M}^{L(T),h=\pm }_{V}|^2  \,. \label{eq:ang_Ga}
\end{align}
 Using Eq.~(\ref{eq:ang_Ga}), we can investigate various interesting physical quantities, such as BR, lepton-helicity asymmetry, lepton forward-backward asymmetry (FBA), and polarization distributions of $V$-meson. We thus introduce these observables in the following discussions. 
 
 When the polar angle is integrated out, the differential decay rate with each lepton helicity as a function of $q^2$  can be obtained as follows: For the $\bar B \to P \ell \bar \nu$ decay,  they can be expressed as:
 \begin{align}
 \frac{d \Gamma^{h=\pm}_{P \ell}}{dq^2 } & = \frac{G^2_F |V_{qb}|^2 \sqrt{\lambda_P} \beta^4_\ell}{256 \pi^3 m^3_B} H^{\pm}_{P}\,, \label{eq:Gah_P}\\
 H^+_{P} &= \frac{2 m^2_\ell }{3 q^2 }\lambda_P  (f^{BP}_1)^2  + 2  q^2 |X^{0\ell}_P|^2\,, ~
 H^-_{P}  =\frac{4}{3} \lambda_P (f^{BP}_1)^2 \,; \nonumber 
 \end{align}
and for the $\bar B\to V \ell \nu$ decay, they are shown as:
  \begin{align}
 \frac{d \Gamma^{\lambda, h=\pm}_{V\ell}}{dq^2 } & = \frac{G^2_F |V_{qb}|^2 \sqrt{\lambda_{V}} \beta^4_\ell}{256 \pi^3 m^3_B} H^{\lambda, \pm}_{V}\,, \label{eq:Gah_V} \\
 H^{L,+}_{V} &= \frac{2 m^2_\ell  }{3 } |h^0_{V}|^2 +  \frac{2 \lambda_{V} }{q^2}  |X^{0\ell}_{V}|^2\,, ~ H^{L,-}_{V}  =\frac{4 q^2 }{3}  |h^0_{V}|^2\,, \nonumber \\
 H^{T=\pm,+}_{V} &= \frac{2 m^2_\ell  }{3 } |h^{\pm}_{V}|^2 \,, ~ H^{T=\pm,-}_{V} = \frac{4 q^2 }{3 } |h^{\pm}_{V}|^2\,. \nonumber
 \end{align}
 Accordingly, the partial decay rates  for $\bar B \to (P, V) \ell \bar\nu$ can be directly obtained as:
  \begin{align}
  \Gamma_{P\ell}& = \frac{G^2_F |V_{qb}|^2 }{256 \pi^3 m^3_B}  \int^{q^2_{\rm max}} _{m^2_\ell}  dq^2    \sqrt{\lambda_P} \beta^4_\ell \left( H^{+}_{P} + H^{-}_{P} \right)\   \,, \nonumber \\
   \Gamma_{V\ell}& =  \frac{G^2_F |V_{qb}|^2 }{256 \pi^3 m^3_B}  \int^{q^2_{\rm max}} _{m^2_\ell}  dq^2  \sqrt{\lambda_{V}} \beta^4_\ell \sum_{\lambda=L,T=\pm}\left( H^{\lambda, +}_{V} + H^{\lambda,-}_V \right)\,. \label{eq:GammaBPV}
  \end{align}
  Moreover, the $q^2$-dependent longitudinal polarization and transverse polarization fractions can be defined as:
  \begin{align}
  f^{L}_{V\ell}(q^2) & = \frac{\sum_{h} d\Gamma^{L,h}_{V\ell}/dq^2 }{\sum_{\lambda, h}  d\Gamma^{\lambda,h}_{V\ell} /dq^2} = \frac{H^{L,+}_V + H^{L,-}_V}{\sum_{\lambda,h} H^{\lambda,h}_V} \,, \label{eq:fL}\\
  f^{T}_{V\ell} (q^2)&=  \frac{\sum_{T,h} d\Gamma^{T,h}_{V\ell}/dq^2 }{\sum_{\lambda, h}  d\Gamma^{\lambda,h}_{V\ell} /dq^2} = \frac{\sum_{T=\pm} \left( H^{T,+}_V + H^{T,-}_V \right)}{\sum_{\lambda,h} H^{\lambda,h}_V}\,. \label{eq:fT}
  \end{align}

Based on Eqs.~(\ref{eq:Gah_P}) and (\ref{eq:Gah_V}), we define the $q^2$-dependent lepton helicity asymmetry as:
 \begin{equation}
 {\cal P}^\ell_{M} (q^2) = \frac{d\Gamma^{h=+}_{M\ell}/dq^2  - d\Gamma^{h=-}_{M\ell}/q^2}{d\Gamma^{h=+}_{M\ell} /dq^2+ d\Gamma^{h=-}_{M\ell}/dq^2}\,,
 \end{equation}
 where  the sum of $V$ polarizations is indicated in $d\Gamma^{h=\pm }_{V\ell}$. Thus, the results for the pseudoscalar and vector meson processes can be respectively formulated as:
  \begin{align}
   {\cal P}^\ell_{P} (q^2) &  = \frac{ \frac{2}{3} \left(m^2_{\ell} -2 q^2 \right) \lambda_P  ( f^{BP}_1)^2 /q^2 + 2  q^2  |X^{0\ell}_P|^2 }{ \frac{2}{3}   \left( m^2_{\ell}+ 2 q^2 \right) \lambda_P   (f^{BP}_1)^2/q^2 + 2  q^2 |X^{0\ell}_P|^2}\,,   \\
   {\cal P}^\ell_V (q^2) & = \frac{ \frac{2}{3} (m^2_\ell -2 q^2) \left( \sum_{\lambda=L,\pm} |h^{\lambda}_{V}|^2\right) + 2 \lambda_{V}/q^2 |X^{0\ell}_{V}|^2 }{ \frac{2}{3} (m^2_\ell + 2 q^2) \left( \sum_{\lambda=L,\pm} |h^{\lambda}_{V}|^2\right) + 2 \lambda_{V}/q^2 |X^{0\ell}_{V}|^2 }\,.
  \end{align}
In addition, using the helicity decay rates, the $q^2$-independent lepton helicity asymmetry can be defined as~\cite{Kalinowski:1990ba,Tanaka:2010se,Tanaka:2012nw,Datta:2012qk,Chen:2017eby}:
 \begin{align}
 P^{\ell}_{M} & = \frac{\Gamma^{h= +}_{M\ell} - \Gamma^{h=-}_{M\ell}}{ \Gamma^{h= +}_{M\ell} +  \Gamma^{h=-}_{M\ell} }\,,
  \end{align}
 where   the formulations for $\bar B\to (P, V) \ell \bar\nu$ with charged Higgs effects  can be found as: 
\begin{align}
 P^{\ell}_{P} & = \frac{\int^{q^2_{\rm max}}_{m^2_\ell} dq^2 \sqrt{\lambda_P} \beta^4_\ell \left[ \frac{2}{3} \left(m^2_{\ell} -2 q^2 \right) \lambda_P  ( f^{BP}_1)^2 /q^2 + 2  q^2  |X^{0\ell}_P|^2\right]}{\int^{q^2_{\rm max}}_{m^2_\ell} dq^2 \sqrt{\lambda_P} \beta^4_\ell \left[ \frac{2}{3}   \left( m^2_{\ell}+ 2 q^2 \right) \lambda_P   (f^{BP}_1)^2/q^2 + 2  q^2 |X^{0\ell}_P|^2\right]}\,,  \label{eq:PtauP}  \\
 P^{\ell}_{V} & =  \frac{\int^{q^2_{\rm max}}_{m^2_\ell} dq^2 \sqrt{\lambda_{V}} \beta^4_\ell\left[ \frac{2}{3} (m^2_\ell -2 q^2) \left( \sum_{\lambda=L,\pm} |h^{\lambda}_{V}|^2\right) + 2 \lambda_{V}/q^2 |X^{0\ell}_{V}|^2 \right]}{\int^{q^2_{\rm max}}_{m^2_\ell} dq^2 \sqrt{\lambda_{V}} \beta^4_\ell \left[ \frac{2}{3} (m^2_\ell + 2 q^2) \left( \sum_{\lambda=L,\pm} |h^{\lambda}_{V}|^2\right) + 2 \lambda_{V}/q^2 |X^{0\ell}_{V}|^2 \right]}\,. \label{eq:PtauV}
 \end{align}

From the angular differential decay rates shown in Eq.~(\ref{eq:ang_Ga}), the lepton FBA can be defined as:
\begin{equation}
 A^{M,\ell}_{FB} (q^2) = \frac{\int^{1}_{0} dz (d\Gamma_{M\ell}/dq^2dz) - \int^{0}_{-1} dz (d\Gamma_{M\ell}/dq^2dz)}{\int^{1}_{0} dz (d\Gamma_{M\ell}/dq^2dz) + \int^{0}_{-1} dz (d\Gamma_{M\ell}/dq^2dz)}\,,
 \end{equation}
where $z=\cos\theta_\ell$ and $d\Gamma_{M\ell}/(dq^2 dz)$ have included all possible lepton helicities and polarizations of the $V$-meson. The FBAs mediated by the  charged Higgs and $W$-boson  in $\bar B \to (P, V) \ell \bar \nu_\ell$ are obtained as:
 \begin{align}
 A^{P,\ell}_{FB} (q^2)&=  - \frac{2 m_{\ell} \sqrt{\lambda_P} f^{BP}_1 Re(X^{0\ell *}_P)}{H^+_{P} + H^-_{P}}\,, \nonumber \\
  A^{V, \ell}_{FB} (q^2) &=  \frac{1}{\sum_{\lambda=L,\pm} (H^{\lambda, +}_{V} +H^{\lambda, -}_{V})} \left[ -2 m_\ell \frac{\sqrt{\lambda_{V}}}{\sqrt{q^2}}
 Re(h^0_{V} X^{0\ell *}_{V})   +4 q^2 \sqrt {\lambda_{V}} A^{BV}_1 V^{BV} \right]. \label{eq:fba}
 \end{align}
From the above equations, it can be seen that   $A^{P,\ell}_{FB}$ and the longitudinal part of $A^{V,\ell}_{FB}$  depend on $m_\ell$ and are chiral suppressed.  Since  $m_\tau/m_b \sim 0.4$  is not highly suppressed, it can be expected that   $\bar B \to P \tau \bar \nu$ can have a sizable FBA. $A^{V,\ell}_{FB}$ does not vanish in the chiral limit; therefore,  it can be sizable for a light lepton.  

 The observations of the tau polarization and FBA  rely on tau-lepton reconstruction. Due to  the involvement of one invisible neutrino in the final state, it is experimentally  challenging to measure these observables.  As an alternative to the $\tau$ reconstruction, the extraction of $\tau$ polarization and FBA through an angular asymmetry of visible particles in  a tau decay was recently proposed in~\cite{Nierste:2008qe, Alonso:2017ktd}, where  the $\tau \to \pi \nu_\tau$ decay is the most sensitive channel. Using this approach,  a statistical precision of $10\%$ can be reached at  Belle II with an integrated luminosity of 50 ab$^{-1}$. The detailed study  can be found in~\cite{Alonso:2017ktd}. 

\section{ $\Delta B=2$ and $b\to s \gamma$ processes in the generic 2HDM}

It is known that  tree-level FCNCs can occur in the generic 2HDM; therefore, the measured mass difference $\Delta M_{q'}$ $(q'=d,s)$ of neutral $B_{q'}$-meson will give a strict limit on the  parameters $X^d_{q'b, bq'}$. In our approach, due to the   Cheng-Sher ansatz, the $\Delta B=2$ process,  mediated by the neutral scalars at the tree level,  is proportional to $m_{q'}m_b t^2_\beta/ v^2 (\chi^d_{q' b})^2$. Although the tree-level effect has a suppression factor $m_{q'}/v$, the  factor $t^2_\beta$ can largely enhance its contribution; hence,  $\Delta M_{q'}$ will  severely  bound the  $\chi^d_{q'b, bq'}$ parameters.  

 In addition to the tree-level effects, we find through box diagrams that  the charged-Higgs contributions to $\Delta B=2$  can be significant when $t_\beta$ is large, and $\chi^u_{tt,ct}$ and $\chi^d_{bb}$ are of $O(0.1)$-$O(1)$.  The same charged-Higgs effects also contribute to  the radiative $b\to s(d) \gamma$ decay via penguin diagrams. Since $b\to s \gamma$ is measured well in experiments, in this study, we only focus on the $b\to s \gamma$ decay. It is of interest to investigate whether the sizable new parameters $\chi^{u}_{tt,ct}$ and $\chi^d_{bb}$  in the generic 2HDM can accommodate the  $\Delta M_{q'}$ and $b\to s \gamma$ data.  Hence,  in this section, we formulate the contributions of charged-Higgs and neutral Higgses to the $B_{d, s}$-$\bar B_{d,s}$ mixings and $b\to s \gamma$ process.

\subsection{ Charged-Higgs contributions to the $\Delta M_{q'}$ }

\begin{figure}[phtb]
\includegraphics[scale=0.85]{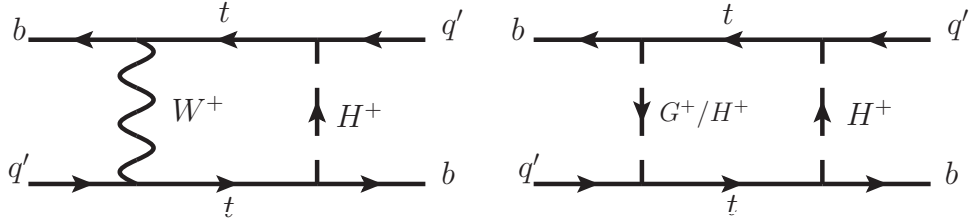}
\caption{The representative box diagrams for the $ B_{q'}$-$\bar{B}_{q'}$ mixing with the intermediates of $W^{+}$-$H^{+}$, $G^+$-$H^+$, and $H^{+}$-$H^+ $, where $G^+$ is the charged Goldstone boson.}
\label{fig:boxes}
\end{figure}

We first consider the charged-Higgs contributions to the $\Delta B=2$ processes, where the typical Feynman diagrams mediated by $W^+$-$H^+$, $G^+$-$H^+$, and $H^+$-$H^+$ are sketched in Fig.~\ref{fig:boxes}, and $G^+$ is the charged Goldstone boson. Since  the  Yukawa couplings of $H^\pm$ to the quarks are associated with the quark masses, the vertices that involve heavy quarks can enhance the loop $H^\pm$ effects. Thus, we only consider the top-quark loop contributions in the $B$-meson system. Accordingly, the relevant charged-Higgs interactions  are shown as:
 \begin{align}
 {\cal L}^{H^\pm}_Y & \supset \frac{\sqrt{2}}{v} V_{tb} \bar t \left( m_t \zeta^u_{tt} P_L + m_b \zeta^d_{bb} P_R \right) b  H^+  \nonumber \\
 & + \frac{\sqrt{2}}{v} V_{tq'} \bar t  \left( m_t \zeta^u_{tq'} P_L  -  m_b \zeta^d_{tq'} P_R\right) q'  H^+ +H.c.,  \label{eq:btoqpp}
 \end{align}
where  the parameters $\zeta^{f}_{ij}$ are defined as:
\begin{align}
\zeta^u_{tt} & \approx  \frac{1}{t_\beta} -\frac{\chi^{u*}_{tt} }{s_\beta}   \,, \  \ \zeta^d_{bb}  \approx  t_\beta \left( 1- \frac{\chi^d_{bb}}{ s_\beta} \right) \,, \nonumber \\
\zeta^u_{tq'} & \approx  \frac{1}{t_\beta} - \frac{\chi^L_{tq'} }{s_\beta}\,, \ \zeta^{d}_{tq'} = t_\beta\left( \sqrt{\frac{m_{q'}}{m_b}} \frac{\chi^d_{bq'}}{s_\beta} \frac{V_{tb}}{V_{tq'}}\right) \,, \nonumber \\
\chi^L_{tq'} & =\chi^{u*}_{tt} + \sqrt{\frac{m_c}{m_t}} \frac{V_{cq'}}{V_{tq'}} \chi^{u*}_{ct}\,.
  \label{eq:zetas}
\end{align}
Detailed discussions for  the couplings of $tq'H^\pm$ can be found in the Appendix. 
From Eqs.~(\ref{eq:btoqpp}) and (\ref{eq:zetas}),  when $\chi^f_{ij}=0$,  the vertices in the type-II 2HDM are reproduced. Unlike the type-II model, where $\zeta^u_{tt, tq'}\ll 1$ for $t_\beta \sim m_t/m_b$,  $\zeta^u_{tt, tq'}$  in the type-III model can be of order unity even at small $t_\beta$.  We will show the impacts of  these new 2HDM parameters  on the flavor physics in the following analysis. 

Based on the convention in~\cite{Becirevic:2001jj}, the effective Hamiltonian for $B_{q'}$-$\bar B_{q'}$ mixing can be written as:
 \begin{equation}
 H^{\Delta B=2}_{\rm eff} = \frac{ G^2_F \left( V^*_{tb} V_{tq'}\right)^2 }{16 \pi^2} m^2_W  \left( \sum^5_{i=1}  C_i (\mu) Q_{i}  + \sum^3_{i=1} \tilde C_i (\mu) \tilde{Q}_i \right)\,, 
 \end{equation}
where the effective operators with the color indices $\alpha,\beta$ are given as:
 \begin{align}
 Q_1 & = \left(\bar b^\alpha \gamma_\mu P_L q'^\alpha \right)  \left(\bar b^\beta \gamma^\mu P_L q'^\beta \right) \,, \nonumber \\
 Q_2 &=  \left(\bar b^\alpha P_L q'^\alpha \right)  \left(\bar b^\beta P_L q'^\beta  \right) \,,  \ Q_3  = \left(\bar b^\beta P_L q'^\alpha \right)  \left(\bar b^\alpha P_L q'^\beta \right) \,, \nonumber \\
 Q_4 &= \left( \bar b^\alpha P_L q'^\alpha \right) \left( \bar b^\beta P_R q'^\beta \right)\,, \  Q_5 = \left( \bar b^\beta P_L q'^\alpha \right) \left( \bar b^\alpha P_R q'^\beta \right)\,.
 \end{align}
 The operators $\tilde O_{j}$ can be obtained from $O_{j}$  using $P_R$ instead of $P_L$. The Wilson coefficients at the scale $\mu=m_b=4.6$ GeV can be related to those at $\mu_H$ scale and are given as~\cite{Becirevic:2001jj}:
 \begin{equation}
 C_i(m_b) \approx \sum_{k,j} \left( b^{(i,j)}_k + \eta c^{(i,j)}_k \right) \eta^{a_k} C_j(\mu_H)\,, \label{eq:mbWCs}
 \end{equation}
 where $\mu_H = m_{H^\pm}$, $\eta = \alpha_s(\mu_H)/\alpha_s(m_t)$, $C_j(\mu_H)$ are the Wilson coefficients at $\mu_H$ scale, and the magic numbers for $a_k$, $b^{(i,j)}_k$, and $c^{(i,j)}_k$ can be found in~\cite{Becirevic:2001jj}. To obtain $C_j(\mu_H)$, we adopt the 't Hooft-Feynman gauge for the propagator of $W$-gauge boson;  therefore,  the charged Goldstone $G^\pm$ boson effects have to be taken into account. To show the results of the box diagrams, we define some useful parameters as:  $x_t=m^2_t/m^2_W$, $y_t=m^2_t/m^2_{H^\pm}$, $y_W=m^2_W/m^2_{H^\pm}$, and $y_b = m^2_b/m_{H^\pm}^2$. Thus, the effective Wilson coefficients at $\mu_H$ scale can be  formulated as:
  \begin{subequations}
 \begin{align}
 C_1(\mu_H) & = 4 \zeta^u_{tq'} \zeta^{u*}_{tt} \left( 2 y^2_t I^{WH}_1(y_t,y_W) + x_t y_t I^{WH}_{2}(y_t,y_W) \right) + 2 \left( \zeta^{u}_{tq'} \zeta^{u*}_{tt} \right)^2  x_t y_t I^{HH}_1(y_t) \,,  \\
 C_2(\mu_H) & = -4 \left( \zeta^{u}_{tq'} \zeta^{d*}_{bb} \right)^2   x_b y^2_t  I^{HH}_{2}(y_t) \,,  \\
 C_4(\mu_H) & = 8 \frac{m^2_b}{m^2_t} \zeta^d_{tq'} \zeta^{d*}_{bb} \left( 2 y_t I^{WH}_{2}(y_t, y_W) + x_t y^2_t I^{WH}_1(y_t,y_W)\right)  \nonumber \\
 &+ 8 (\zeta^d_{tq'} \zeta^{u*}_{tt} )(\zeta^u_{tq'} \zeta^{d*}_{bb}) x_b y^2_t I^{HH}_2 (y_t) \,,  \label{eq:C4muH}\\
 C_5(\mu_H) & = -8 (\zeta^u_{tq'} \zeta^{u*}_{tt}) ( \zeta^d_{tq'} \zeta^{d*}_{bb}) x_b y_t I^{HH}_1(y_t)\,,
 \end{align}
 \end{subequations}
 where the loop integral functions are defined as:
 \begin{subequations}
  \begin{align}
  I^{WH}_1 (y_t, y_W) & =  \int^1_0 dx_1 \int^{x_1}_0 dx_2 \frac{x_2}{(1-x_1 +y_t x_2 + y_W (x_1 -x_2))^2} \,, \\
  I^{WH}_2 (y_t, y_W) & = \int^1_0 dx_1 \int^{x_1}_0 dx_2 \frac{x_2}{1-x_1 +y_t x_2 + y_W (x_1 -x_2)}\,, \\
  I^{HH}_1(y_t) & = \int^1_0 dx \frac{x(1-x)}{1-x +y_t x }\,, \\
   I^{HH}_2(y_t) & = \int^1_0 dx \frac{x(1-x)}{(1-x +y_t x)^2}\,.
  \end{align}
  \end{subequations}
  The effective Wilson coefficients for the $\tilde O_{1,2}$ operators at $\mu_H$ scale are given as:
 \begin{align}
 \tilde{C}_1 (\mu_H) & = 2 \left( \zeta^d_{tq'} \zeta^{d*}_{bb}\right)^2 x_b y_b I^{HH}_1(y_t) \,, \nonumber \\
 \tilde{C}_2(\mu_H) &= -4 \left( \zeta^d_{tq'} \zeta^{u*}_{tt} \right)^2 x_b y^2_t I^{HH}_2(y_t) \,.
 \end{align}
We have checked that  our results are the same as those obtained in~\cite{Urban:1997gw}  when $y_b=\chi^{u,d}_{ij}=0$.
Using  Eq.~(\ref{eq:mbWCs}) and the magic numbers shown in~\cite{Becirevic:2001jj}, we obtain the Wilson coefficients  $C_i(m_b)$ at $\mu=m_b$ scale as:
 \begin{align}
& C_1(m_b) \approx 0.848 C_1(\mu_H)\,, \ C_2(m_b) \approx 1.708 C_2(\mu_H)\,, \ C_3(m_b) \approx -0.016 C_2(\mu_H)\,, \nonumber \\
 & C_{4}(m_b) \approx 2.395 C_{4}(\mu_H) + 0.431 C_{5} (\mu_H)\,, \  C_{5}(m_b) \approx 0.061 C_4(\mu_H) + 0.904 C_5(\mu_H)\,.
 \end{align}

The matrix elements of the renormalized operators for $\Delta B=2$ are defined as~\cite{Becirevic:2001jj}:
 \begin{subequations}
 \begin{align}
 \langle B_{q'} | \hat Q_1(\mu)  |  \bar B_{q'} \rangle   &= \frac{1}{3} f^2_{B_{q'}} m_{B_{q'}} B_{1q'}(\mu) \,,  \\
  \langle B_{q'} | \hat Q_2(\mu)  | \bar B_{q'}\rangle  &= -\frac{5}{24} \left(\frac{m_{B_{q'}}}{m_b(\mu) + m_{q'} (\mu)} \right)^2  f^2_{B_q} m_{B_{q'}} B_{2{q'}}(\mu) \,,  \\
   \langle B_{q'} | \hat Q_3(\mu)  | \bar B_{q'}\rangle  &= \frac{1}{24} \left(\frac{m_{B_{q'}}}{m_b(\mu) + m_{q'} (\mu)} \right)^2  f^2_{B_{q'}} m_{B_{q'}} B_{3q'}(\mu) \,,  \\
   \langle B_{q'} | \hat Q_4(\mu)  | \bar B_{q'}\rangle  &= \frac{1}{4} \left(\frac{m_{B_{q'}}}{m_b(\mu) + m_{q'} (\mu)} \right)^2  f^2_{B_{q'}} m_{B_{q'}} B_{4q'}(\mu) \,,   \\
   \langle B_{q'} | \hat Q_5(\mu)  | \bar B_{q'}\rangle  &= \frac{1}{12} \left(\frac{m_{B_{q'}}}{m_b(\mu) + m_{q'} (\mu)} \right)^2  f^2_{B_{q'}} m_{B_{q'}} B_{5q'}(\mu) \,,
 \end{align}
 \label{eq:BBmatrix}
 \end{subequations}
 where $B_{iq'}$ denote the nonperturbative QCD bag parameters, and the mixing matrix elements in the SM are related to  $B_{1q'}$. Using the results obtained by HPQCD~\cite{Gamiz:2009ku}, FNAL-MILC~\cite{Bazavov:2012zs}, and RBC-UKQCD~\cite{Aoki:2014nga} collaborations, the  lattice QCD results  with $N_f=2+1$ averaged by the flavor lattice averaging group (FLAG) can be found as $B_{1d} \approx 0.80$ and $B_{1s}\approx 0.84$~\cite{Aoki:2016frl}. In our numerical calculations,  the quark masses and $B_{iq'}$ parameters at the $m_b$ scale  in the Landau RI-MOM scheme~\cite{Becirevic:2001jj,Becirevic:2001xt,Becirevic:2001yv,Carrasco:2013zta} and the decay constants of $B_{q'}$ are shown in Table~\ref{tab:BPs}, where for self-consistency, all $B_{iq}$ values are quoted from~\cite{Carrasco:2013zta}. Due to $B_{is}\approx B_{id}$, we adopt $B_{is}= B_{id}\equiv B_{iq'}$.  As a result,  $\langle B_{q'}| H^{\rm \Delta B=2}_{\rm eff} | \bar B_{q'} \rangle $ can be written as:
 \begin{equation}
\langle B_{q'}| H^{\rm \Delta B=2}_{\rm eff} | \bar B_{q'} \rangle =  \langle B_{q'}| H^{\rm \Delta B=2}_{\rm eff} | \bar B_{q'} \rangle^{\rm SM} \left( 1 + \Delta^{H^\pm}_{q'} \right)\,.
\end{equation}
The SM result and the charged-Higgs contributions can be formulated as:
 \begin{align}
\langle B_{q'}| H^{\rm \Delta B=2}_{\rm eff} | \bar B_{q'} \rangle ^{\rm SM} & = \frac{G^2_F (V^*_{tb} V_{tq'})^2 }{48\pi^2} m^2_W f^2_{B_{q'}}  m_{B_{q'}} \hat{\eta}_{1B} B_{1q'} ( 4S_0(x_t) )\,, \nonumber \\
 \Delta^{H^\pm}_{q'}  & = \frac{1}{4 S_0(x_t)}   \left\{ C_1 (\mu_H)+ \tilde{C}_1(\mu_H) +  \frac{ m^2_{B_{q'}}   }{8 (m_b + m_{q'})^2 \hat{\eta}_{1B} B_{1q'} }  \right. \nonumber \\
 & \times \left[ \left( -5 \hat{\eta}_{2B} B_{2q'} + \hat{\eta}_{3B} B_{3q'} \right) \left( C_2(\mu_H) +  \tilde{C}_2(\mu_H)  \right)  \right. \nonumber \\
 & + \left( 6 \hat\eta_{44B} B_{4q'} + 2 \hat\eta_{45B} B_{5q'}\right) C_4(\mu_H)  \nonumber \\
 & \left. \left.    + \left( 6 \hat\eta_{54B} B_{4q'} + 2\hat\eta_{55B} B_{5q'}\right) C_5(\mu_H)  \right]
 \right\}\,, 
  \end{align}
where $4 S_0(m^2_t/m^2_W)=3.136 (m^2_t/m^2_W)^{0.76} \approx 9.36$~\cite{Buchalla:1995vs}; $\hat{\eta}_{iB}$ are the QCD corrections, and their values are shown in Table~\ref{tab:BPs}. Accordingly, the mass difference between the physical $B_{q'}$ states can be obtained by:
 \begin{equation}
 \Delta M^{H^\pm}_{q'} = 2 | \langle B_{q'}| H^{\rm \Delta B=2}_{\rm eff} | \bar B_{q'} \rangle|=\Delta M^{\rm SM}_{q'} |1+\Delta^{H^\pm}_{q'}|\,. 
 \label{eq:DMq}
 \end{equation}
 
 \begin{table}[htp]
\caption{Values of quark masses , $B_{iq}$ parameters, and $\hat\eta_{iB}$  at $m_b$ scale in the RI-MOM scheme,  where  the  $B_{iq}$ results are quoted from~\cite{Carrasco:2013zta}. The decay constants of the $B_{d,s}$ mesons are from~\cite{Lenz:2010gu}, and $f_{B_c}$ is from~\cite{Colquhoun:2015oha}.}
\begin{tabular}{ccccccccc}  \hline \hline
 $m_b$ & $m_s$ & $m_{q'}$ & $f_{B_s}$ & $f_{B_d}$ & $f_{B_c}$ & $B_{1q'}$ & $B_{2q'}$ & $B_{3q'} $   \\ \hline 
 ~ 4.6GeV~& ~0.10GeV~ & 5.4MeV & 0.231GeV &  0.191GeV  & 0.434 GeV & ~0.84~ & ~ 0.88 ~ & ~ 1.10 ~ \\ \hline
  $B_{4q'}$ & $B_{5q'}$  & $\hat\eta_{1B}$  & $\hat\eta_{2B}$ & $\hat\eta_{3B}$  & $\hat\eta_{44B}$ & $\hat\eta_{45B}$ & $\hat\eta_{54B}$ & $\hat\eta_{55B}$\\ \hline
 ~1.12 ~& ~1.89~& 0.848 & 1.708 & $-0.016$ & 2.395 & 0.061  & 0.431  & 0.094 \\ \hline \hline 
  
  \end{tabular}
\label{tab:BPs}
\end{table}%

 Taking $V_{td} \approx 0.0082 e^{- i \phi_1}$ with $\phi_1 \approx 21.9^{\circ}$, $V_{ts} \approx -0.04$, and $m_t= \bar m_t (\bar m_t) \approx 165$ GeV, the $B_{q'}$-meson oscillation parameters $\Delta M_{d,s}$ in the SM are respectively estimated as:
 \begin{align}
 \Delta M^{\rm SM}_d &  \approx 3.20\times 10^{-13}\ \text{GeV} = 0.487\; {\rm ps}^{-1} \,, \nonumber \\
 \Delta M^{\rm SM}_s & \approx  1.13 \times 10^{-11}\ \text{GeV} = 17.22\; {\rm ps}^{-1}\,, 
 \end{align}
where the current data are $\Delta M^{\rm exp}_d = (0.5065 \pm 0.0019)$ ps$^{-1}$ and $\Delta M^{\rm exp}_s = (17.756 \pm 0.021)$ ps$^{-1}$~\cite{PDG}.  In order to include the new physics contributions, when we use the $\Delta M^{\rm exp}_{q'}$ to bound the free parameters, we take the SM predictions to be $\Delta M^{\rm SM}_d= 0.555^{+0.073}_{-0.046}$ ps$^{-1}$ and $\Delta M^{\rm SM}_s=16.8^{+2.6}_{-1.5}$ ps$^{-1}$~\cite{Lenz:2010gu}, in which the next-to-leading order (NLO) QCD corrections~\cite{Buras:1990fn,Ciuchini:1997bw,Buras:2000if} and the uncertainties from various parameters, such as CKM matrix elements, decay constants, and top-quark mass, are taken into account.  Hence, from Eq.~(\ref{eq:DMq}), the bounds from $\Delta B=2$ can be used as:
 \begin{align}
 0.76 \lesssim |1+ \Delta^{H^\pm}_d | \lesssim 1.15 \,, \nonumber \\
 0.87 \lesssim |1+ \Delta^{H^\pm}_s | \lesssim 1.38 \,. \label{eq:DB2_CH}
 \end{align}

\subsection{ $\Delta M_{q'}$ from the tree FCNCs }

 To formulate the scalar boson contributions to $\Delta M_{q'}$ at the tree level, we write  the Yukawa couplings of scalars $H$ and $A$ to the quarks with Cheng-Sher ansatz as~\cite{Benbrik:2015evd}:
 \begin{align}
-{\cal L}^{H, A}_{Y} &=   \frac{t_\beta}{v} \bar d_{i L} \left[  m_{d_i} \delta_{ij} - \frac{\sqrt{m_{d_i} m_{d_j}}}{ s_\beta} \chi^d_{ij} \right] d_{j R} (H -i A) + H.c. \label{eq:Yu_HA}
\end{align}
The effective Hamiltonian  for $\Delta B=2$ process  mediated by the neutral scalar bosons $H$ and $A$ at $\mu_H$ scale  can then be straightforwardly obtained  as:
 \begin{align}
{\cal H}^{\Delta B=2}_{S} & = - \left( \frac{m_b t_\beta}{v s_\beta}\right)^2  \frac{m_{q'}}{ 4 m_b} \left[ (\chi^{d*}_{q' b})^2  \left( \frac{1}{m^2_H} - \frac{1}{m^2_A} \right) Q_2  \right. \nonumber \\
& \left. + (\chi^d_{bq'} )^2 \left( \frac{1}{m^2_H} - \frac{1}{m^2_A} \right) \tilde Q_2 + 2 \chi^d_{b q' } \chi^{d*}_{q' b}\left( \frac{1}{m^2_H} + \frac{1}{m^2_A} \right) Q_4\right]\,. \label{eq:HDB2S}
 \end{align}
It can be seen that when $m_H= m_A$, the contributions from the operators $Q_2$ and $\tilde{Q_2}$ vanish. We note that the box diagrams, mediated by $Z$-$H(A)$, $G^0$-$H(A)$, and $H(A)$-$H(A)$, involve the  $q_i$-$b$-$H(A)$ FCNC couplings, which are the same as the tree contributions.  Thus, it is expected that the box contributions will be smaller than the tree; therefore, we do not further discuss such box diagrams and neglect their contributions. 

Using Eq.~(\ref{eq:mbWCs}) and the hadronic matrix elements shown in Eq.~(\ref{eq:BBmatrix}), the $\Delta M_{q'}$, which combines the SM and $S=H+A$ effects, can be found as:
\begin{equation}
 \Delta M^{S}_{q'} = 2 | \langle B_{q'}| H^{\rm \Delta B=2}_{S} | \bar B_{q'} \rangle|=\Delta M^{\rm SM}_{q'} |1+\Delta^{S}_{q'}|\,,
 \label{eq:DMq_S}
 \end{equation}
 where    the $H$ and $A$ contributions are expressed as:
 \begin{align}
 \Delta^S_{q'} &= - \frac{1}{4 S_0(x_t)} \left( \frac{\sqrt{2} \pi^2 t^2_\beta \sqrt{ x_b x_{q'} } }{2G_F (V^*_{tb} V_{tq'} )^2 s_\beta}\right) \frac{m^2_{B_{q'}}}{ (m_b + m_{q'} )^2 \hat\eta_{1B} B_{1q'}}\nonumber \\
 & \times \left[ \left(-5 \hat\eta_{2B} B_{2 q'} + \hat\eta_{3B} B_{3q'}  \right) \left(C^S_2  + \tilde{C}^S_2 \right) \right. \nonumber \\
 & \left. + \left(6\hat\eta_{44B} B_{4q'} + 2 \hat\eta_{45B} B_{5q'} \right) C^S_4  \right]\,;
 \end{align}
$x_{b(q')}=m^2_{b(q')}/m^2_W$, the $\hat\eta_{iB}$ are the QCD factors as shown in Table~\ref{tab:BPs}, and the factors $C^S_2$, $\tilde{C}^S_2$, and $C^S_4$ are defined as:
\begin{align}
 C^S_{2}   & =  (\chi^{d*}_{q' b})^2  \left( \frac{1}{m^2_H} - \frac{1}{m^2_A}\right)  \,, \nonumber \\
 \tilde{C}^S_2 & =  (\chi^d_{bq'})^2 \left( \frac{1}{m^2_H} - \frac{1}{m^2_A}\right) \,, \nonumber \\
 C^S_4 & = 2 \chi^{d*}_{q' b} \chi^d_{b q'} \left( \frac{1}{m^2_H} + \frac{1}{m^2_A}\right)\,.
\end{align}
Since  Eq.~(\ref{eq:DMq_S})  is directly related to $\chi^d_{bq',q'b}$, in order to show the $\Delta M_{q'}$ constraint on the different parameters,  here we do not combine the neutral scalar with  the charged-Higgs contributions.  According to Eq.~(\ref{eq:DB2_CH}), the bounds on $\Delta^S_{d,s}$ can be given as:
\begin{align}
 0.76 \lesssim |1+ \Delta^{S}_d | \lesssim 1.15 \,, \nonumber \\
 0.87 \lesssim |1+ \Delta^{S}_s | \lesssim 1.38 \,. 
 \end{align}

\subsection{ Charged-Higgs contributions to the $b\to s \gamma$ process} 

In addition to the $\Delta B=2$ processes,  the penguin induced $b\to s \gamma$ decay is also sensitive to new physics. The current experimental value is $BR(\bar B \to X_s \gamma)^{\rm exp}=(3.32 \pm  0.15) \times 10^{-4}$ for $E_\gamma > 1.6$ GeV~\cite{Amhis:2016xyh}, and the  SM prediction with next-to-next-to-leading oder (NNLO) QCD corrections  is  $BR(\bar B \to X_s \gamma)^{\rm SM}=(3.36 \pm  0.23) \times 10^{-4}$~\cite{Czakon:2015exa,Misiak:2015xwa}.  Since the SM result is close to  the experimental data, we can use the $\bar B \to X_s \gamma$ decay to  give a strict bound on the new physics effects. The effective Hamiltonian arisen from the $W^\pm$ and $H^\pm$ bosons for $b\to s \gamma$ at $\mu_H$ scale can be written as:
\begin{align}
{\cal H}_{b\to s \gamma} = -\frac{4G_F}{\sqrt{2}} V^*_{ts} V_{tb} \left(C_{7\gamma} (\mu_H)Q_{7\gamma}  + C_{8\gamma} (\mu_H)Q_{8G} + C'_{7\gamma} (\mu_H)Q'_{7\gamma}  + C'_{8\gamma} (\mu_H)Q'_{8G}\right)\,, \label{eq:Hbtosga}
\end{align}
where the electromagnetic and gluonic dipole operators are given as:
 \begin{equation}
 Q_{7\gamma} = \frac{e}{16\pi^2} m_b \bar s \sigma^{\mu\nu} P_R b F_{\mu\nu}\,, \ Q_{8G} = \frac{g_s}{16\pi^2} m_b \bar s_\alpha \sigma^{\mu\nu}  T^a_{\alpha \beta} P_R b_\beta G^a_{\mu\nu}\,,
 \end{equation}
 and the $Q'_{7\gamma, 8G}$ operators can be obtained from the unprimed operator using $P_L$ instead of $P_R$.  We note that  $C'_{7\gamma,8G}$ from the SM contributions are suppressed by $m_s$ and are negligible; therefore, the main primed operators are from the new physics effects.

\begin{figure}[phtb]
\includegraphics[scale=0.85]{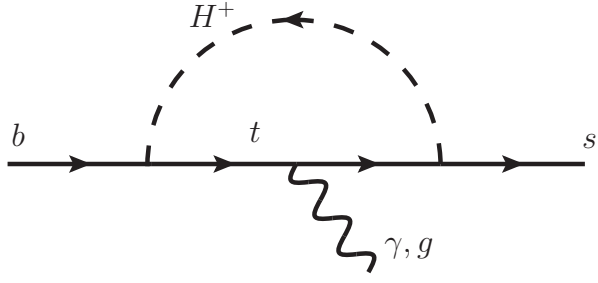}
\caption{ Penguin diagrams for $b\to s (\gamma, g)$ with the intermediate of $H^\pm$.}
\label{fig:pen_CH}
\end{figure}

According to the charged-Higgs  interactions in Eq.~(\ref{eq:btoqpp}), the relevant Feynman diagrams for $b\to s (\gamma, g)$ are sketched in Fig.~\ref{fig:pen_CH}, and the $H^\pm$ contributions to $C^{H^\pm}_{7\gamma,8G}$ at $\mu_H$ scale can be derived  as : 
 \begin{align}
 C^{H^\pm}_{7\gamma(8G)} (\mu_H)& = \zeta^{u*}_{ts} \zeta^u_{tt}  C^{H^\pm}_{7(8),LL}(y_t) +  \zeta^{u*}_{ts} \zeta^d_{bb}  C^{H^\pm}_{7(8),RL}(y_t)  \,, \nonumber \\
 C'^{H^\pm}_{7\gamma(8G)} (\mu_H)& = \zeta^{d*}_{ts} \zeta^{d}_{bb} C^{H^\pm}_{7(8),RR} (y_t)+  \zeta^{d*}_{ts} \zeta^{u}_{tt} C^{H^\pm}_{7(8),LR}(y_t)\,, 
\label{eq:CCH7}
 \end{align}
 where the loop integral functions are defined as:
\begin{subequations}
 \begin{align}
 C^{H^\pm}_{7,LL} (y_t)& = \frac{y_t}{72} \left[ \frac{8 y^2_t + 5 y_t -7}{(1-y_t)^3} - \frac{6 y_t (2-3y_t)}{(1-y_t)^4} \ln(y_t)\right]\,,  \\
 C^{H^\pm}_{8,LL} (y_t)& = \frac{y_t}{24} \left[ \frac{ y^2_t - 5 y_t -2}{(1-y_t)^3} - \frac{6 y_t }{(1-y_t)^4} \ln(y_t)\right]\,,  \\
 C^{H^\pm}_{7,RL} (y_t)& = \frac{y_t}{12}  \left[ \frac{3-5y_t}{(1-y_t)^2} + \frac{2(2-3y_t)}{(1-y_t)^3 }\ln(y_t)  \right]\,,  \\
 C^{H^\pm}_{8,RL} (y_t)& = \frac{y_t}{4}  \left[ \frac{3-y_t}{(1-y_t)^2} + \frac{2}{(1-y_t)^3 } \ln(y_t)\right]\,,
 \end{align}
 \end{subequations}
 $C'^{H^\pm}_{7(8),RR} (y_t)= -(m^2_b/m^2_t ) C^{H^\pm}_{7(8),LL}(y_t)$, and  $C'^{H^\pm}_{7(8),LR} (y_t)= - C^{H^\pm}_{7(8),RL}(y_t)$. From Eq.~(\ref{eq:CCH7}), we can easily understand  the effects of the type-II 2HDM as follows: taking $\chi^u_{tt,ct}=\chi^d_{bb,sb}=0$ in Eq.~(\ref{eq:CCH7}),  $(\zeta^{u*}_{ts} \zeta^u_{tt} )_{\rm type-II}$  is suppressed by $1/t^2_\beta$, and  $(\zeta^u_{bb} \zeta^{u*}_{ts})_{\rm type-II}=1$ becomes  $t_\beta$-independence. As a result,  the mass of charged-Higgs  in type-II 2HDM is limited to be $m_{H^\pm} > 580$ GeV at 95\% confidence level (CL) when  NNLO QCD corrections are taken into account~\cite{Misiak:2017bgg}. In the generic 2HDM,  since the new parameters $\chi^u_{tt,ct}/c_\beta$ and $\chi^d_{bb}/s_\beta$ are involved in Eq.~(\ref{eq:CCH7}), we have more degrees of freedom to reduce $\zeta^u_{bb} \zeta^{u*}_{ts}$ away from unity; thus, the charged-Higgs mass can be lighter than 580 GeV. 
 
 To calculate the BR of $\bar B \to X_s \gamma$, we employ the results  in~\cite{Blanke:2011ry,Buras:2011zb}, which are  shown as:
 \begin{equation}
  BR(\bar B \to X_s \gamma)  = 2.47\times 10^{-3} \left( |C_{7\gamma}(\mu_b) |^2 + |C'_{7\gamma}(\mu_b)|^2 + N(E_\gamma) \right)\,, \label{eq:brbsga}
  \end{equation}
where $N(E_\gamma) =(3.6 \pm 0.6)\times 10^{-3}$ denotes a nonperturbative effect;  $C_{7\gamma}(\mu_b) = C^{\rm SM}_{7\gamma}(\mu_b) + C^{H^\pm}_{7\gamma}(\mu_b)$ and $C'_{7\gamma}(\mu_b) = C'^{H^\pm}_{7\gamma}(\mu_b)$ are the Wilson coefficients at  the $\mu_b$ scale, and their relations to the initial conditions at the higher energy scalar $\mu_H$ occur  through renormalization group (RG) equations. Using Eq.~(\ref{eq:brbsga}) and $BR(\bar B\to X_s \gamma)^{\rm SM} \approx 3.36\times 10^{-4}$, we obtain $C^{\rm SM}_{7\gamma}\approx -0.364$ at $\mu_b\approx 2.5$ GeV. The NLO~\cite{Ciuchini:1997xe,Borzumati:1998tg,Borzumati:1998nx} and NNLO~\cite{Hermann:2012fc} QCD corrections to the $C_{7\gamma}(\mu_b)$ in the 2HDM have been calculated. In this study,  the charged-Higgs effects  with RG running are taken from~~\cite{Blanke:2011ry,Buras:2011zb}, and they are written as:
 \begin{align}
 C^{(\prime)H^\pm}_{7\gamma} (\mu_b) & =  \kappa_7 C^{(\prime)H^\pm}_{7\gamma} (\mu_H) + \kappa_8 C^{(\prime)H^\pm}_{8G}(\mu_H)\,, \label{eq:Wsatmub}
 \end{align}
 where $\kappa_{7,8}$ are the LO QCD effects, for which their values with different values of $\mu_H$ can be found in~\cite{Blanke:2011ry,Buras:2011zb}.

 \subsection{$H/A$ contributions to the $b\to s \gamma$ process}
 
\begin{figure}[phtb]
\includegraphics[scale=0.85]{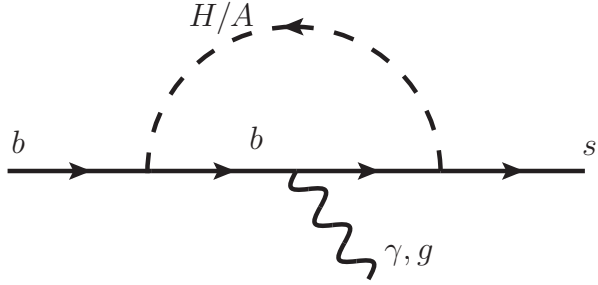}
\caption{The same as Fig.~\ref{fig:pen_CH} but with the intermediates of neutral scalar bosons $H$ and A.  }
\label{fig:pen_S}
\end{figure}

 In addition to the charged currents, the $b\to s \gamma$ process can be generated through the FCNCs in the type-III 2HDM, where the corresponding Feynman diagrams for $b\to s (\gamma, g)$ are shown in Fig.~\ref{fig:pen_S}. From the diagrams, it can be seen that unlike the   $m^2_t/m^2_{H^\pm}$ result from the $H^\pm$  and top-quark loops,  the $b$-quark loops are suppressed by $m^2_b/m^2_{H,A}$.  Therefore, it is expected that the radiative $b$ decay induced by the neutral currents will be much smaller than the charged currents. 

Using the Yukawa couplings in Eq.~(\ref{eq:Yu_HA}), we can derive the Wilson coefficients of $C_{7\gamma}$ and $C'_{7\gamma}$ at  the $\mu_H$ scale, defined in Eq.~(\ref{eq:Hbtosga}), as:
 \begin{align}
 C^S_{7\gamma} & = -\frac{t^2_\beta Q_b}{4 V^*_{ts} V_{tb}} \sqrt{\frac{m_s}{m_b}}  \frac{\chi^d_{sb}}{s_\beta} {\cal N}_{S}  \,, \quad C'^S_{7\gamma} = -\frac{t^2_\beta Q_b}{4 V^*_{ts} V_{tb}} \sqrt{\frac{m_s}{m_b}} \frac{\chi^{d*}_{bs} }{s_\beta} {\cal N}^*_{S} \,, \label{eq:C7S} \\
 %
 %
 {\cal N}_{S} & = - \left( 1 -\frac{\chi^{d*}_{bb}}{s_\beta}\right)  \left[ J_1 \left( \frac{m^2_b}{m^2_A}\right)+ J_1 \left( \frac{m^2_b}{m^2_H}\right)\right] \nonumber \\
 & + \left( 1 -\frac{\chi^{d}_{bb}}{s_\beta}\right)  \left[ J_2 \left( \frac{m^2_b}{m^2_A}\right)- J_2\left( \frac{m^2_b}{m^2_H}\right)\right]\,, \label{eq:NS}
 \end{align}
where the superscript $S$ denotes the scalar contributions; $Q_b=-1/3$ is the electric charge of $b$-quark, and the functions $J_{1,2}$ are defined as:
 \begin{align}
 J_{1}(y) &= \frac{y}{6} \left[ \frac{1}{1-y} + \frac{y \ln(y)}{(1-y)^2} \right] \,, \nonumber \\
 J_2(y) & = \frac{y}{2} \left[ -\frac{1}{1-y} + \frac{(y-2) \ln(y)}{(1-y)^2}\right]\,.
 \end{align}
  The contributions of $H$ and $A$ bosons to the chromomagnetic dipole operators can be related to the electromagnetic dipole operators, and the relations can be easily found as $C^{(\prime)S}_{8G} = C^{(\prime)S}_{7\gamma}/Q_b$.
 We can apply the result in Eq.~(\ref{eq:Wsatmub}) to get the Wilson coefficients at $\mu_b$ scale as:
 \begin{align}
 C^{(\prime) S}_{7\gamma} (\mu_b) & =  \kappa_7 C^{(\prime)S}_{7\gamma} (\mu_H) + \kappa_8 C^{(\prime)S}_{8G}(\mu_H)\,.
 \end{align}
Using Eq.~(\ref{eq:brbsga}), we can directly obtain the $S$-mediated $BR(\bar B \to X_s \gamma)$.
  
  \section{Numerical analysis and discussions}
  
  \subsection{Numerical and theoretical inputs}
  
  In addition to the parameter values shown in Table~\ref{tab:BPs}, the  values of the CKM matrix elements used in the following analysis are taken as~\cite{Amhis:2016xyh}:
   \begin{align}
   &V_{ub} \approx 0.0037 e^{-i \phi_3}\,, \ \phi_3=73.5^\circ\,, \ V_{cd(s)}\approx -0.22 (0.973)\,, \  V_{cb} \approx 0.0393 \,,  \nonumber \\
   & V_{td} \approx 0.0082  e^{-i \phi_1 }\,, \ \phi_1=21.9^\circ\,, \ V_{ts} \approx -0.040\,, \ V_{tb} \approx 1.0\,.
   \end{align}
    To study the semileptonic $\bar B \to ( P ,V) \ell \bar \nu$ decays,  we need the information for the $\bar B \to ( P, V)$ transition form factors. For the $\bar B \to \pi$ decay, we use the results obtained by the LCSRs and express them as ~\cite{Ball:2004ye,Ball:2006jz}:
  \begin{align}
  f^{B\pi}_1(q^2) &=\frac{ f_1(0) }{1-q^2/5.32^2 } \left( 1 + \frac{r_{BZ}\; q^2/5.32^2}{1 - \alpha_{BZ} \; q^2/m^2_B} \right)\,,
  \nonumber \\
  f^{B\pi}_0(q^2) & =  \frac{f_1(0)}{1 -q^2 /33.81}\,,
  \end{align}
  where we take $f_1(0)=0.245$, $\alpha_{BZ}=0.40$, and $r_{BZ}=0.64$.  It is worth mentioning that lattice QCD results with $N_f=2+1$ for the $\bar B\to \pi$ form factors, calculated by HPQCD~\cite{Dalgic:2006dt}, FNAL-MILC~\cite{Lattice:2015tia}, and RBC-UKQCD~\cite{Flynn:2015mha} collaborations, recently have  significant progress. The detailed summary of the lattice QCD results can be found in~\cite{Aoki:2016frl}. We checked that the results of LCSRs are consistent with the values of Table IV in~\cite{Flynn:2015mha}.  For the $\bar B\to \rho$ decay, the form factors based on the LCSRs are given as~\cite{Ball:2004rg}:
  \begin{align}
  V^{B\rho}(q^2) & = \frac{1.045}{1- q^2/(5.32)^2} - \frac{0.721}{1- q^2/38.34}\,, \nonumber \\
  A^{B\rho}_0(q^2) & =\frac{1.527}{1-q^2/(5.28)^2} - \frac{1.220}{1- q^2/33.36} \,, \nonumber \\
  A^{B\rho}_1(q^2) & = \frac{0.220}{1 - q^2/37.51} \,, \nonumber \\
  A^{B\rho}_2(q^2) & = \frac{0.009}{1-q^2/40.82} - \frac{0.212}{(1-q^2/40.82 )^2}   \,.
  \end{align}

 Recently, the $B\to D^{(*)}$ form factors associated with various types of currents, which are formulated in the heavy quark effective theory (HQET)~\cite{Falk:1992wt}, were studied up to $O(\Lambda_{QCD}/m_{b,c})$ and $O(\alpha_s)$ in \cite{Bernlochner:2017jka}, where several fit scenarios were shown. We summarize the relevant results of Ref.~\cite{Bernlochner:2017jka} with ``th:$\rm L_{w\geq 1}$+SR''  scenario in the appendix, where the ``th:$\rm L_{w\geq 1}$+SR" scenario combines the QCD sum rule constraints and the QCD lattice data~\cite{Lattice:2015rga}.  The parametrizations of HQET form factors are different from those shown in Eqs.~(\ref{eq:ffBP}) and (\ref{eq:ffBV}), and their relations can be straightforwardly found as follows: For $B\to D$, they are:
  \begin{align}
  f^{BD}_1 (q^2) & = \frac{1}{2\sqrt{m_B m_D}} \left[ (m_B + m_D) h_+ (w) - (m_B - m_D) h_{-}(w)\right]\,, \nonumber \\
  f^{BD}_0(q^2) & = \frac{1}{2\sqrt{m_B m_D}} \left[ \frac{(m_B + m_D)^2-q^2}{m_B + m_D} h_{+}(w) + \frac{q^2- (m_B-m_D)^2}{m_B-m_D} h_{-}(w)\right]\,,
  \end{align}
   while for $B\to D^*$, they can be written as:
   \begin{align}
   V^{BD^*}(q^2) & = \frac{m_B + m_{D^*} }{ 2\sqrt{m_B m_{D^*} }} h_V(w)\,,  \nonumber \\
   A^{BD^*}_0(q^2) & = \frac{1}{2\sqrt{m_B m_{D^*}}} \left[ \frac{(m_B + m_{D^*})^2-q^2}{2m_{D^*}} h_{A_1}(w) 
   -\frac{m^2_B -m^2_{D^*} + q^2}{2 m_B} h_{A_2}(w) \right. \nonumber \\
   & \left. -\frac{m^2_B -m^2_{D^*} -q^2 }{2m_{D^*}} h_{A_3} (w)
   \right]\,, \nonumber \\
   A^{BD^*}_1(q^2) & =\frac{1}{2\sqrt{m_B m_{D^*}}} \frac{(m_B + m_{D^*})^2 - q^2}{m_B + m_{D^*}}h_{A_1} (w)\,, \nonumber \\
    A^{BD^*}_2(q^2) & = \frac{m_B + m_{D^*}}{2\sqrt{m_B m_{D^*} }} \left[ \frac{m_{D^*}}{m_B} h_{A_2} (w) + h_{A_3} (w) \right]\,,
   \end{align}
  where $w=(m^2_B + m^2_{D^{(*)}} -q^2 )/(2m_B m_{D^{(*)}})$, and  the $h_i$ functions and their relations to the leading and subleading Isgur-Wise functions can be found in the Appendix.

  \subsection{ Case with $\chi^d_{bq'}\neq 0$ and $\chi^{u}_{tt,ct}=\chi^d_{bb}=\chi^\ell_\ell=0$}
  
  The free parameters involved in this study are: $\chi^u_{tt}$, $\chi^u_{ct,tc}$, $\chi^{u}_{ut,tu}$, $\chi^d_{bb}$, $\chi^d_{bs,sb}$,  $\chi^d_{bd,db}$, $t_\beta$, and the scalar masses $m_{H,A,H^\pm}$.  To reduce the number of free parameters without loss of generality, we adopt $\chi^{q}_{ij}=\chi^q_{ji}$ and take the new free parameters to be real numbers with the exception of $\chi^u_{tu,ut}$.  Thus, the parameters $\chi^d_{db,sb}$ and $\chi^u_{tc}$ in leptonic $B^-_q \to \ell \bar \nu$ become correlated to $\chi^d_{bd,bs}$ and $\chi^u_{ct}$ in the $\Delta B=2$ and $b\to s \gamma$ processes.  
 
\begin{figure}[phtb]
\includegraphics[scale=0.5]{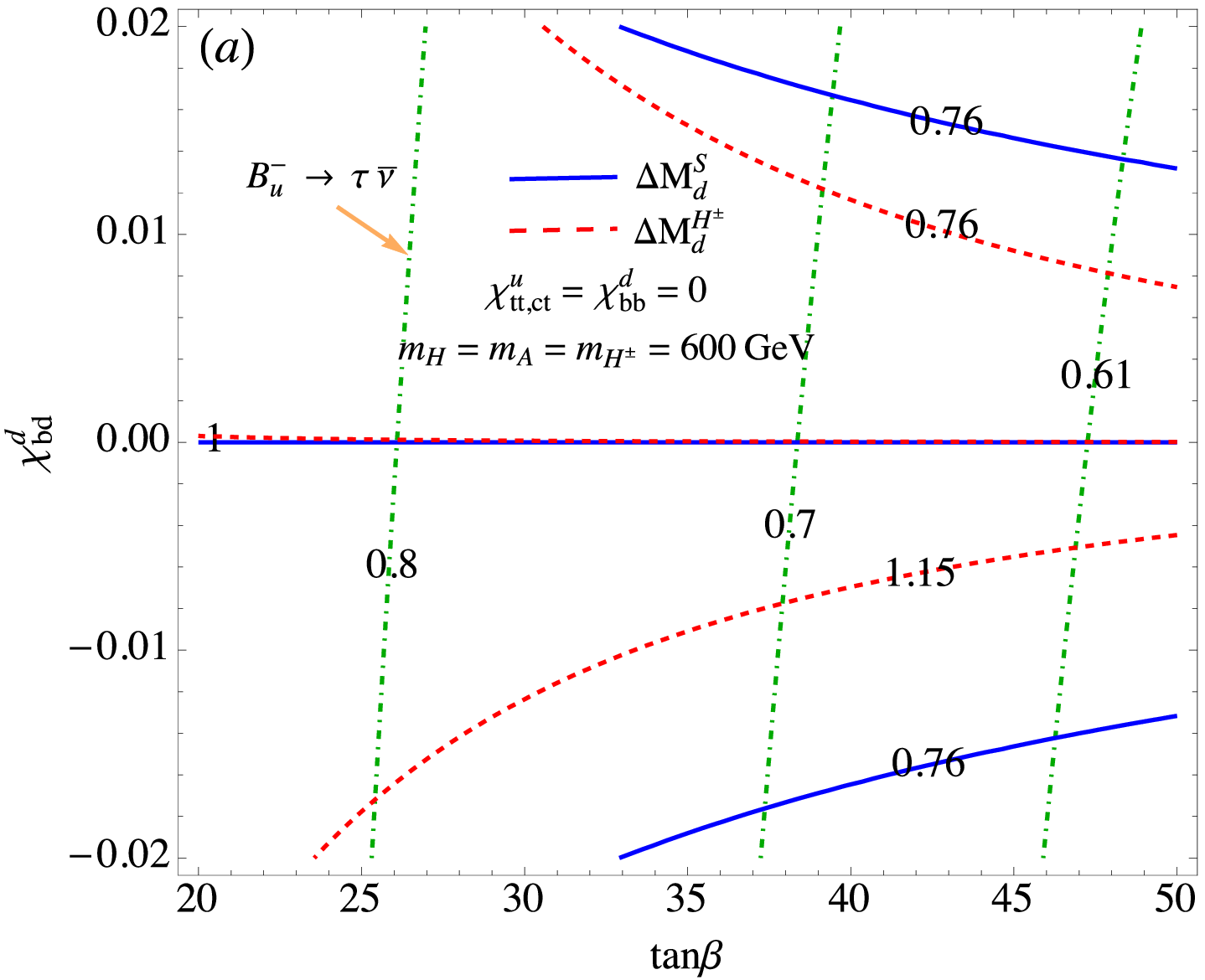}
\includegraphics[scale=0.5]{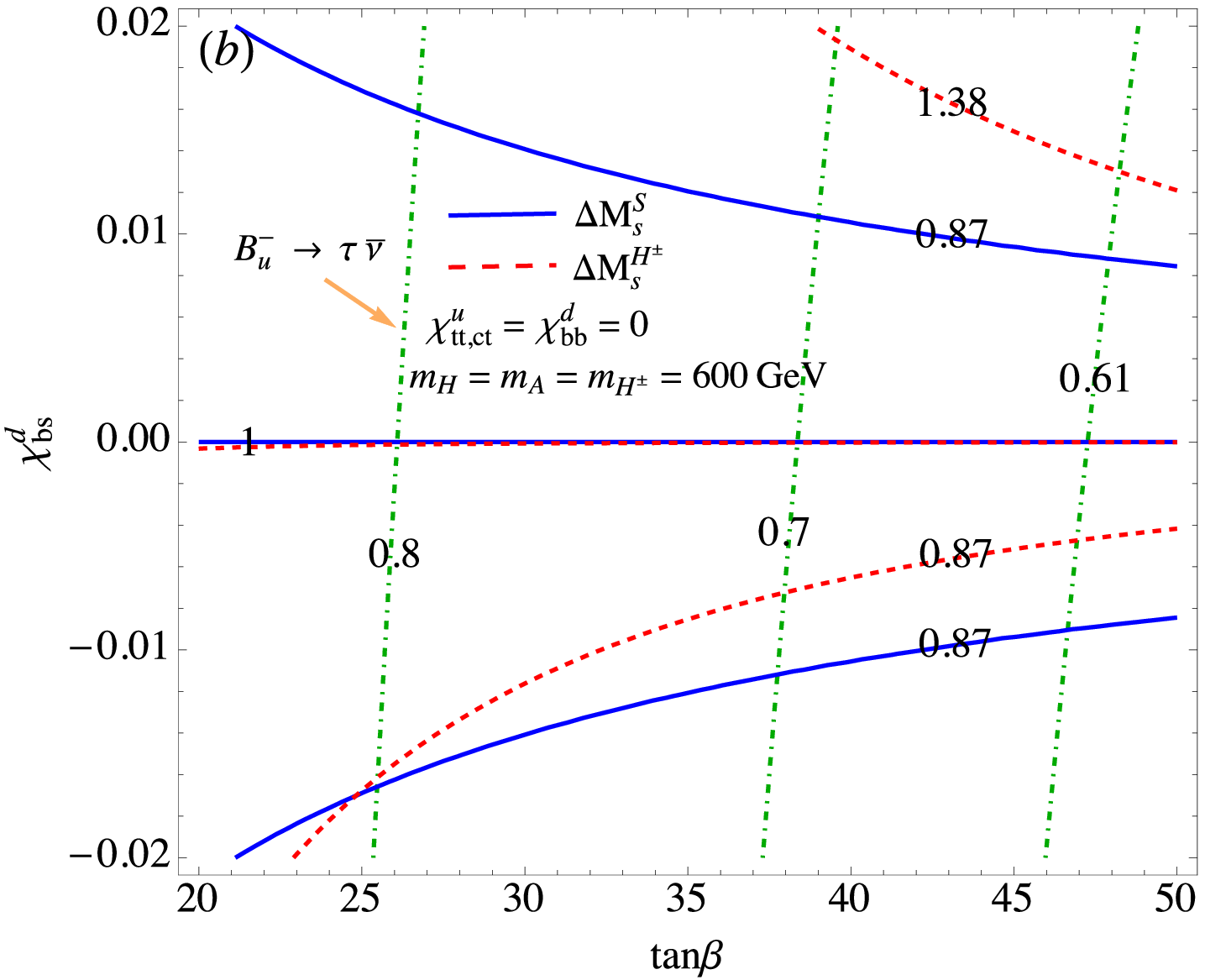}
\caption{  Contours of $|1+\Delta^{S,H^\pm}_{d(s)}|$  and $BR(B^-_u \to \tau \bar\nu)10^{4}$ as a function of $\chi^d_{bd(s)}$ and $\tan\beta$ for $\chi^u_{tt,ct}=\chi^d_{bb}=0$, where $m_{H}=m_{A}=m_{H^\pm}=600$ GeV is used.}
\label{fig:chidbq}
\end{figure}

  According to Eq.~(\ref{eq:HDB2S}), it can be seen that the involving parameters in $S$-mediated $\Delta B=2$ processes are only related to $\chi^d_{bs}$ and $\chi^d_{bd}$.  To understand how strict the experimental bounds on the  $\chi^d_{bq'}$ are,  we first discuss the simple situation with $\chi^u_{tt,ct}=\chi^d_{bb}=0$.  Thus, the  contours of  $|1+\Delta^{S,H^\pm}_{d[s]}|$   as a function of $\chi^{d}_{bd[s]}$ and $\tan\beta$ are shown in Fig.~\ref{fig:chidbq}(a)[(b)], where the solid and dashed lines denote the tree-level $S$-mediated and loop $H^\pm$-mediated effects, respectively, and $m_H=m_A=m_{H^\pm}=600$ GeV is used.  From the plots, we can see that the tree-induced $\Delta M^S_{s}$ gives a  stronger constraint  in the region of $\chi^d_{bs}>0$. However, in the regions of $\chi^d_{bd}>0$ and $\chi^d_{bq'}<0$, the $H^\pm$ contributions to $B_{q'}$ mixings become dominant. In addition to the $\sqrt{x_b x_{q'}}$ suppression in $\Delta M^S_{q'}$, the loop effect can be over the tree effect because $\chi^d_{bq'}$ in $\Delta M^{H^\pm}_{q'}$ is linear dependent, but it is quadratic  in $\Delta M^S_{q'}$; as a result, when $\chi^d_{bq'}$ is of $O(10^{-2})$, the $\Delta M^{H^\pm}_{q'}$ can be larger than $\Delta M^{S}_{q'}$.  
  
  As mentioned earlier, the charged-Higgs contributions to the $b\to q \ell \bar \nu$ processes are destructive in the type-II 2HDM.  From Eq.~(\ref{eq:Cqb}),  when $\chi^u_{tc}=\chi^d_{bb}=\chi^\ell_\ell=0$, the sign change of $C^{R,\ell}_{qb}$ relies on the magnitude of $\chi^R_{qb}$; however, the feasibility is excluded by the $\Delta M_{q'}$ constraint  due to the result of $\chi^d_{bq'}\sim O(10^{-2})$. Hence, in such cases, the charged-Higgs effect in the type-III model is also destructive to the SM result. To illustrate  the  $H^\pm$ influence   on the leptonic decays, we show the contours of $BR(B^-_u \to \tau \bar \nu)$ (dot-dashed lines) in units of $10^{-4}$ in Fig.~\ref{fig:chidbq}(a) and (b). Since $\chi^d_{bd}$ and $\chi^d_{bs}$ both appear in $\chi^R_{ub}$, as shown in Eq.~(\ref{eq:chiRub}), when we focus on one of them, the other is set to vanish. From the plot, it can be seen that  $BR(B^-_u \to \tau \bar\nu)$ is always smaller than the SM result:
  \begin{equation}
  BR(B^-_u \to \tau \bar \nu)^{\rm SM} \approx 0.89\times 10^{-4}\,.
  \end{equation}
 In addition,  the resulted $BR(B^-_u \to \tau \bar\nu)$ is even smaller than  the experimental lower bound of $1\sigma$ errors.  Since similar behavior also occurs in $B^-_c \to \tau \bar\nu$, here,  we just show the $B^-_u \to \tau \bar\nu$ decay. Hence, only considering the $\chi^d_{bq'}$ effect will not cause interesting implications in the leptonic $B^-_q$ decay.
  
\begin{figure}[phtb]
\includegraphics[scale=0.6]{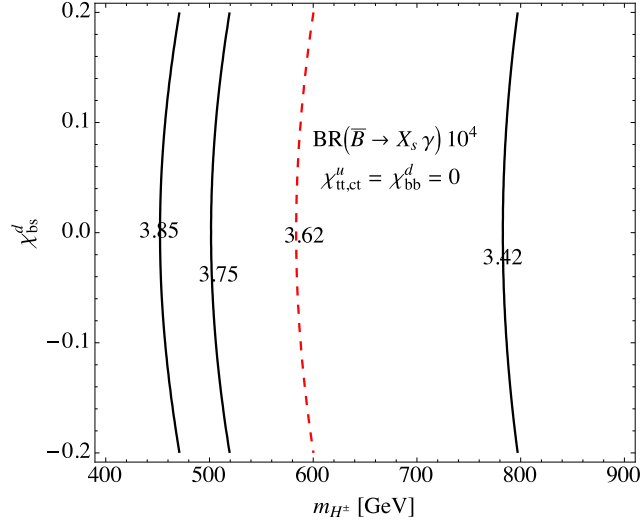}
\caption{  Contours of  $BR(\bar B \to X_s \gamma)$ (in units of $10^{-4}$) as a function of $\chi^d_{bs}$ and $m_{H^\pm}$ for $\chi^u_{tt,ct}=\chi^d_{bb}=0$, where the dashed line denotes the $2\sigma$ upper limit of data.}
\label{fig:bsga_chidbs}
\end{figure}

  The $\chi^d_{bs}$ also affects the radiative $b \to s \gamma$ decay through the intermediates of   $H^\pm$ and $S$ shown in Figs.~ \ref{fig:pen_CH} and \ref{fig:pen_S}.  Since the quark in the $S$-mediated penguin diagram is the $b$ quark, due to the suppression of $m^2_b/m^2_{H(A)}$, the contribution of $|\chi^d_{bs}|=0.02$ to $C^{(\prime)S}_{7\gamma}$ in Eq.~(\ref{eq:C7S}) is  of $O(10^{-4})$ and is thus negligible. According to Eq.~(\ref{eq:CCH7}), the  $\chi^d_{bs}$ of  the $H^\pm$ contribution only appears in $C'^{H^\pm}_{7\gamma(8G)}$ and shows up by means of  $\zeta^{d*}_{ts} \zeta^d_{bb}$ and $\zeta^{d*}_{ts} \zeta^u_{tt}$. Although the former has a $t^2_\beta$ enhancement, due to the $m^2_b/m^2_t$ suppression in $C^{H^\pm}_{7(8),RR}$, the associated contribution is much smaller than the latter, which is insensitive to $t_\beta$. We find that with $|\chi^d_{bs}|=0.02$, the result is $|C'^{H^\pm}_{7\gamma}| \approx 0.012$ and is still much less than  $|C^{\rm SM}_{7\gamma}|$.  We note that the situation with $\chi^u_{tt,ct}=\chi^d_{bb}=0$ is similar to the type-II model; therefore, with $|\chi^d_{bs}| <  O(0.1)$, the charged-Higgs effect on $b\to s \gamma$ is insensitive to $t_\beta$ and $\chi^d_{bs}$, but is sensitive to $m_{H^\pm}$. To numerically show the result, we plot the contours of $BR(\bar B \to X_s \gamma)$ in units of $10^{-4}$  in Fig.~\ref{fig:bsga_chidbs}, where the dashed line denotes the $2\sigma$ upper limit of experimental data, and the lower bound on the charged-Higgs mass is  given by $m_{H^\pm} > 580$ GeV.

  According to above analysis, we learn that when $\chi^u_{tt,ct}=\chi^d_{bb}=0$ is taken in the type-III 2HDM, due to the strict limits of $\Delta M_{d}$ and $\Delta M_s$, the $\chi^d_{bd}$ and $\chi^d_{bs}$ effects contributing to $B^-_q \to \ell \bar \nu$ and $b\to s \gamma$ are small and have no interesting implications on the phenomena of interest.  For simplicity, we thus take $\chi^d_{bd}= \chi^d_{bs} = 0$  in the following analysis; that is, we only consider the charged-Higgs contributions. 
  
   \subsection{Correlation with the constraint from the  $H/A \to \tau^+ \tau^-$ limits }
 
 In the 2HDM, $m_{H^\pm}$ indeed correlates with $m_{H(A)}$. According to the study in~\cite{Benbrik:2015evd}, the allowed mass difference can be $m_H - m_{H^\pm} \sim 100$ GeV if $m_A = m_H$ is used.  Since $m_{H^\pm}=300$ GeV is taken in our numerical analysis, the effects arisen from $m_S\equiv m_{H(A)}\sim 400$ GeV   in the 2HDM cannot be arbitrarily dropped. Using this correlation, it was pointed out  that the upper limit of  tau-pair production through the $pp(b\bar{b})\to H/A\to \tau^+ \tau^-$ processes measured in the LHC can give a strict bound on the parameter space, which is used to explain the $R(D^{(*)})$ anomalies~\cite{Faroughy:2016osc}.

 In order to understand how strict the constraint from the LHC data is, we now write the scalar Yukawa couplings to the quarks, proposed in~\cite{Faroughy:2016osc}, as:
  \begin{equation}
  {\cal L}_{H'} \supset -Y_b \bar Q_3 H' b_R -Y_c \bar Q_3 \tilde H c_R - Y_{\tau} \bar L_3 H' \tau_R + H.c.\,, \label{eq:LHp}
  \end{equation}
where $H'^T=(H^+, (H + i A)/\sqrt{2})$, $Q^T_3=(V^*_{jb} u^j_{L}, b_L)$, and $j$ denotes the  flavor index. 
  It can be seen that the  parameters shown in the $b\bar{b} \to H/A \to \tau^+ \tau^-$ processes are associated with  $Y_b$ and $Y_\tau$.  In our model, the parameters $Y_{b,\tau}$  are  given as:
   \begin{align}
   Y_b & = \frac{\sqrt{2} m_b t_\beta}{v} \left( 1-\frac{\chi^d_{bb}}{s_\beta} \right)\,, \nonumber \\
   Y_\tau & = \frac{\sqrt{2} m_\tau t_\beta}{v} \left(1- \frac{\chi^{\ell}_{\tau}}{s_\beta} \right)\,. \label{eq:Ybtau}
   \end{align}
  Comparing with Eq.~(\ref{eq:Clepton}), it can be seen that the lepton couplings to $H(A)$ are the same as those to $H^\pm$. Due to the FCNC and CKM matrix effects, the $H^\pm c_L b_R$ coupling shown in~Eq.~(\ref{eq:CLR}) is generally different  from $Y_b$; however, when we take $\chi^d_{bb}=\chi^d_{sb}=\chi^d_{db}=0$, they become the same and are $Y_b = \sqrt{2} m_b t_\beta/v$. 
 
 According to the ATLAS search for the $\tau$-pair production through the resonant scalar decays,  in which the result was measured at $\sqrt{s}=13$ TeV with a luminosity of 3.2 fb$^{-1}$, it was shown  in~\cite{Faroughy:2016osc} that  the allowed values of $Y_b$ and $Y_\tau$ in Eq.~(\ref{eq:LHp}) should satisfy $|Y_b Y_\tau| v^2/m^2_{S} < 0.3$ for $m_{S}=400$ GeV. Thus,  using $t_\beta=50$, we can obtain the limit from  Eq.~(\ref{eq:Ybtau}) as:
  \begin{equation}
  |(1- \chi^\ell_\tau/s_\beta)(1-\chi^d_{bb}/s_\beta)| < 1.70\,, \label{eq:tautau}
    \end{equation}
    where $m_b(m_{S}) = 3.18$ GeV and $m_\tau = 1.78$ GeV are applied.  Hence, we will take Eq.~(\ref{eq:tautau})  as an input to bound the $\chi^\ell_\tau$ and $\chi^d_{bb}$ parameters. 

  \subsection{ Constraints of $b\to s \gamma$ and  $B_{q'}$ mixings}
  
  From Eq.~(\ref{eq:CCH7}), there are two terms contributing to $C^{H^\pm}_{7\gamma(8G)}$, where the associated charged-Higgs effects are $\zeta^{u*}_{ts}\zeta^u_{tt}$ and $\zeta^{u*}_{ts} \zeta^d_{bb}$. Using the definitions in Eq.~(\ref{eq:zetas}), it can be seen that the new factor $\chi^{L*}_{ts} \chi^{u*}_{tt}/s^2_\beta$ in the first term is insensitive to $t_\beta > 10$; however,
$\zeta^{u*}_{ts} \zeta^d_{bb} \propto 1 - t_\beta (\chi^{L*}_{ts}/s_\beta)(1 - \chi^d_{bb}/s_\beta)$ ( unity denotes the result of type-II model) formed in the 2nd term can be largely changed by a large $t_\beta$. In addition, we see that $C^{H^\pm}_{7(8),LL}$ and $C^{H^\pm}_{7(8),RL}$ are negative values, and the magnitude of the former is approximately one order smaller than that of the latter; that is, $\zeta^{u*}_{ts} \zeta^d_{bb}$ indeed dominates. Due to the negative  loop integral value, it can be understood that the Wilson coefficient $C^{H^\pm}_{7\gamma}(\mu_b)$ in the type-II model is the same sign as $C^{\rm SM}_{7\gamma}(\mu_b)$; thus, $m_{H^\pm}$ is severely limited and the low bound is $m_{H^\pm}> 580$ GeV, as shown in~\cite{Misiak:2017bgg} and confirmed in Fig.~\ref{fig:bsga_chidbs}. 
  
\begin{figure}[phtb]
\includegraphics[scale=0.5]{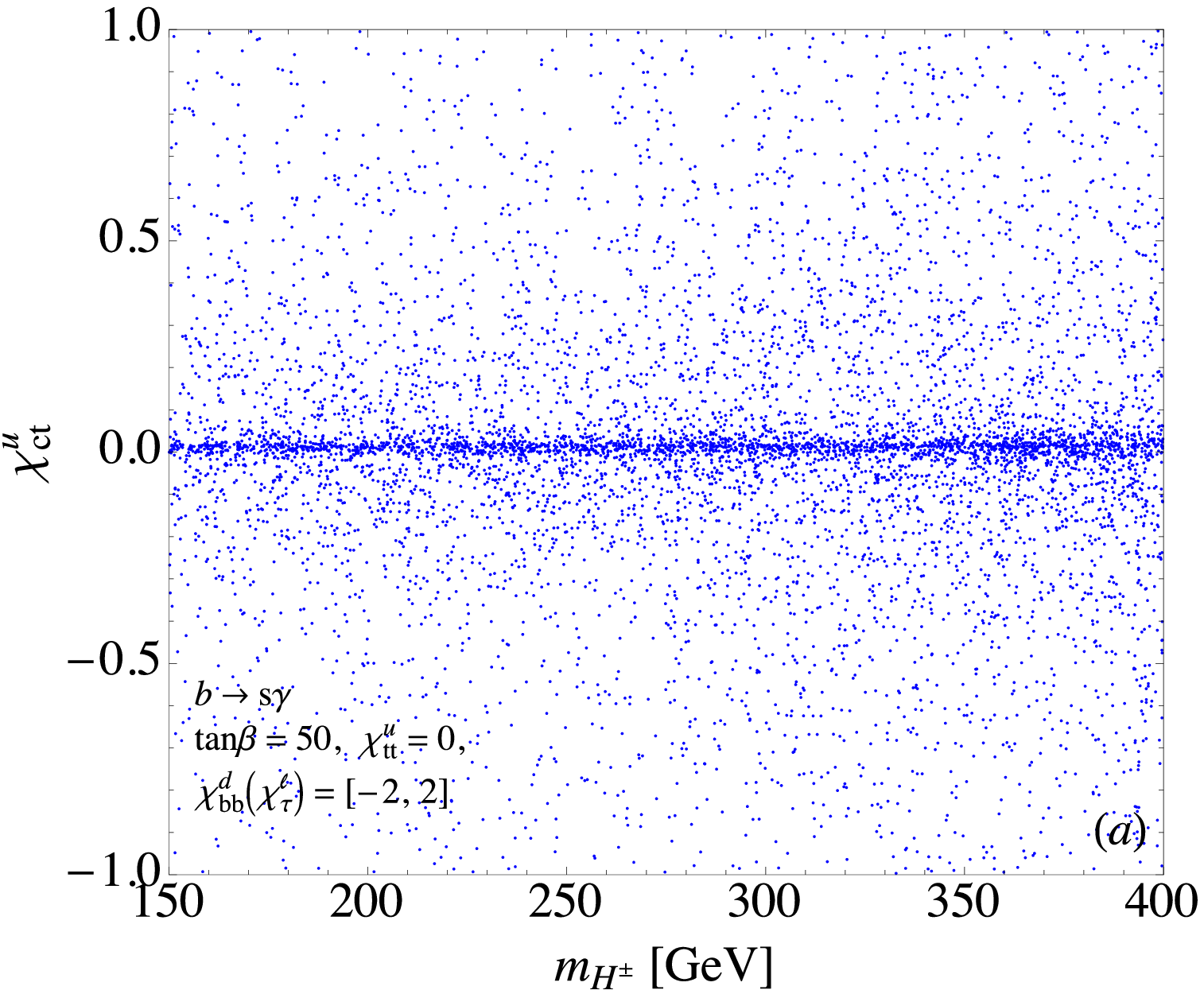}
\includegraphics[scale=0.5]{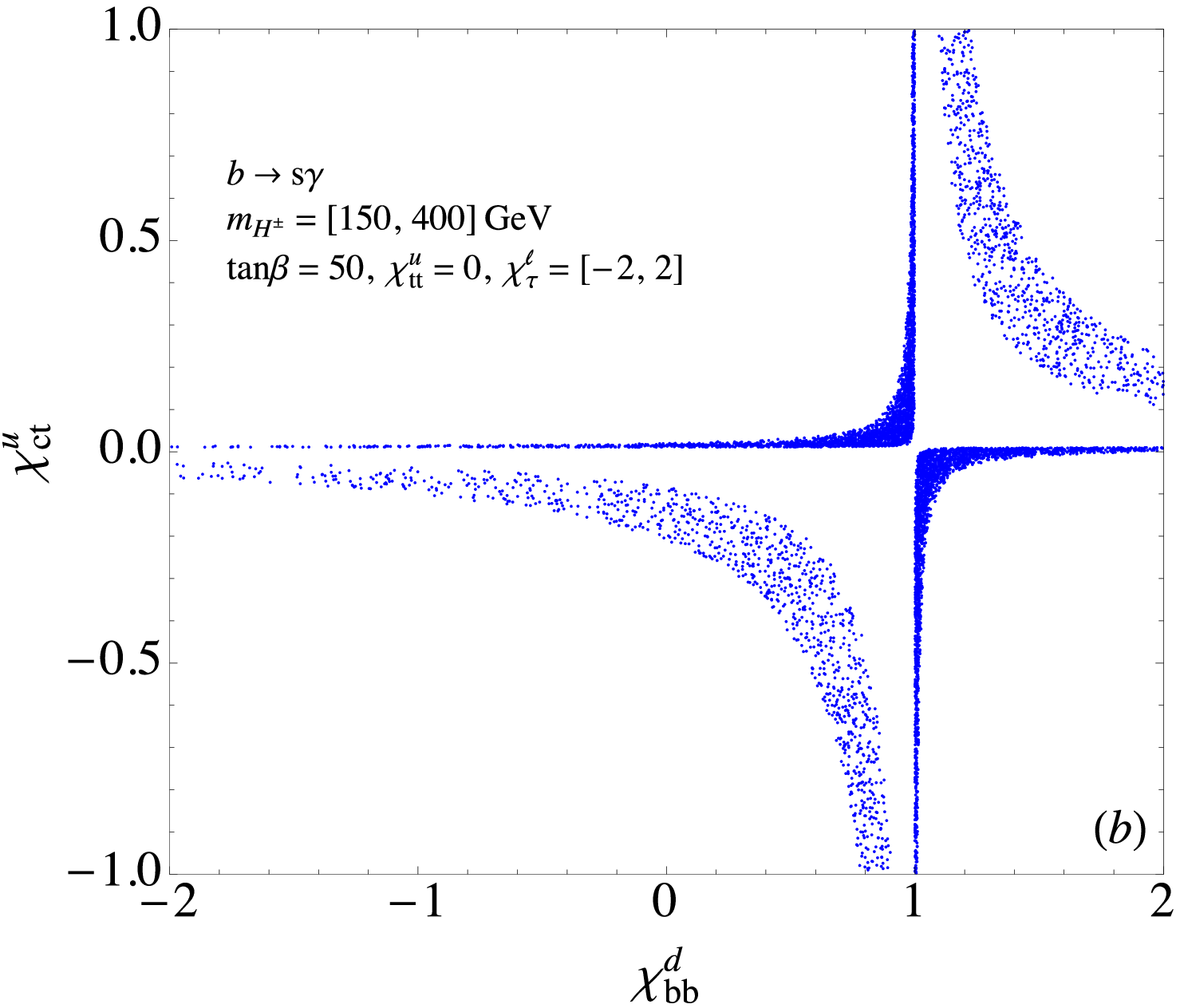}
\caption{  Allowed parameter spaces by the $\bar B \to X_s \gamma$ constraint, where $\chi^u_{tt}=0$ and $\tan\beta=50$ are fixed.}
\label{fig:limit_ab}
\end{figure}

 Due to new Yukawa couplings involved in the type-III model, e.g. $\chi^u_{tt,ct}$ and $\chi^d_{bb}$, the $b\to s\gamma$ constraint on $m_{H^\pm}$ can be relaxed. To see the  $b\to s \gamma$ constraint,   we scan the parameters with the sampling points of $5\times 10^{5}$, for which the results are shown in Fig.~\ref{fig:limit_ab}(a) and~\ref{fig:limit_ab}(b), where in both plots, $t_\beta=50$ is fixed, and the scanned regions of parameters are  set as: $m_{H^\pm}=[150, 400]$ GeV, $\chi^u_{ct}=[-1,1]$,  $\chi^d_{bb}=[-2,2]$, and $\chi^\ell_{\tau}=[-2, 2]$. Since $\chi^u_{tt}$ and $\chi^u_{ct}$  in $\chi^L_{ts}$ appear in addition form, we take $\chi^u_{tt}=0$ for simplicity, although it is not necessary.  From the results, it can be clearly seen that due to the new charged-Higgs effects, the bound on $m_{H^\pm}$ is much looser than that in the type-II model.  From the plot (b), the sampling points are condensed at $\chi^d_{bb}\approx 1$ because $(1 - \chi^d_{bb}/s_\beta)$ becomes small when $\chi^d_{bb}$ approaches one. 
  
\begin{figure}[phtb]
\includegraphics[scale=0.5]{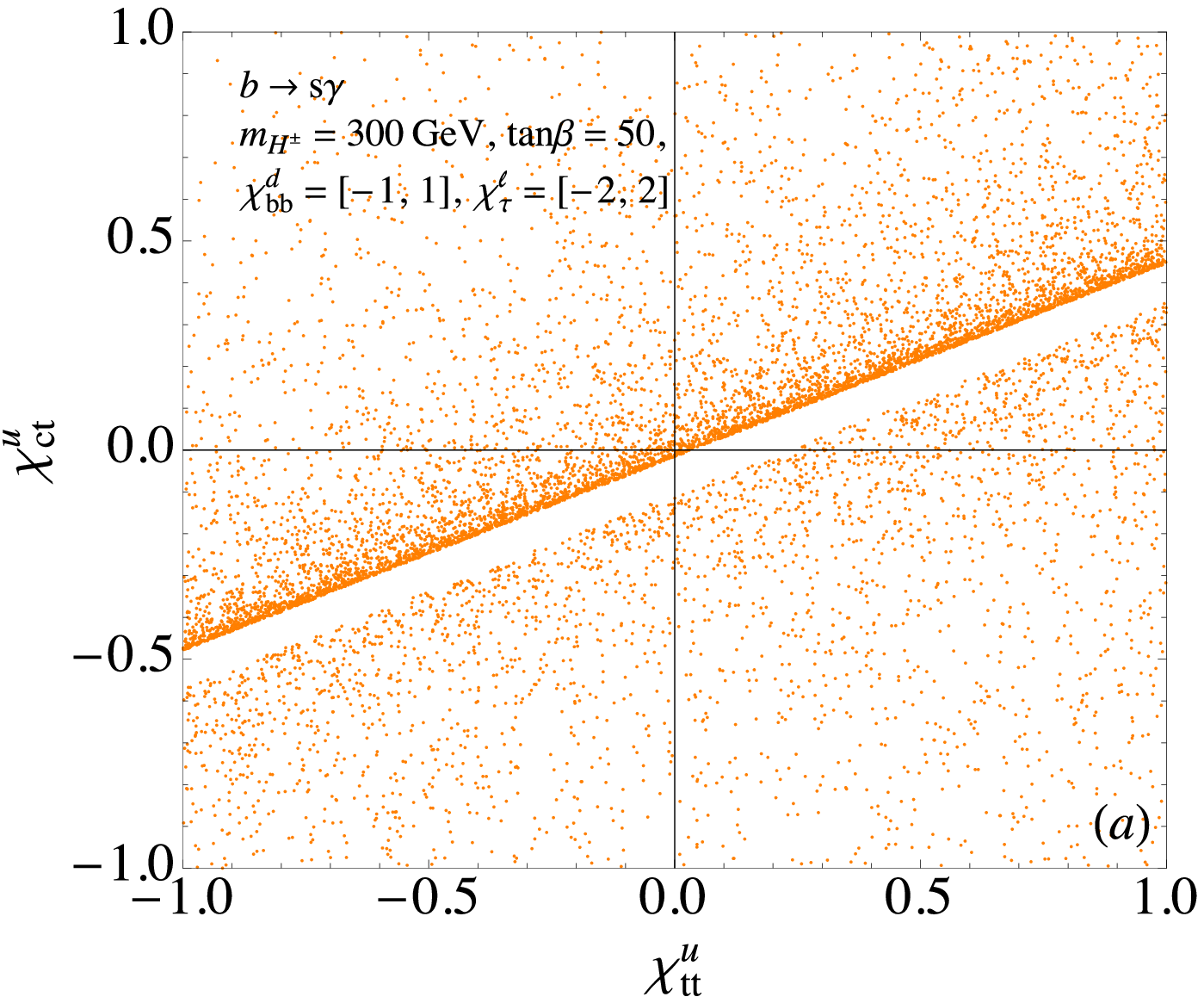}
\includegraphics[scale=0.5]{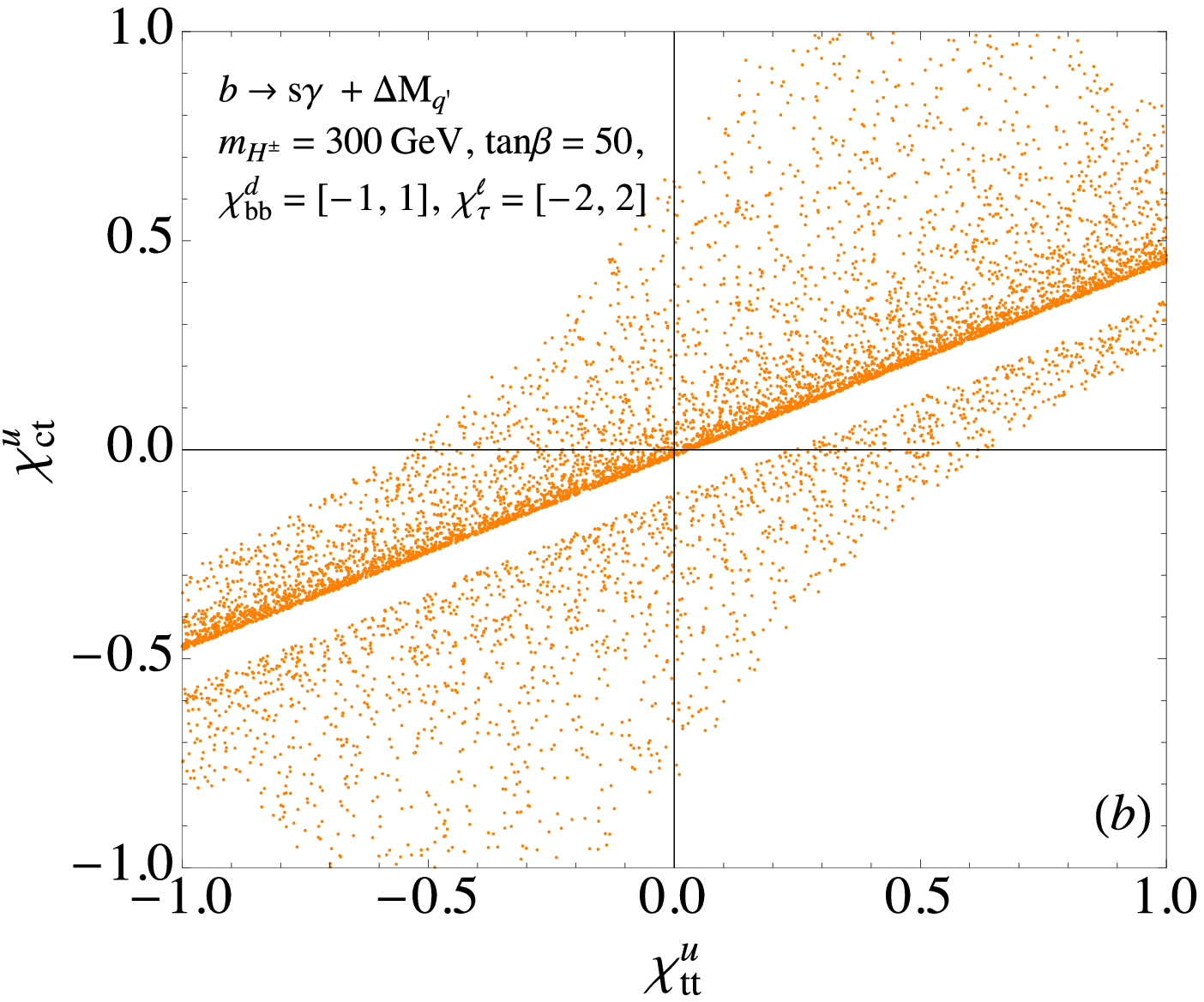}

\caption{  Allowed parameter spaces by the constraint  from (a) $\bar B\to X_s \gamma$ and (b) $\bar B\to X_s \gamma + \Delta M_{q'}$, where $\chi^d_{bb}=[-1,1]$ and $\chi^{\ell}_{\tau}=[-2,2]$ are taken, and $\tan\beta=50$ and  $m_{H^\pm}=300$ GeV are used.}
\label{fig:bound_ab}
\end{figure}

 We now know that $H^\pm$ can be as light as a few hundred GeV in the type-III model. In order to include the contributions of all $\chi^d_{tt,ct}$ and $\chi^d_{bb}$ with large $t_\beta$ and combine the constraints from  the $\Delta B=2$ processes shown in Eq.~(\ref{eq:DB2_CH}) altogether,  we fix $t_\beta=50$ and $m_{H^\pm}=300$ GeV and use the sampling points of $5\times 10^{5}$ to scan the involving parameters. The allowed parameter spaces, which only consider the $\bar B\to X_s \gamma$ constraint, are shown in Fig.~\ref{fig:bound_ab}(a), and  those  of combining the $\bar B\to X_s \gamma$ and $\Delta M_{d,s}$ constraints are given in~Fig.~\ref{fig:bound_ab}(b), where  $|\chi^u_{tt,ct}|\leq 1$, $|\chi^d_{bb}|\leq 1$, and $|\chi^\ell_\tau|\leq 2$ have been used. Comparing Fig.~\ref{fig:bound_ab}(a) and~\ref{fig:bound_ab}(b), it can be obviously seen that $\Delta B=2$ processes can further exclude some free parameter spaces. 
  
  \subsection{ Charged-Higgs on the leptonic $ B^-_q \to \ell \bar \nu$ decays}
  
\begin{figure}[tb]
\includegraphics[scale=0.5]{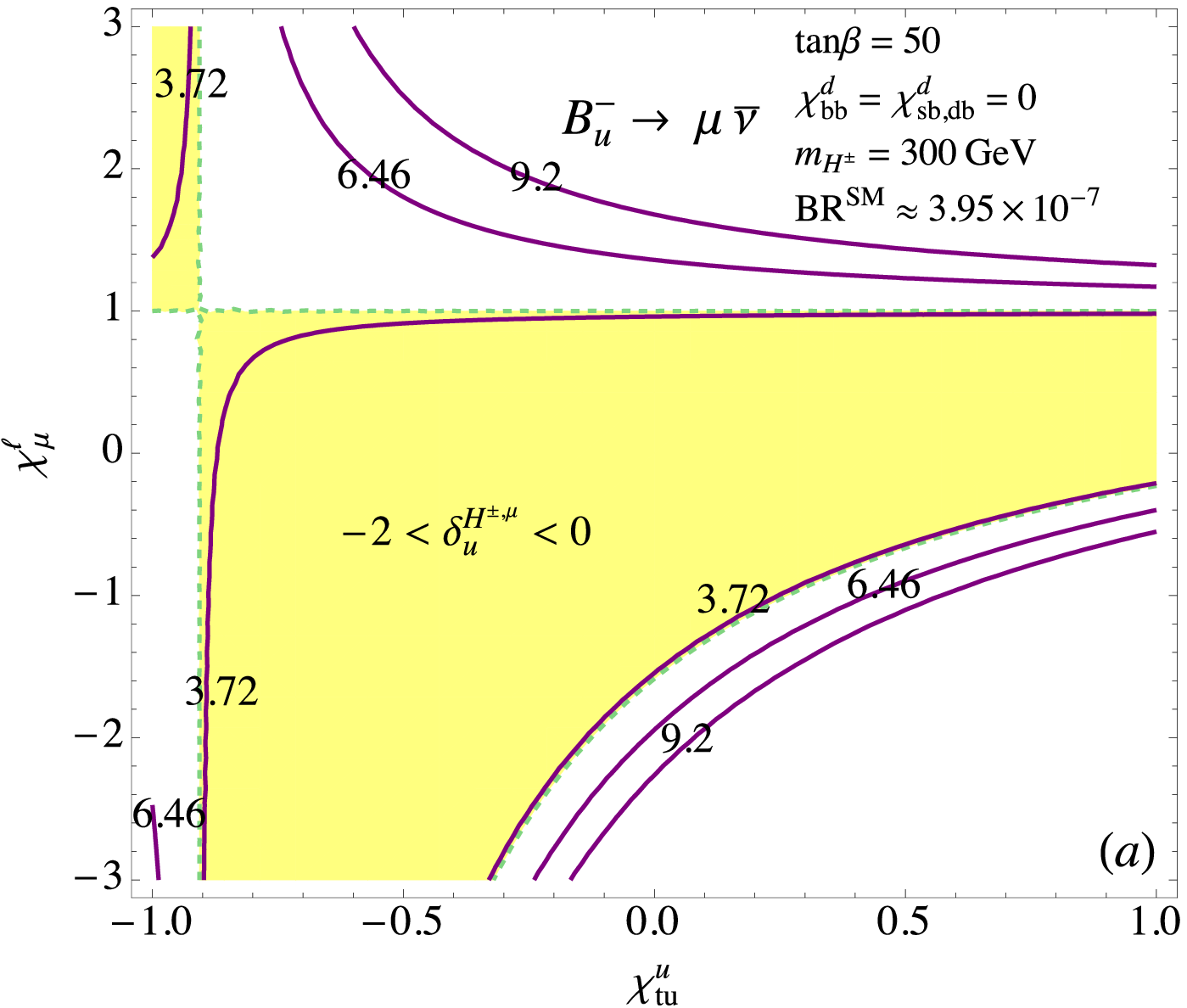}
\includegraphics[scale=0.5]{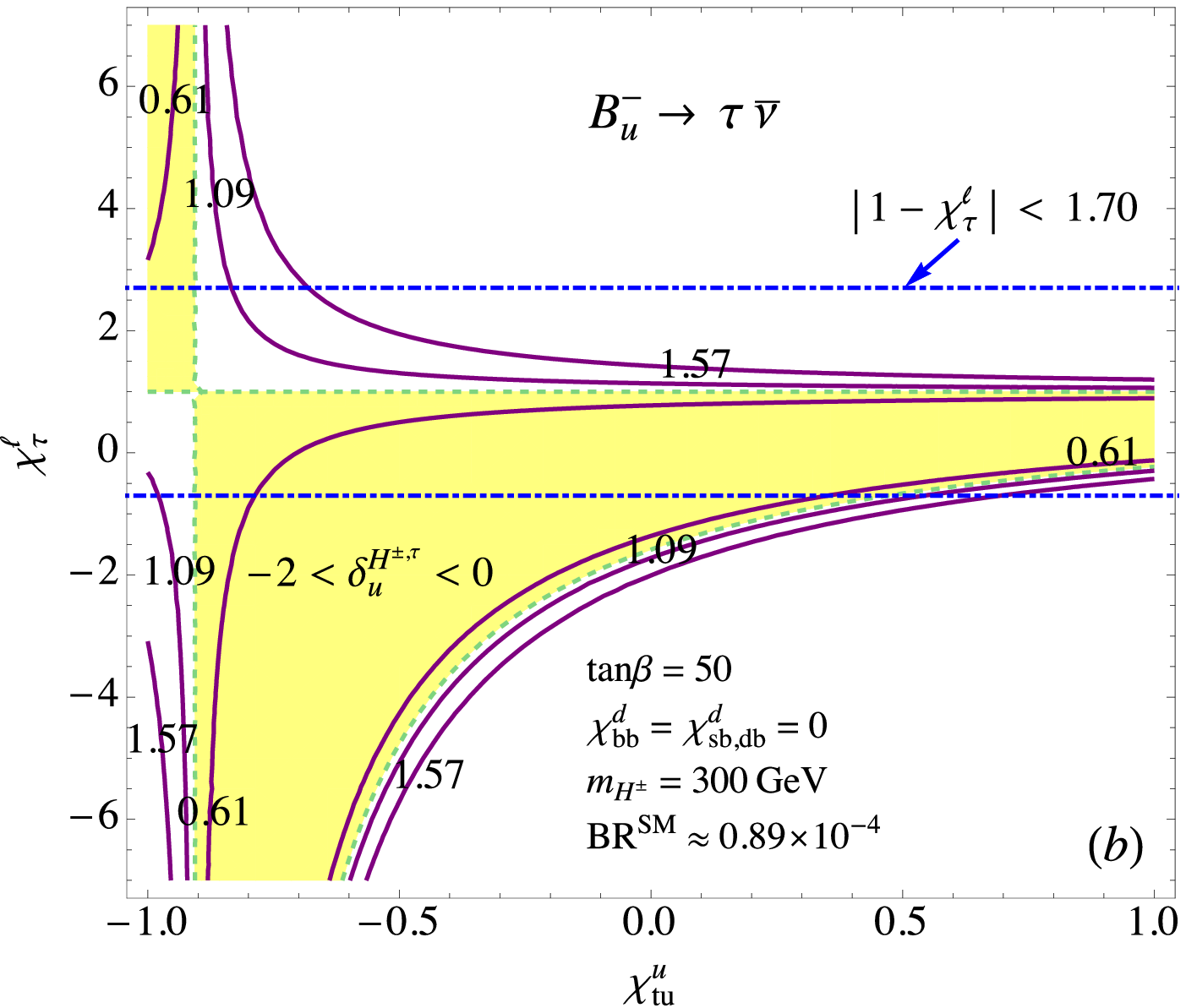}
\includegraphics[scale=0.5]{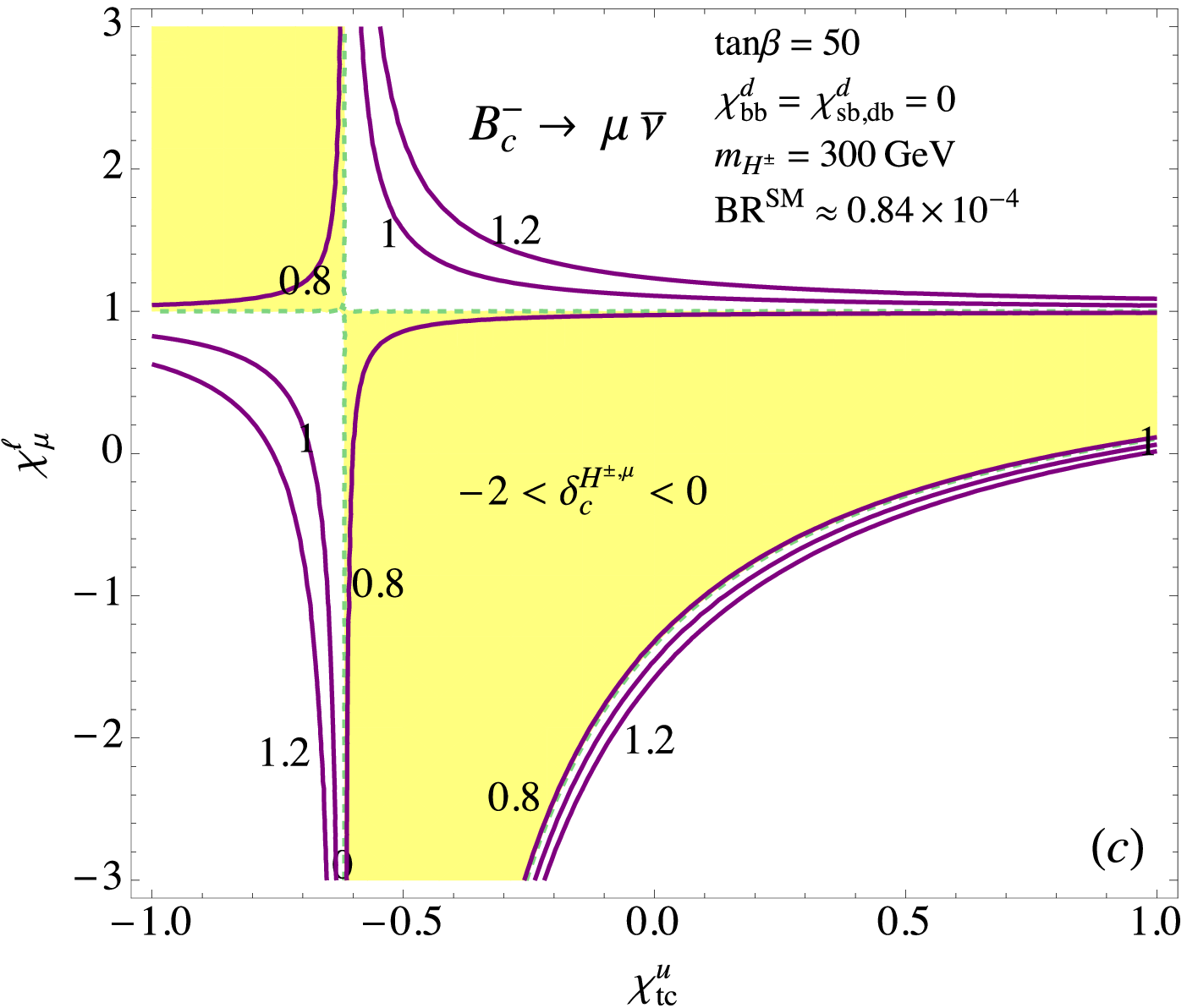}
\includegraphics[scale=0.5]{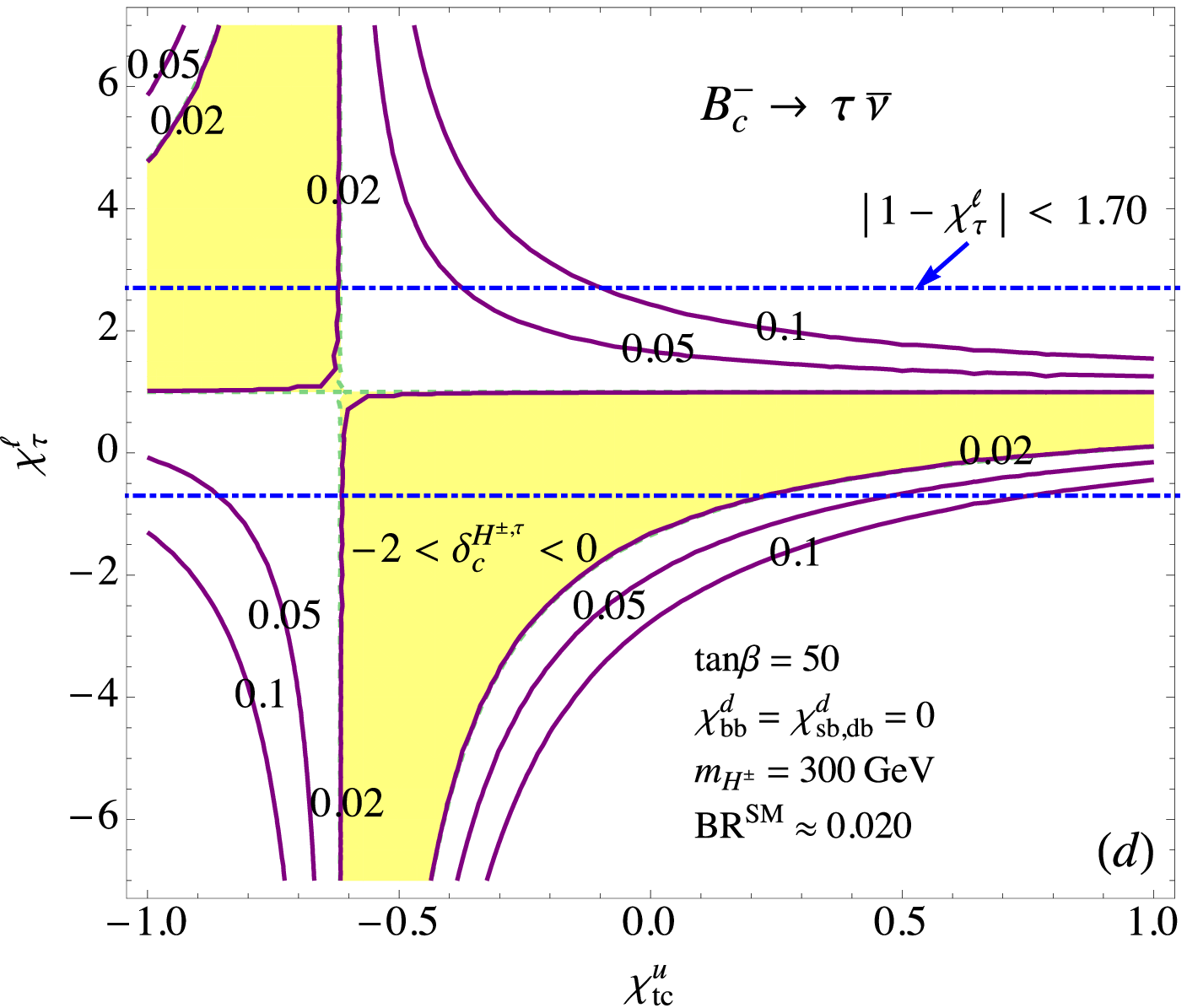}
\caption{  Contours for  (a) $BR(B^-_u\to \mu \bar \nu)$ in units of $10^{-7}$, (b) $BR(B^-_u\to \tau \bar \nu)$ in units of $10^{-4}$, (c) $BR(B^-_c\to \mu \bar \nu)$ in units of $10^{-4}$, and (d) $BR(B^-_c\to \tau \bar \nu)$, where the hatched region denotes $-2 < \delta^{H^\pm, \ell}_{q} < 0$. The dot-dashed lines in plots (b) and (d) are the constraint from Eq.~(\ref{eq:tautau}).}
\label{fig:Bellnu}
\end{figure}

  After analyzing the $b\to s \gamma$ and $\Delta B=2$ constraints, we study the charged-Higgs contributions to the leptonic and semileptonic $B$ decays in the remaining part of the paper.  In order to focus on the  $\chi^u_{tc, tu}$ and $\chi^\ell_\ell$ effects,  we fix  $\chi^d_{bb}=\chi^d_{db,sb}=0$, $t_\beta=50$, and $m_{H^\pm}=300$ in the following numerical analyses, unless stated otherwise. With the numerical inputs, the BRs of leptonic $B^-_{u,c}$ decays in the SM are estimated as:
  \begin{align}
  BR(B^-_u \to \mu \bar \nu)^{\rm SM} & \approx 3.95 \times 10^{-7}\,, \quad BR(B^-_u \to \tau \bar \nu)^{\rm SM}\approx 0.98 \times 10^{-4}\,, \nonumber \\ 
  BR(B^-_c \to \mu \bar \nu)^{\rm SM} & \approx 0.84 \times 10^{-4}\,, \quad BR(B^-_c \to \tau \bar \nu)^{\rm SM} \approx 0.02\,.
  \end{align}  
  From Eqs.~(\ref{eq:GaBqellnu}) and (\ref{eq:RH}), there are two ways to enhance the BR of $B^-_{q} \to \ell \bar \nu$: one is $\delta^{H^\pm,\ell}_q >0$, and the other is $\delta^{H^\pm,\ell}_{q} < -2 $. For clarity, the contours for $B^-_u \to (\mu, \tau) \bar \nu$ and $B^-_c \to (\mu, \tau) \bar\nu$ as a function of $\chi^\ell_{\mu ,\tau}$ and $\chi^u_{tu,tc}$ are shown in Figs.~\ref{fig:Bellnu}(a)-(d), where we have chosen the weak phase of $\chi^{u*}_{tu}$ to be the same as $V_{ub}$ so that $\delta^{H^\pm,\ell}_u$ is  real, where  the  hatched regions denote $-2 < \delta^{H^\pm, \ell}_{q} < 0$,  and the dot-dashed lines are the constraint from the $H/A\to \tau^+ \tau^-$ decays, shown in Eq.~(\ref{eq:tautau}). $\delta^{H^\pm, \ell}_{q}>0$ occurs in the up-right and down-left unhatched regions while other unhatched regions are for $\delta^{H^\pm, \ell}_{q}< -2$. From the results,  if we do not further require the values of $\delta^{H^\pm,\ell}_q$, both  $\delta^{H^\pm,\ell}_q >0$ and $\delta^{H^\pm,\ell}_q< -2$ can significantly enhance the $BR(B^-_q \to \ell \bar \nu)$. From Figs.~\ref{fig:Bellnu}(b) and \ref{fig:Bellnu}(d), although the measured values of  $BR(B^-_u \to \tau \bar\nu)$ and  the indirect upper bound of $BR(B^-_c \to \tau \bar\nu)< 10\%$~\cite{Akeroyd:2017mhr} can constrain the parameters to be a small region, the constraint from the $pp\to H/A\to \tau^+\tau^-$ processes further excludes the region of $\chi^\ell_\tau < -0.7$.  If $BR(B^-_u \to \mu \bar\nu)$ can be measured at Belle II, the $\chi^\ell_\mu$ parameter can be further constrained.

  \subsection{ Charged-Higgs on the $\bar B_d \to (\pi^+, \rho^+) \ell \bar \nu$ decays}
  
 Compared to the charged $B$-meson decays,  $\bar B_d \to (\pi^+, \rho^+) \ell \bar \nu$ have larger BRs;  thus, we discuss the neutral $B$-meson decays.  With the LCSR form factors, the BRs of these decays in the SM are given in Table~\ref{tab:Bpi_rho}, where the current measurements of light lepton channels  are also shown. From the table, we can see that the BRs for $\bar B_d \to (\pi^+,\rho^+) \ell \bar \nu$ (here $\ell=e, \mu)$ in the SM  are close to the observed values.  Due to the $H^\pm$ Yukawa coupling to the lepton being proportional to $t_\beta m_\ell /v$, the charged-Higgs contributions to the light lepton channels are small. Thus, we can conclude that the consistency between the data and the SM verifies the reliability of the LCSR  form factors in the $\bar B \to (\pi, \rho)$ transitions. In the following analysis, we  study the charged-Higgs influence on the $\tau$-lepton modes and their associated observables.  
  
  \begin{table}[htbp]
 \caption{Branching ratio for $\bar B_d \to ( \pi^+, \rho^+) \ell \bar \nu$ based on the LCSR form factors and the measured data. }
  \label{tab:Bpi_rho}
  \begin{tabular}{ c|cccc}\hline \hline
Model & ~~$\bar B_d \to \pi^+ e(\mu)\bar \nu$~~ & ~~$\bar B_d \to \pi^+ \tau \bar \nu$~~ &  ~~$\bar B_d \to \rho^+ e(\mu) \bar \nu$~~ & ~~$\bar B_d \to \rho^+ \tau \bar \nu$~~ \\  \hline
SM  & $1.43 \times 10^{-4}$  & $1.05 \times 10^{-4}$ & $2.87 \times 10^{-4}$ &  $1.68 \times 10^{-4}$\\  \hline
 Exp~\cite{PDG} & $(1.45\pm 0.05)\times 10^{-4}$ & $<2.5 \times 10^{-4} $ & $(2.94\pm 0.21)\times 10^{-4}$ &  none\\  \hline \hline

  \end{tabular}
 \end{table}
    
\begin{figure}[phtb]
\includegraphics[scale=0.5]{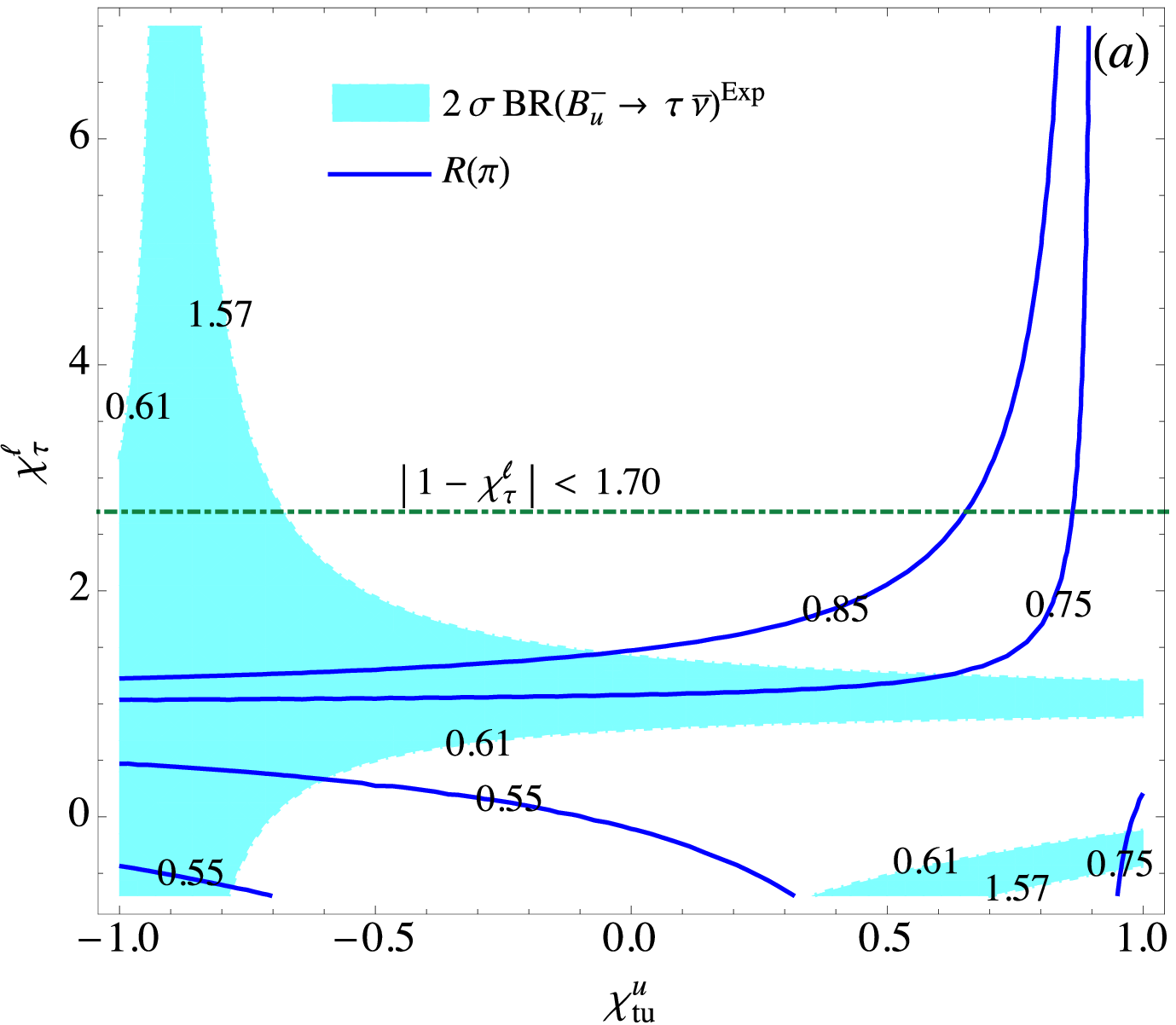}
\includegraphics[scale=0.5]{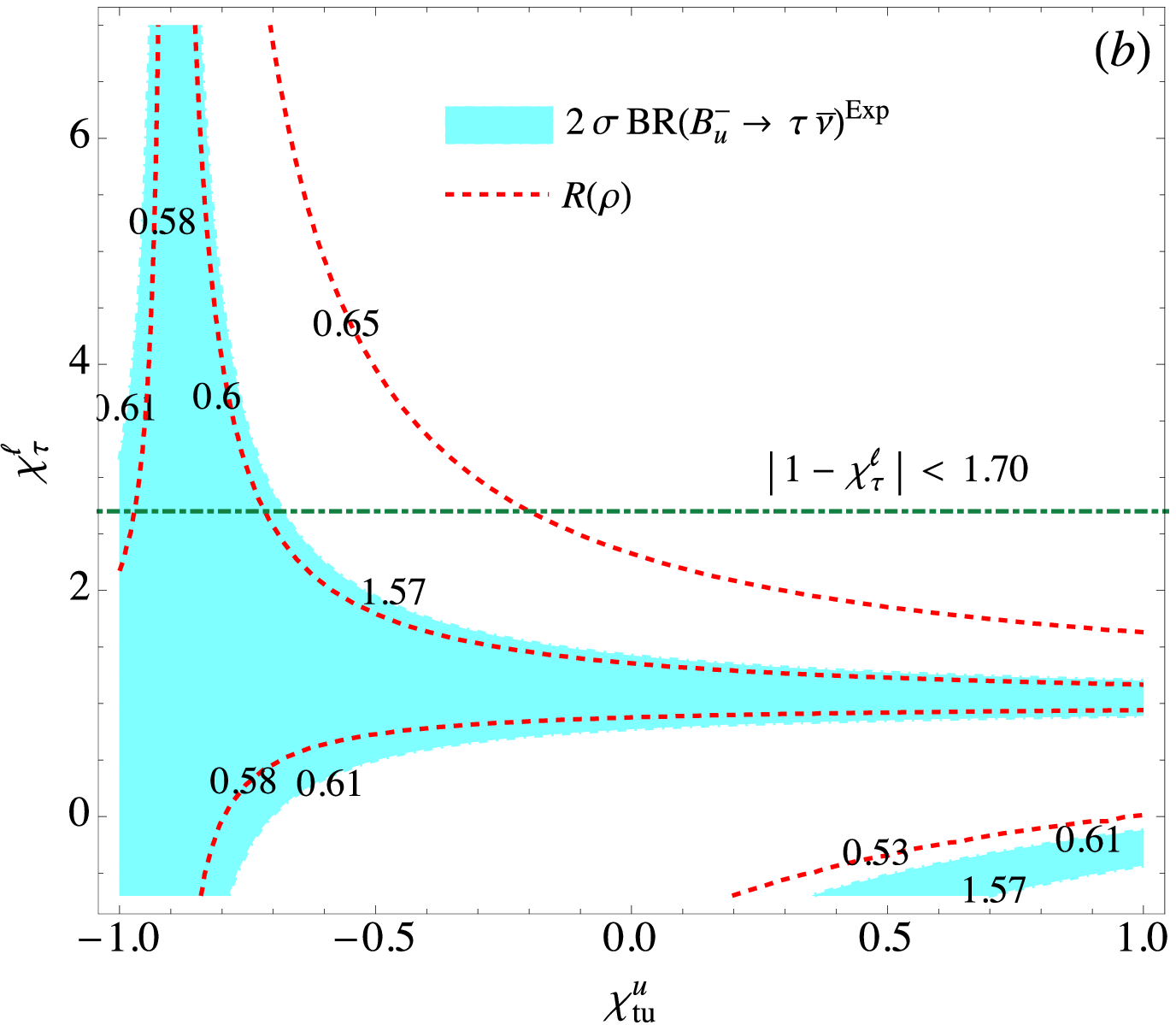}
\caption{ Contours for (a) $R(\pi)$ and (b) $R(\rho)$ as a function of $\chi^u_{tu}$ and $\chi^\ell_\tau$, where the hatched regions are the observed  $B^-_u \to \tau \bar\nu$ within $2\sigma$ errors, and the dat-dashed lines denote the constraint from the $pp\to H/A\to \tau^+ \tau^-$ processes. }
\label{fig:Rpi_Rrho}
\end{figure}

  From Table~\ref{tab:Bpi_rho}, the ratios of branching fractions for $\bar B_d \to (\pi^+, \rho^+) \ell \bar\nu$ in the SM can be estimated as:
   \begin{align}
   R(\pi) & = \frac{BR(\bar B_d \to \pi^+ \tau \bar\nu)}{BR(\bar B_d \to \pi^+ e(\mu) \bar\nu)} \approx 0.731 \,, \nonumber \\
   R(\rho) & = \frac{BR(\bar B_d \to \rho^+ \tau \bar\nu)}{BR(\bar B_d \to \rho^+ e(\mu) \bar\nu)} \approx 0.585\,.
   \end{align}
    Using Eq.~(\ref{eq:GammaBPV}), the contours for $R(\pi)$ and $R(\rho)$ as a function of $\chi^u_{tu}$ and $\chi^\ell_\tau$  are shown in Fig.~\ref{fig:Rpi_Rrho}(a) and (b), respectively, where the hatched regions denote $BR(B^-_u \to \tau \bar\nu)^{\rm exp}$ with $2\sigma$ errors. According to the results, it can be found that due to the constraint of $B^-_u \to \tau \bar\nu$, the allowed $R(\rho)$ is limited to being a very narrow range of $\sim(0.58, 0.60)$. From Fig.~\ref{fig:Rpi_Rrho}(b), since $R(\rho)$ and $BR(B^-_u \to \tau \bar\nu)$ do not  overlap at the down-right region;  basically,  this $\delta^{H^\pm,\tau}_u < 0$ region has been excluded by the data of $B^-_u \to \tau \bar \nu$. The reason, why $B^-_u \to \tau \bar\nu$ gives a strict limit on $\bar B_d \to \rho^+ \tau \bar \nu$ can be understood from Eq.~(\ref{eq:amps}), where  both decays share the same $C^{R,\tau}_{ub}-C^{L,\tau}_{ub}$ charged-Higgs effect. On the contrary, $\bar B_d \to \pi^+ \tau \bar \nu$ is related to $C^{R,\tau}_{ub}+C^{L,\tau}_{ub}$, so $R(\pi)$ can have a wider range of values.  Although the $pp\to H/A \to \tau^+ \tau^-$ constraint (dot-dashed) does not affect the allowed values of $R(\pi)$ and $R(\rho)$, it can reduce the allowed region of $\chi^\ell_\tau$. 
    
\begin{figure}[phtb]

\includegraphics[scale=0.5]{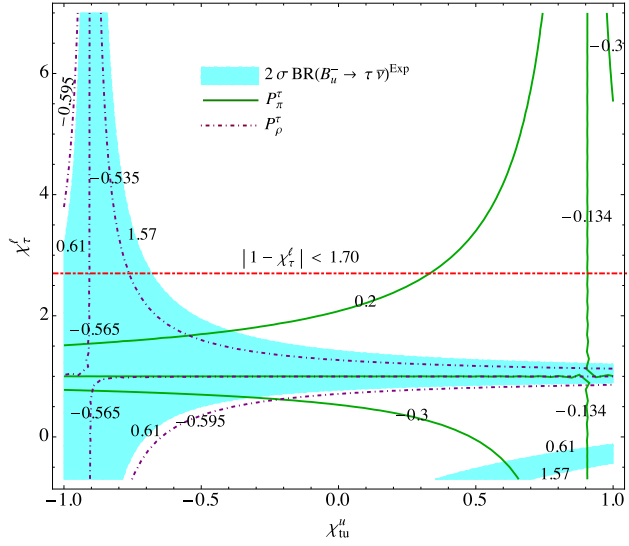}
\caption{ Contours for $P^\tau_\pi$ and $P^\tau_{\rho}$.   }
\label{fig:Ptau_pirho}
\end{figure}

   Although it is difficult to measure the lepton polarization in the $\bar B_d \to (\pi^+, \rho^+) \ell \bar\nu$, we theoretically investigate the charged-Higgs contributions to the semileptonic $B$ decays.  Using Eqs.~(\ref{eq:PtauP}) and (\ref{eq:PtauV}), the lepton helicity asymmetries  in the SM can be found as:
     \begin{align}
  P^{e(\mu)}_\pi & \approx -1 (-0.986) \,, \quad P^{\tau}_\pi  \approx  -0.134\,, \nonumber \\
  P^{e(\mu)}_{\rho} & \approx -1 ( -0.992)\,, \quad P^{\tau}_{\rho} \approx -0.565\,.
  \end{align}
Due to the fact that the helicity asymmetry is strongly dependent on $m_\ell$,  it can be understood that only $\tau\bar\nu$ modes can be away from unity. All lepton polarizations show negative values because the $V-A$ current in the SM dominates. The sign of $\tau$-lepton polarization in $\bar B \to D \tau\bar\nu$ can be flipped to be a  positive sign. In order to show the $H^\pm$ influence, the contours for $P^{\tau}_{\pi}$ and $P^{\tau}_{\rho}$ as a function of $\chi^u_{tu}$ and $\chi^\ell_\tau$ are given in Fig.~\ref{fig:Ptau_pirho},  where the constraint from $pp\to H/A\to \tau^+ \tau^-$ (dot-dashed) with $\chi^d_{bb}=0$ is also shown. With the $B^-_u \to \tau \bar\nu$ constraint, the allowed values of $P^{\tau}_{\rho}$ are limited in a  narrow region around the SM value. However, the allowed values of $P^\tau_\pi$ are wider and can have both negative and positive signs.

\begin{figure}[phtb]
\includegraphics[scale=0.5]{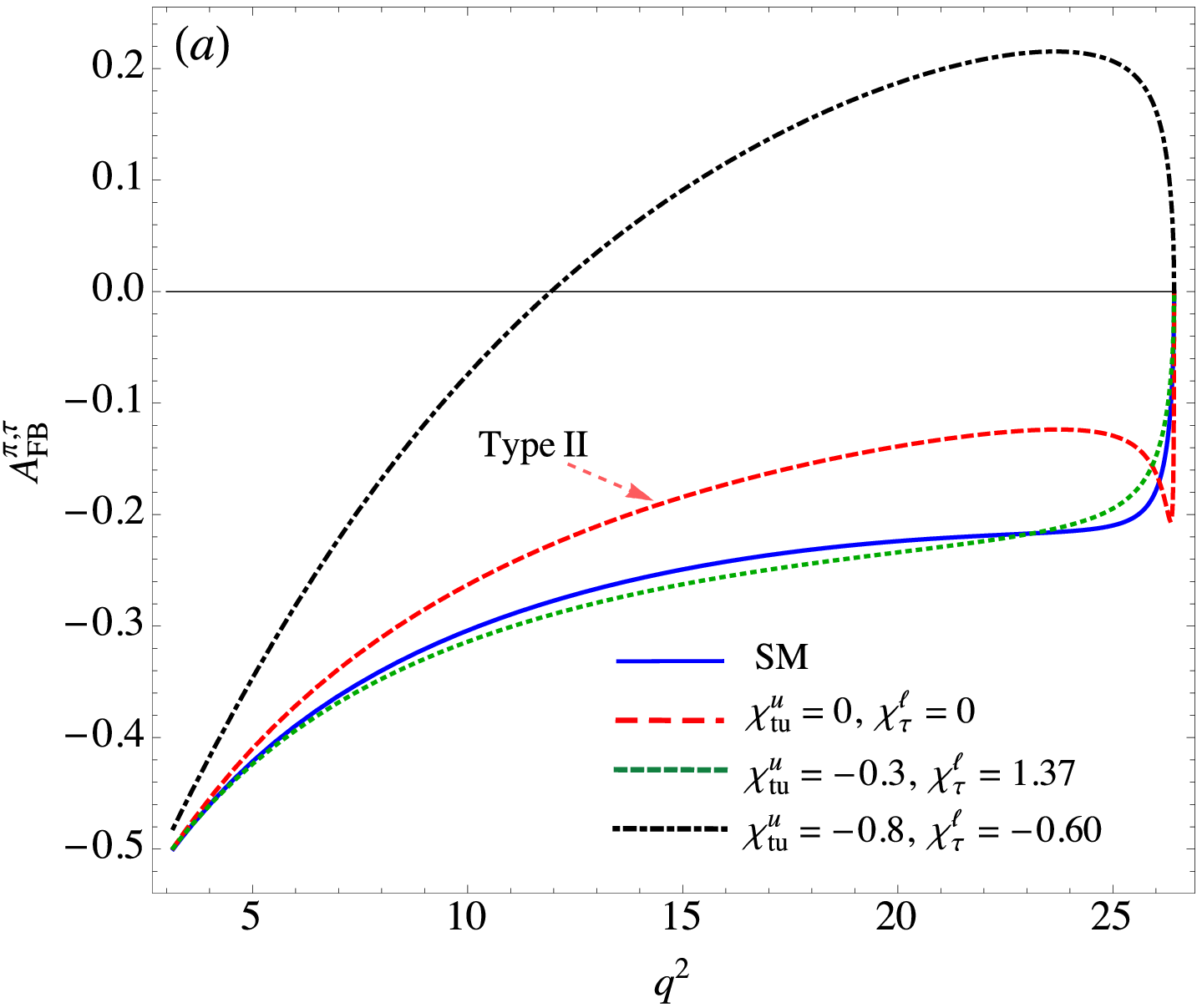}
\includegraphics[scale=0.5]{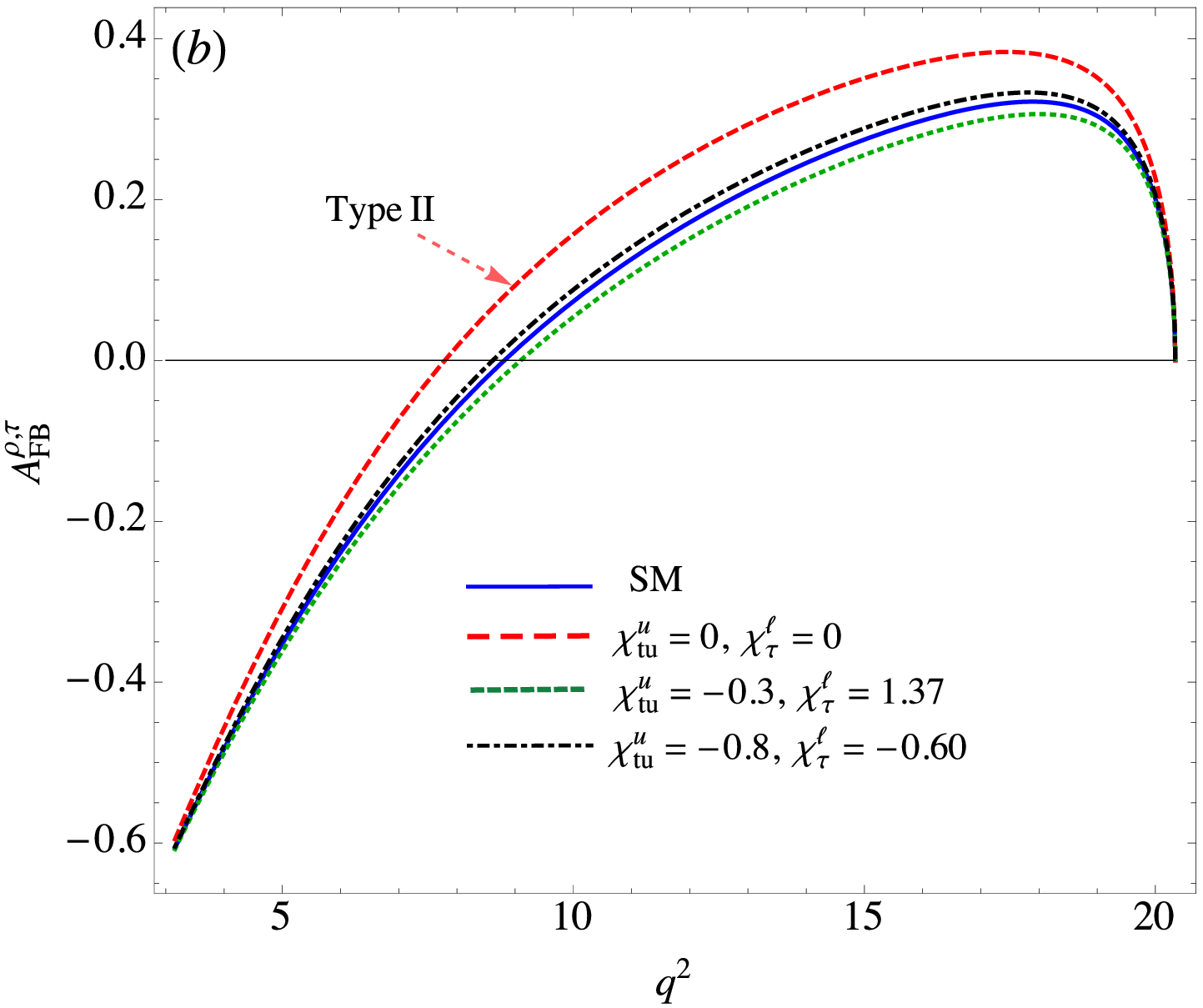}
\caption{ $q^2$-dependent lepton forward-backward asymmetry for (a) $\bar B_d \to \pi^+ \tau \bar\nu$ and (b) $\bar B_d \to \rho^+ \tau \bar \nu$ in the SM (solid), type-II (dashed), and type-III 2HDM (dotted, dot-dashed).    }
\label{fig:FBA_pirho}
\end{figure}

The lepton FBAs are also interesting observables in the semileptonic $B$ decays. Following the formulae in Eq.~(\ref{eq:fba}), we show the FBAs of  $\bar B_d \to \pi^+ \tau \bar\nu$ and $\bar B_d \to \rho^+ \tau \bar\nu$ as a function of $q^2$ in Fig.~\ref{fig:FBA_pirho}(a) and (b), respectively, where the solid line is the SM and  the dashed line is the type-II model. For the type-III 2HDM, we select two benchmarks that obey the $B^-_u \to \tau \bar\nu$ constraint as follows: the dotted line is  $\chi^u_{tu}=-0.3$ and $\chi^\ell_\tau=1.37$, which lead to $R(\pi)\approx 0.855$ and $R(\rho)\approx 0.595$; and the dot-dashed line denotes $\chi^u_{tu}=-0.8$ and $\chi^\ell_\tau=-0.60$, which lead to $R(\pi)\approx 0.550$ and $R(\rho)\approx 0.577$. From plot (a), we can see that $A^{\pi,\tau}_{FB}$ can be largely  changed  by the charged-Higgs effect; in other words, a zero-point  can occur in $A^{\pi,\tau}_{FB}$, where the zero point usually occurs in the $\rho^+$ channel, as shown in plot (b). Hence, we can use the characteristics of FBA to test the SM by examining the shape of $A^{\pi,\tau}_{FB}$.  From the plot (b), due to the strict limit of $B^-_u \to \tau \bar\nu$, the shape change of $A^{\rho,\tau}_{FB}$ in the type-III model is small.

   \subsection{ Charged-Higgs on the $B^-_u \to (D^0, D^{*0})\ell \bar \nu$ decays}
   
   From Eq.~(\ref{eq:GammaBPV}) and the HQET form factors introduced previously, the BRs for the $B^-_u \to (D^0, D^{*0}) \ell \bar \nu$ decays in the SM can be estimated, as shown in Table~\ref{tab:BD_Dv}, where  the current experimental results are also included~\cite{PDG}. It can be seen that  the BRs of the light lepton channels in the SM are consistent with the experimental data; however, the  $\tau\bar\nu$ mode results are somewhat smaller than  those in the current data. The  ratios of branching fractions are obtained as $R(D)^{\rm SM}\approx 0.309$ and $R(D^*)^{\rm SM}\approx 0.257$, which are consistent with the results obtained in the literature.  
 
   \begin{table}[htbp]
 \caption{ Branching ratios for $B^-_u \to (D^0, D^{*0}) \ell \bar\nu$ in the SM and the associated experimental data. }
  \label{tab:BD_Dv}
  \begin{tabular}{ c|cccc}\hline \hline
Model & ~~$ B^-_u \to D^0 \ell \bar \nu$~~ & ~~$ B^-_u \to D^0 \tau \bar \nu$~~ &  ~~$B^-_u \to D^{*0} \ell \bar \nu$~~ & ~~$B^-_u \to D^{*0} \tau \bar \nu$~~ \\  \hline
SM  & $2.10\; \%$  & $6.48\times 10^{-3}$ & $5.74\; \%$ &  $1.48\; \%$\\  \hline
 Exp \cite{PDG} & $(2.27\pm 0.11)\%$ & $(7.7\pm 2.5)\times 10^{-3}$ & $(5.69\pm 0.19)\%$ & $(1.88 \pm 0.20)\%$\\  \hline \hline

  \end{tabular}
 \end{table}
  
\begin{figure}[phtb]
\includegraphics[scale=0.5]{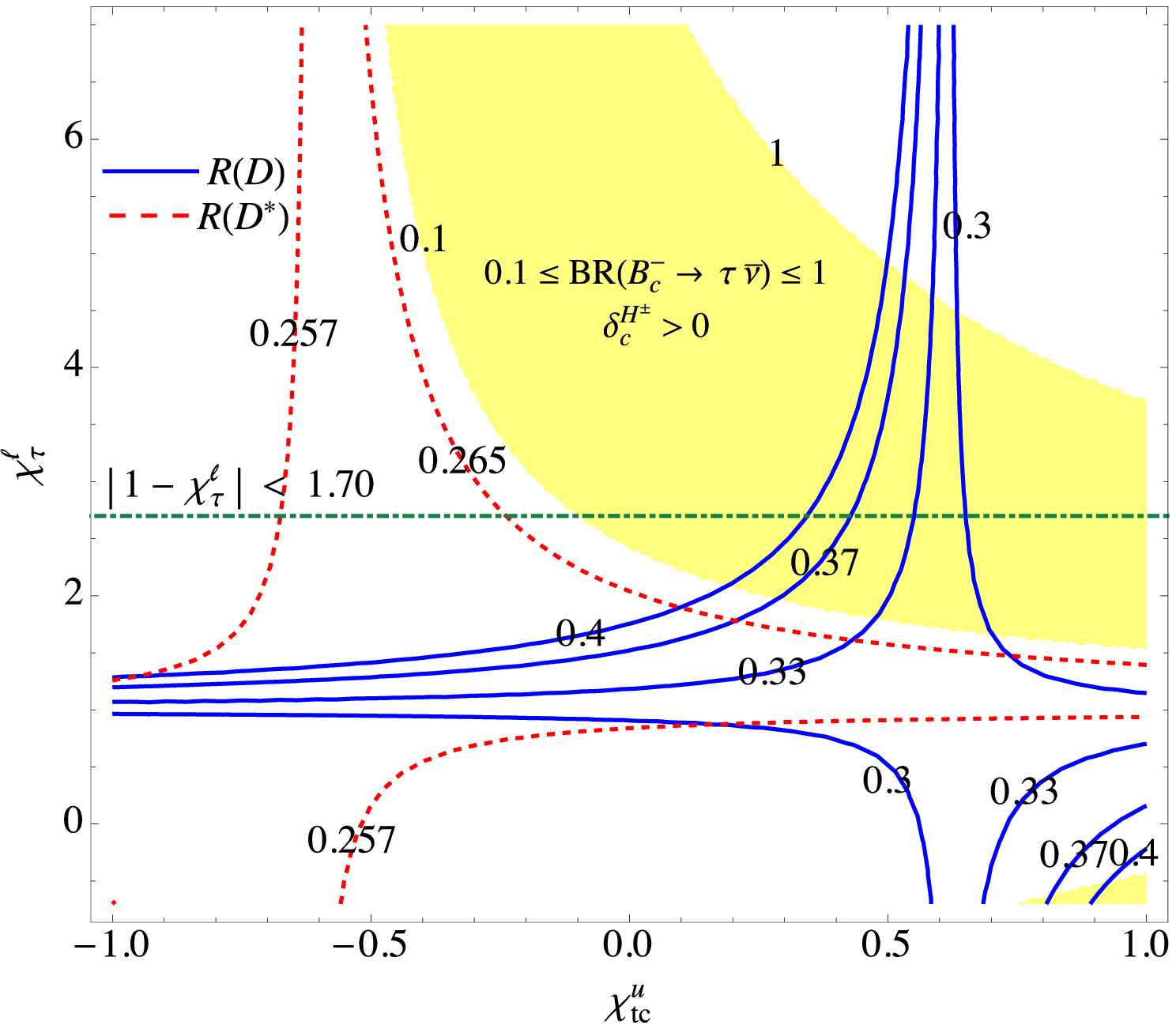}
\includegraphics[scale=0.5]{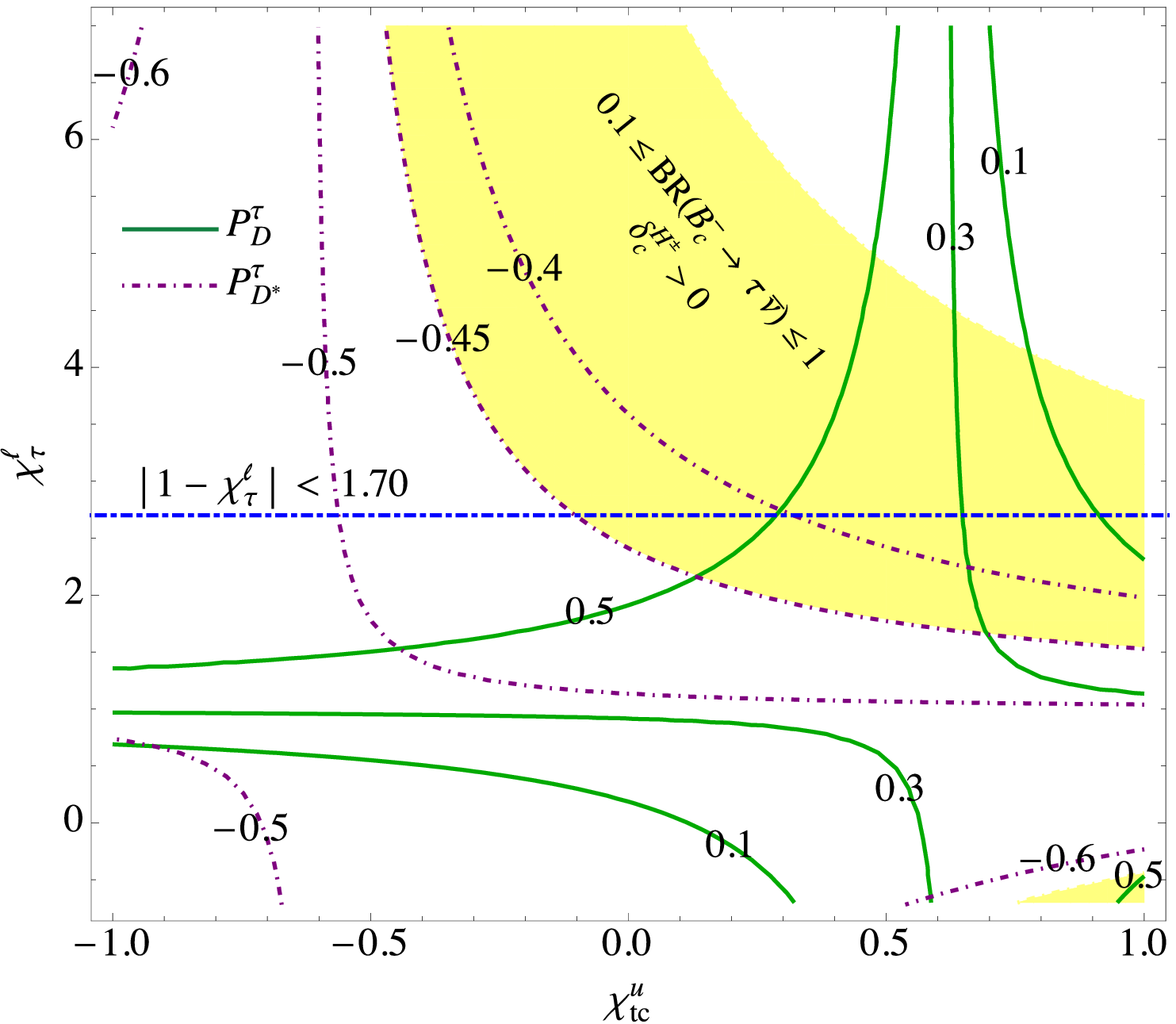}
\caption{ Left panel: $R(D)$, $R(D^*)$, and $BR(B^-_c \to \tau \bar \nu)$ as a function of $\chi^u_{tc}$ and $\chi^\ell_\tau$, where the shaded regions denote  the situation of $0.1\leq BR(B^-_c \to \tau \bar \nu)  \leq 1$. Right panel: Contours for $P^\tau_D$ and $P^\tau_{D^*}$.  The dot-dashed lines are the constraint shown in Eq.~(\ref{eq:tautau}) with $\chi^d_{bb}=0$. }
\label{fig:RD_RDv}
\end{figure}

  As discussed before, the  $H^\pm$ contributions to $B^-_u \to D^0 \ell \bar \nu$ and $B^-_u \to D^{*0} \ell \bar \nu$ are associated with   $C^{R,\ell}_{cb} + C^{L,\ell}_{cb}$ and $C^{R,\ell}_{cb} - C^{L,\ell}_{cb}$, respectively, and the same factor  $C^{R,\ell}_{cb} - C^{L,\ell}_{cb}$ also appears in the $B^-_c \to \ell \bar \nu$ decay;  that is,  $R(D^*)$ and $BR(B^-_c \to \tau \bar\nu)$  have a strong correlation~\cite{Li:2016vvp,Alonso:2016oyd,Akeroyd:2017mhr}. Although there is no direct measurement of the $B^-_c \to \tau \bar\nu$ decay, the indirect upper limit  on the $BR(B^-_c \to \tau \bar\nu)$ can be obtained by the lifetime of $B_c$ with a result of  $30\%$~\cite{Alonso:2016oyd} and the LEP1 data~\cite{Akeroyd:2017mhr} with a result of $10\%$. We show $R(D)$ and $R(D^*)$ as a function of $\chi^{u}_{tc}$ and $\chi^{\ell}_{\tau}$  in  Fig.~\ref{fig:RD_RDv} (left panel), where the shaded regions denote the  results for  $0.1 \leq  BR(B^-_c \to \tau \bar \nu)\leq 1$,  and the dot-dashed line is the upper bound from the $pp\to H/A\to \tau^+ \tau^-$ processes with $\chi^d_{bb}=0$. For clarity, we also show  the  regions for $\delta^{H^\pm}_c >0$ and $\delta^{H^\pm}_c < -2$ in the plot. From the results, we can clearly see that due to the limit of $BR(B^-_c \to \tau \bar \nu)<10\%$, the maximal value of the charged-Higgs contribution to $R(D^*)$ can be only approximately $0.265$; however, the values of $R(D)$ can be  within a $1\sigma$  world average. 

 According to Eqs.~(\ref{eq:PtauP}) and (\ref{eq:PtauV}), it is expected that the  helicity asymmetry of a light lepton will  negatively approach unity, and that only $\tau$-lepton polarizations can  significantly deviate  from one. With HQET form factors, the lepton polarizations 
 in the SM are estimated as:
  \begin{align}
  P^{e(\mu)}_D & \approx -1 (-0.962) \,, \quad P^{\tau}_D  \approx  0.320\,, \nonumber \\
  P^{e(\mu)}_{D^*} & \approx -1 ( -0.986)\,, \quad P^{\tau}_{D^*} \approx -0.506\,,
  \end{align}
where the Belle's current measurement is $P^\tau_{D^*}=-0.38\pm 0.51^{+0.21}_{-0.16}$~\cite{Hirose:2016wfn}. Intriguingly, the sign of $P^{\tau}_{D}$ is opposite to that of $P^{\tau}_{D^*}$, and  the situation is different from the negative sign in $P^{\tau}_{\pi}$.  We find that the origin of the difference in sign between $P^{\tau}_{\pi}$ and $P^{\tau}_{D}$ is from the meson mass.  Due to $m_D \gg m_\pi$, the positive helicity becomes dominant in $B^-_u \to D^0 \tau \bar\nu$. To see the influence of the charged-Higgs on the $\tau$ polarizations, we show the contours for $P^\tau_D$ and $P^\tau_{D^*}$ in the right-panel of Fig.~\ref{fig:RD_RDv}.  With the limit of $BR(B^-_c \to \tau \bar \nu) < 10\%$,  it is found that $P^\tau_D$ can be largely changed by the charged-Higgs effect, and the allowed range of $P^\tau_{D^*}$ is narrow and can be changed by $\sim 10\%$, where the change  in $R(D^*)$ from the same $H^\pm$ effects is only $\sim 3\%$.

\begin{figure}[phtb]
\includegraphics[scale=0.5]{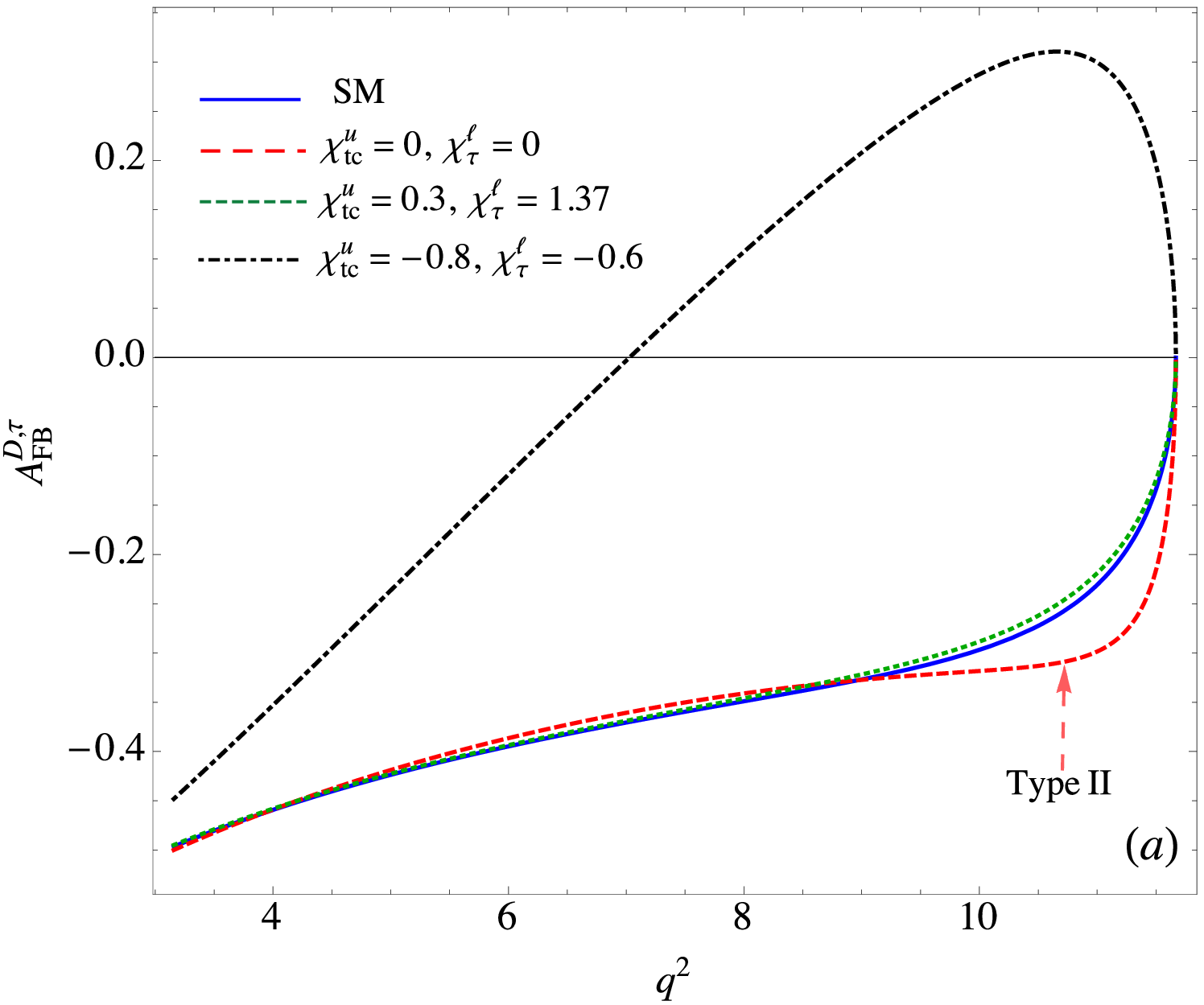}
\includegraphics[scale=0.5]{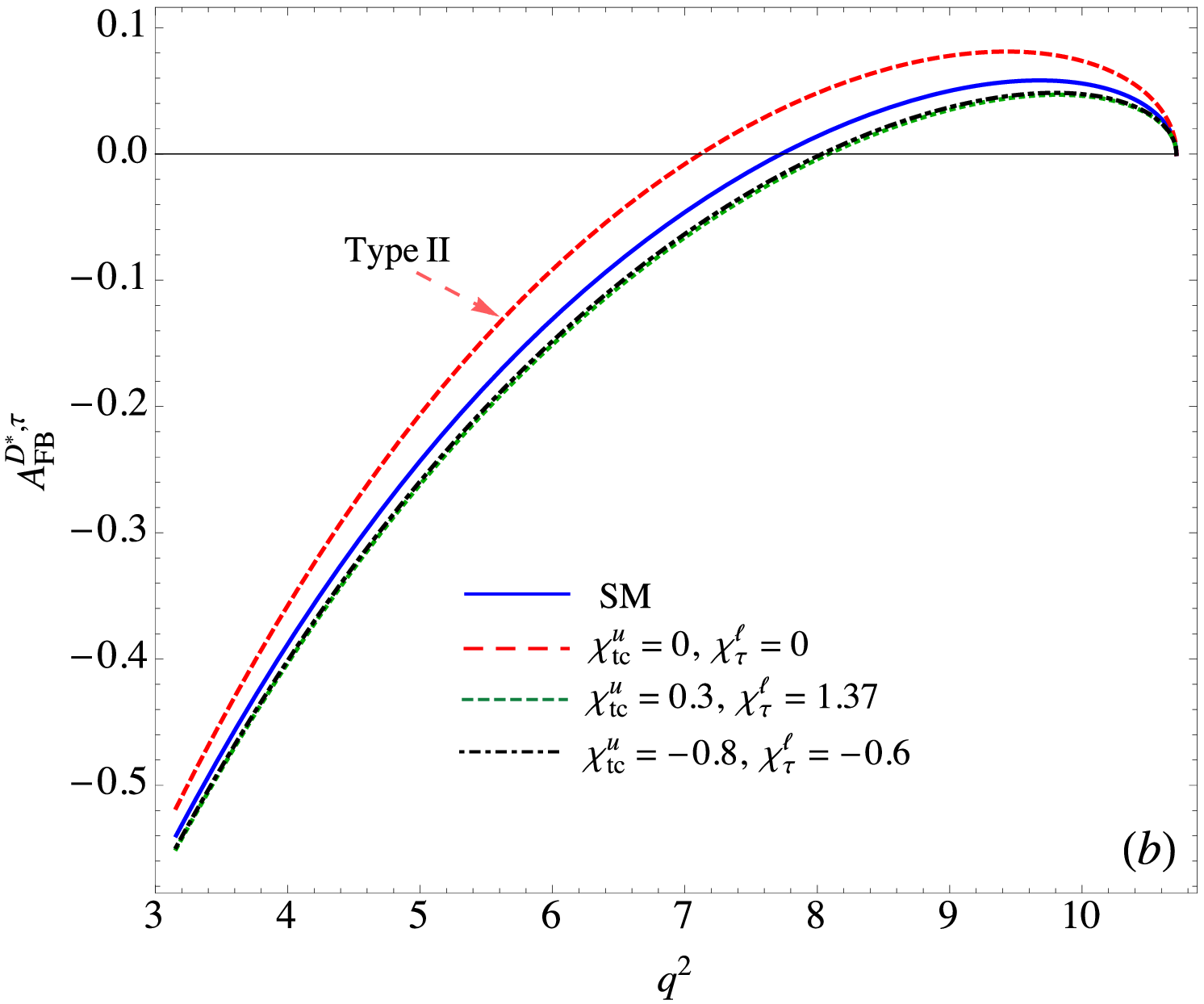}
\caption{ $\tau$-lepton forward-backward asymmetry as a $q^2$-dependence for (a) $B^-_u \to D^0 \tau \bar\nu$ and (b) $B^-_u \to D^{*0} \tau \bar\nu$,  where the solid line is from the SM; the dashed line is the type-II model with $R(D^{(*)})\approx 0.220(0.252)$; the dotted (dotdashed) line is from  $\chi^u_{tc}=0.3 (-0.8)$ and $\chi^\ell_\tau=1.37(-0.60)$, and the corresponding results are $R(D) \approx 0.331(0.145)$ and $R(D^*)\approx 0.262(0.261)$.   }
\label{fig:FBA}
\end{figure}

 Finally, we discuss the lepton FBAs in the $B^-_u \to (D^0, D^{*0})\ell \bar\nu$ decays. As discussed in the $\bar B_d \to ( \pi^+, \rho^+) \ell \bar\nu$ decays, only $A^{D^{(*)},\tau}_{FB}$ are sensitive to the charged-Higgs effects. Thus, we show the $A^{D^{[*]}, \tau}_{FB}$ as a function of $q^2$ in Fig.~\ref{fig:FBA}(a)[(b)], where the solid line denotes the SM result and  the dashed line is the type-II model with $R(D^{(*)})=0.220(0.252)$. We use two benchmarks to show the effects of the type-III 2HDM:  the dotted line  is   the result of $\chi^u_{tc}=0.3$ and  $\chi^\ell_\tau=1.37$ which cause  $R(D^{(*)}) \approx 0.331 (0.262)$, and the dot-dashed line denotes $\chi^u_{tc}=-0.8$ and  $\chi^\ell_\tau=-0.60$ which causes $R(D^{(*)}) \approx 0.145 (0.261)$. From plot (a), similar to the case in $A^{\pi,\tau}_{FB}$, $A^{D,\tau}_{FB}$ can have a vanishing point  in the type-III model when it crosses the $q^2$ axis. Usually,   the zero-point occurs in $B^-_u \to D^{*0} \ell \bar\nu$, and the position of zero-point is sensitive to the new physics, as shown in plot (b). Hence, based on our analysis, we can use this characteristics of FBA  to test the SM.   
  
  \section{Conclusion}
  
  We studied the constraints of the $b\to s \gamma$ and $\Delta B=2$ processes in the type-III 2HDM with the Cheng-Sher ansatz, where the detailed analyses included the neutral scalars $H$ and $A$ (tree + loop) and charged-Higgs (loop) effects. It was found that the tree-induced $\Delta B=2$ processes produce strong constraints on the parameters $\chi^d_{db}$ and $\chi^d_{sb}$, and due to the $m_b/m_{H(A)}$ suppression, the loop-induced $b\to s \gamma$ process by the same $H, A$ effects is small. When we ignore the $\chi^d_{db, sb}$ effects, the dominant contributions to the rare processes are the charged-Higgs.  
  
  We demonstrated  that due to the new parameters involved, i.e. $\chi^u_{tt,tc}$ and $\chi^d_{bb}$, the mass of charged-Higgs in the type-III model can be much lighter than that in the type-II model when  the $b\to s \gamma$ constraint is satisfied. Taking $m_{H^\pm}=300$ GeV and $\tan\beta=50$, we comprehensively studied the charged-Higgs contributions to the leptonic $B^-_{u,c} \to \ell \bar \nu$ and semileptonic $\bar B_{u,d} \to (P ,V) \ell \bar\nu$ ($P=\pi^+, D^0; V= \rho^+, D^{*0}$) decays in the generic 2HDM.
  
  In addition to the constraints from the low energy flavor physics, such as $B_{d,s}$-$\bar B_{d,s}$ mixings and $B_s \to X_s \gamma$, we also consider the constraint from the upper limit of  $pp\to H/A\to \tau^+ \tau^-$ measured in LHC. It was found that the tau-pair production cross section can further constrain the $\chi^\ell_{\tau}$ parameter to be $|1-\chi^\ell_\tau|<1.70$  with $\chi^d_{bb}=0$.

 The main difference in the $b \to (u,c) \ell \bar\nu$ decays between type-II and type-III is that the former is always destructive to the SM results, and the latter can make the situation  constructive. Therefore,  $BR(B^-_u \to (\mu, \tau) \bar \nu)$  can be enhanced from the SM results to  the current experimental observations. Although $B^-_c \to \tau \bar\nu$ has not yet  been observed, the charged-Higgs can also enhance its branching ratio from $2\%$ to the upper limit of $10\%$, where the upper limit is  obtained from the LEP1 data. 
  
  Since heavy lepton can be significantly affected by the charged-Higgs, we analyzed the potential observables in the $\bar B_d \to (\pi^+, \rho^+) \tau \bar \nu$ and $B^-_u \to (D^0, D^{*0}) \tau \bar\nu$ decays. It was shown that since $B^-_{u(c)} \to \tau \bar\nu$ and $B^-_{u} \to \rho^+ (D^{*0}) \tau \bar\nu$ are strongly correlated to the same charged-Higgs effects, the allowed  $R(\rho^+)$, $R(D^*)$, $P^{\tau}_{\rho}$, $P^{\tau}_{D^*}$, and $A^{\rho,\tau}_{FB}$  are very limited in terms of  deviating from the SM. Although the change in $A^{D^*,\tau}_{FB}$ is not large, the deviation is still sizable.  In contrast, the observables in the $\pi^+$ and $D^0$ channels are sensitive to the charged-Higgs effects and exhibit significant changes. \\


\appendix 
\section{}
 
 \subsection{$H^\pm$ Yukawa couplings to the quarks}
 
 According to Eq.~(\ref{eq:YuCH}), we write the charged-Higgs Yukawa couplings to the quarks as
  \begin{align}
 {\cal L}^{H^\pm}_{Y,q} & = \frac{\sqrt{2}}{v}  \bar u_{i R} C^L_{u_i d_k} d_{kL} H^+ +   \frac{\sqrt{2}}{v}  \bar u_{i L} C^R_{u_i d_k} d_{k R} H^+ + H.c.\,, \nonumber \\
 C^L_{u_i d_k} & =  \left(  \frac{m_{u_i}}{ t_\beta }  \delta_{ij} - \frac{\sqrt{m_{u_i} m_{u_j}}}{ s_\beta}  \chi^{u*}_{ji} \right) V_{u_j d_k} \,, \nonumber \\
C^R_{u_i d_k} &= V_{u_i d_j} \left( t_\beta  m_{d_j}  \delta_{j k} - \frac{\sqrt{m_{d_j} m_{d_k}}}{  c_\beta} \chi^d_{jk}\right)\,, \label{eq:CLRud}
 \end{align}
where $u_j$ and $d_j$ denote the sum of all possible up- and down-type quarks, respectively. We showed the $b$-quark related Yukawa couplings in the texts. Here, we discuss the $H^\pm$ Yukawa couplings to $d$- and $s$-quark. In the numerical discussions, we used $m_{u(d)}\approx 5.4$ MeV, $m_s\approx 0.1$ GeV, $m_c\approx 1.3$ GeV, and $m_t\approx 165$ GeV. 
 
  \underline {$ ud H^+$ vertex:}  Following Eq.~(\ref{eq:CLRud}), we write the $C^L_{ud}$ coupling as:
 \begin{align}
 \frac{\sqrt{2}}{v} C^{L}_{ud} = \frac{\sqrt{2}}{v} \left[ \left(\frac{1}{t_\beta} -\frac{\chi^{u*}_{uu}}{s_\beta} \right) m_u V_{ud} - \frac{\sqrt{m_u m_c}}{s_\beta} \chi^{u*}_{cu} V_{cd} - \frac{\sqrt{m_u m_t}}{s_\beta} \chi^{u*}_{tu} V_{td}\right].
 \end{align}
 It can be seen that the first and third terms are negligible due to the suppressions of $m_u/v$ and $\sqrt{m_u/m_t} V_{td}$, respectively. Although the second term is somewhat larger, it is also negligible based on the result of $\sqrt{2 m_u m_c} V_{cd}/v \sim -1.0 \times 10^{-4}$.  Hence, it is a good approximation to take $C^L_{ud} \sim 0$.  For the $C^R_{ud}$ coupling, it can be decomposed to be:
  \begin{align}
 \frac{\sqrt{2}}{v} C^{R}_{ud} &= \frac{\sqrt{2}}{v} \left[ t_\beta m_d V_{ud} \left( 1 -\frac{\chi^{d}_{dd}}{s_\beta} \right)  - \frac{\sqrt{m_s m_d}}{c_\beta} \chi^{d}_{sd} V_{us} - \frac{\sqrt{m_b m_d}}{c_\beta} \chi^{d}_{bd} V_{ub}\right] \nonumber \\
 & \approx \sqrt{2} \frac{m_d t_\beta}{v} V_{ud} \left( 1 - \frac{\chi^R_{ud} }{s_\beta} \right)\,, \quad \chi^R_{ud} = \chi^d_{dd} + \sqrt{\frac{m_s}{m_d}} \frac{V_{us}}{V_{ud}} \chi^d_{ds}\,,
 \end{align}
 where we have neglected the $V_{td}$ contribution in $\chi^R_{ud}$. 
Taking $t_\beta\sim 50$ and $|1-\chi^R_{ud}/s_\beta|\sim 2$, we obtain $C^R_{ud} \sim  5.7\times 10^{-3}$, and this charged-Higgs coupling indeed is two orders of  magnitude smaller than  the charged $W$-gauge boson coupling of $g/\sqrt{2} \approx 0.467$. Thus, we can also take $C^R_{ud} \sim 0$ as a leading order approximation. 
 
   \underline {$cd H^+$ vertex:} From the definition in Eq.~(\ref{eq:CLRud}), we write $C^L_{cd}$ as:
 \begin{align}
 \frac{\sqrt{2}}{v} C^{L}_{cd} & = \frac{\sqrt{2}}{v} \left[ \left(\frac{1}{t_\beta} -\frac{\chi^{u*}_{cc}}{s_\beta} \right) m_c V_{cd} - \frac{\sqrt{m_c m_u}}{s_\beta} \chi^{u*}_{uc} V_{ud} - \frac{\sqrt{m_c m_t}}{s_\beta} \chi^{u*}_{tc} V_{td}\right] \nonumber \\
 & \approx - \sqrt{2} \frac{m_c}{v s_\beta} V_{cd} \chi^L_{cd}\,, \quad \chi^L_{cd} = \chi^{u*}_{cc} + \sqrt{\frac{m_u}{m_c}}\frac{V_{ud}}{V_{cd}}  \chi^{u*}_{uc}  + \sqrt{\frac{m_t}{m_c}}\frac{V_{td}}{V_{cd}}  \chi^{u*}_{tc} \,,
 \end{align}
 where we have dropped $1/t_\beta$ term in the second line. Numerically, we get $\sqrt{ m_u/m_c} V_{ud}/|V_{cd}|\approx 0.28$ and   $\sqrt{m_t/m_c}|V_{td}/V_{cd}|\approx 0.09$; therefore, $\chi^L_{21}$ is dominated by $\chi^{u*}_{cc}$. Nevertheless, with the result of $\sqrt{2} m_c V_{cd} /v \approx -1.6\times 10^{-3}$, the $C^L_{cd}$ effect is two orders of magnitude smaller than the contribution of the $W$-boson in the SM. This contribution can be ignored for a phenomenological analysis. Similarly, we write $C^R_{cd}$ as:
 \begin{align}
 \frac{\sqrt{2}}{v} C^{R}_{cd} &= \frac{\sqrt{2}}{v} \left[ t_\beta m_d V_{cd} \left( 1 -\frac{\chi^{d}_{11}}{s_\beta} \right)  - \frac{\sqrt{m_s m_d}}{c_\beta} \chi^{d}_{21} V_{cs} - \frac{\sqrt{m_b m_d}}{c_\beta} \chi^{d}_{31} V_{cb}\right]\,.
  \end{align}
Using $t_\beta \sim 50$, we find that the first, second, and third terms in $C^R_{cd}$ are  around $1.3\times 10^{-3}$ with $|1-\chi^d_{11}/s_\beta|=2$, $9.1 \times 10^{-3}$, and $2.5 \times 10^{-3}$, respectively; that is, $C^R_{cd}$ is dominated by the $\chi^d_{sd}$ term and can be one order smaller than the SM gauge coupling of $(g/\sqrt{2}) V_{cd}$.  Taking the case with $1/c_\beta \approx t_\beta$, a simple expression can be given as:
 \begin{equation}
 \frac{\sqrt{2}}{v} C^R_{cd} \approx -\sqrt{2} \frac{m_d t_\beta }{v} V_{cs} \sqrt{\frac{m_s}{m_d}} \chi^d_{sd} \approx 8.3\times 10^{-4} t_\beta \chi^d_{sd} V_{cd}\,.
  \end{equation}

   \underline {$td H^+$ vertex:}  The $C^L_{td}$ coupling is expressed as:
 \begin{align}
 \frac{\sqrt{2}}{v} C^L_{td} & = \frac{\sqrt{2}}{v} \left[ \left( \frac{1}{t_\beta} - \frac{\chi^{u*}_{tt} }{s_\beta}\right) m_t V_{td} - \frac{\sqrt{m_t m_c}}{s_\beta} \chi^{u*}_{ct} V_{cd} - \frac{\sqrt{m_t m_u}}{s_\beta} \chi^{u*}_{ut} V_{ud}\right] \nonumber \\
 & \approx  \sqrt{2} \frac{m_t}{v } V_{td} \left( \frac{1}{t_\beta} - \frac{\chi^L_{td}}{s_\beta}  \right)\,, \quad \chi^L_{td} = \chi^{u*}_{tt} +\sqrt{\frac{m_c}{m_t}} \frac{V_{cd}}{V_{td}} \chi^{u*}_{ct}\,. \label{eq:CLtd}
 \end{align}
Since the coefficient of  $\chi^{u*}_{ut}$ term  is a factor of 4 smaller than that of  $\chi^{u*}_{ct}$, we  dropped the $\chi^{u*}_{ut}$ term. From Eq.~(\ref{eq:CLtd}), it can be seen that the $C^L_{td}$ effect in the generic 2HDM is comparable to the SM coupling of $(g/\sqrt{2})V_{td}$, where $\chi^L_{td}$ is the main parameter. 
Due to $m_d V_{td} \ll \sqrt{m_s m_d} V_{ts}  \ll \sqrt{m_b m_d} V_{tb}$,  the  $C^R_{td}$ Yukawa coupling can be simplified  as:
 \begin{align}
 \frac{\sqrt{2}}{v} C^R_{td} 
 & \approx - \sqrt{2} \frac{m_b t_\beta }{v }  \sqrt{\frac{ m_d}{m_b}}\chi^d_{bd} V_{tb}\,. \label{eq:CRtd}
 \end{align}
 Intriguingly, unlike the case in $C^L_{td}$,  $C^R_{td}$ has no $V_{td}$ suppression; thus, its value  with a large $t_\beta$ scheme can be even larger than $(g/\sqrt{2}) V_{td}$ in the SM. Moreover, when $\chi^{u*}_{tt}$ and $\chi^{u*}_{ct}$ are  in the range of  $O(0.1)-O(1)$, $\chi^{L}_{td}$ in $C^L_{td}$ can be small  if the cancellation occurs between $\chi^{u*}_{tt}$ and $\chi^{u*}_{ct}$. However, since the cancellation cannot occur in Eq.~(\ref{eq:CRtd}), $\chi^d_{bd}$ will be directly bounded by the rare decays. 
 
  \underline {$ u(c) s H^+$ vertex:}  To analyze the $u(c)$-$s$-$H^+$ couplings,   $C^{L,R}_{us}$ and $C^{L,R}_{cs}$ can be reduced to be:
  \begin{align}
 \frac{\sqrt{2}}{v}  C^L_{us} & = \frac{\sqrt{2}}{v} \left[\left(  \frac{1}{ t_\beta }  - \frac{\chi^{u*}_{uu}}{ s_\beta}\right)  m_u V_{u s}  - \frac{\sqrt{m_u m_c}}{s_\beta} \chi^{u*}_{cu} V_{cs}  - \frac{\sqrt{m_u m_t}}{s_\beta} \chi^{u*}_{tu} V_{ts}\right] \sim O(10^{-4})\,,\\
\frac{\sqrt{2}}{v}  C^L_{cs} & = \frac{\sqrt{2}}{v} \left[ \left(\frac{1}{t_\beta} -\frac{\chi^{u*}_{cc}}{s_\beta} \right) m_c V_{cs} - \frac{\sqrt{m_c m_u}}{s_\beta} \chi^{u*}_{uc} V_{us} - \frac{\sqrt{m_c m_t}}{s_\beta} \chi^{u*}_{tc} V_{ts}\right] \nonumber \\
 & \approx - \sqrt{2} \frac{m_c}{v s_\beta} V_{cs} \left(\chi^{u*}_{cc} +  \sqrt{\frac{m_t}{m_c}} \frac{V_{ts}}{V_{cs}} \chi^{u*}_{tc} \right) \approx  - \sqrt{2} \frac{m_c}{v s_\beta} V_{cs} ( \chi^{u*}_{cc} - 0.45 \chi^{u*}_{tc}) \,,
  \end{align}
  where $\sqrt{2}C^L_{us}/v$ is around $10^{-4}$ and is thus  negligible. Although $\sqrt{2} m_c/v V_{cs}\sim 7.4 \times 10^{-4}$, it is still two orders smaller than the gauge coupling in the SM.  In the phenomenological analysis, the $C^L_{cs}$ effect can be neglected. Similarly, the $C^R_{us}$ and $C^R_{cs}$ couplings can be simplified as:
   \begin{align}
 \frac{\sqrt{2}}{v}  C^R_{us} & \approx \sqrt{2} \frac{m_s t_\beta}{v} V_{us} \left(1 - \frac{\chi^R_{us}}{s_\beta} \right)\,, \quad \chi^R_{us} =  \chi^d_{ss} + \sqrt{\frac{m_d}{m_s}} \frac{V_{ud}}{V_{us}} \chi^d_{ds}\,,  \\
  \frac{\sqrt{2}}{v}  C^R_{cs} & \approx \sqrt{2} \frac{m_s t_\beta}{v} V_{cs} \left(1 - \frac{\chi^R_{cs}}{s_\beta} \right)\,, \quad \chi^R_{cs} = \chi^d_{ss} + \sqrt{\frac{m_b}{m_s}} \frac{V_{cb}}{V_{cs}} \chi^d_{bs}\,.
   \end{align}
 
 \underline {$ t s H^+$ vertex:}   using $m_t V_{ts} \sim 6.72 \ {\rm GeV} < \sqrt{m_c m_t} V_{cs} \sim 14.8$ GeV and   $ m_s V_{ts} \ll \sqrt{m_s m_b} V_{tb} \sim 0.66$ GeV, we can simplify $C^{L,R}_{ts}$ to be:
  \begin{align}
\frac{\sqrt{2}}{v}  C^L_{ts} & \approx \sqrt{2} \frac{m_t}{v} V_{ts} \left( \frac{1}{t_\beta} - \frac{\chi^L_{ts}}{s_\beta}\right)\,, \quad \chi^L_{ts} = \chi^{u*}_{tt} + \sqrt{\frac{m_c}{m_t}} \frac{V_{cs}}{V_{ts}} \chi^{u*}_{ct}\,, \label{eq:CLts} \\
 \frac{\sqrt{2}}{v} C^R_{ts} & \approx  -\sqrt{2} \frac{\sqrt{m_b m_s}}{v c_\beta} \chi^d_{bs} V_{tb}\,. \nonumber 
  \end{align}
It can be seen that due to the new factor $\chi^L_{ts}$, $\sqrt{2}C^L_{ts}/v$ can be comparable with the SM coupling of $g V_{ts}/\sqrt{2}$ without relying on the large $t_\beta$ scheme. 

 \subsection{$\bar B\to (D,D^*)$ form factors in the HQET}
 
We summarize the relevant $\bar B\to D^{(*)}$ form factors with the corrections of $\Lambda_{QCD}/m_{b,c}$ and $\alpha_s$, which are shown  in~\cite{Bernlochner:2017jka}.   To describe the $\bar B\to (D, D^*)$ transition form factors based on the HQET, it is convenient to use the dimensionless kinetic variables, defined as:
  \begin{equation}
  v=\frac{p_B}{m_B}\,, \  v'=\frac{p_{D^{(*)}}}{m_{D^{(*)}}}\,, \ w= v\cdot v' = \frac{m^2_B + m^2 _{D^{(*)}} -q^2 }{2 m_B m_{D^{(*)}}}\,.
  \end{equation}
  Thus,   the $\bar B\to D$ form factors  can be defined as:
  \begin{align}
  \langle D| \bar c b| \bar B\rangle &= \sqrt{m_B m_D} h_S (w+1)\,, \nonumber \\
   \langle D| \bar c \gamma^\mu  b| \bar B\rangle &= \sqrt{m_B m_D} \left( h_+ (v+v')^\mu + h_{-} (v-v')^\mu \right)\,, \nonumber \\
    \langle D| \bar c \sigma^{\mu \nu}   b| \bar B\rangle &= i \sqrt{m_B m_D}  h_T \left( v'^{\mu} v^{\nu} - v'^\nu v^\mu \right)\,,
     \label{eq:HQETBP}
  \end{align}
 while the form factors for  $\bar B\to D^*$ are:
  \begin{align}
   \langle D^* | \bar c \gamma^5 b| \bar B\rangle &= -\sqrt{m_B m_{D^*}} h_P \, \epsilon^* \cdot v \,, \nonumber \\
    \langle D^* | \bar c \gamma^\mu  b| \bar B\rangle &= i \sqrt{m_B m_{D^*}} h_V \varepsilon^{\mu \nu \alpha \beta} \epsilon^*_\nu v'_\alpha v_\beta \,, \nonumber \\
     \langle D^* | \bar c \gamma^\mu  \gamma^5 b| \bar B\rangle &= \sqrt{m_B m_{D^*}} \left[ h_{A_1} ( w+1) \epsilon^{*\mu} -h_{A_2} (\epsilon^*\cdot v) v^\mu - h_{A_3} (\epsilon^* \cdot v) v'^\mu   \right]\,, \nonumber \\
   \langle D^* | \bar c \sigma^{\mu\nu}   b| \bar B\rangle &=  - \sqrt{m_B m_{D^*}} \left[ h_{T_1} \epsilon^*_\alpha (v+ v')_\beta + h_{T_2} \epsilon^*_\alpha ( v-v')_\beta + h_{T_3} (\epsilon^*\cdot v ) v_\alpha v'_\beta \right]\,,
  \end{align}
where $h_{-}$, $h_{A_2}$, and $h_{T_{2,3}}$  vanish  in the heavy quark limit,  and the remaining form factors are equal to the leading order Isgur-Wise function $\xi(w)$.
 
We take the parametrization of leading order Isgur-Wise function as~\cite{Bernlochner:2017jka,Caprini:1997mu}:
 \begin{equation}
 \frac{\xi(w)}{\xi(w_0)} \simeq  1 - 8 a^2 \bar\rho^2_{*}  z_{*} + \left[ V_{21} \bar\rho^2_{*} -V_{20} + \Delta(e_b,e_c,\alpha_s) \right] z^2_{*}\,,
 \end{equation}
 where $V_{21}=57.0$, $V_{20}=7.5$; $z_*$ and  $a$  are defined as~\cite{Caprini:1997mu}:
 \begin{equation}
 z_* = \frac{\sqrt{w+1} -\sqrt{2} a}{\sqrt{w+1} + \sqrt{2} a}\,, \ a = \sqrt{\frac{1+r_D}{2\sqrt{r_D} } }\,, 
 \end{equation}
 $r_D=m_D/m_B$, $w_0$ is determined from $z(w_0)=0$; $\bar \rho^2_{*}$ is the slop parameter of $\xi(w)/\xi(w_0)$, and $\Delta(e_b,e_c,\alpha_s)$ denotes the correction effects of $O(e_{b,c})$ with $e_{b(c)}=\bar\Lambda/m_{b(c)}$ and $O(\alpha_s)$. We take the results using  the fit scenario of ``th:$L_{w\geq 1}$+SR" shown in~\cite{Bernlochner:2017jka}. In addition to $\bar\rho^2_{*}=1.24\pm 0.08$, the values of sub-leading Isgur-Wise functions at $w=1$ are given in Table~\ref{tab:subIW}. Using these results, the correction of $O(e_{b,c})$ and $O(\alpha_s)$ can be obtained as:
  \begin{equation}
  \Delta(e_b,e_c,\alpha_s) \approx 0.582 \pm 0.298\,,
  \end{equation}
 where we adopt the $1S$ scheme for $m_b$ and use the value of $m^{1S}_{b}=4.71 \pm 0.05$ GeV~\cite{Bernlochner:2017jka}. In addition,  $\delta m_{bc}=m_b - m_c =3.40\pm 0.02$ GeV and $\bar \Lambda=0.45$ GeV are used. 
 
 \begin{table}[htbp]
   \caption{ The results of sub-leading Isgur-Wise functions using the fit scenario of ``th:$L_{w\geq  1}$+SR".}
  \label{tab:subIW}
   \begin{tabular}{c|ccccc} \hline \hline
   FS &  $\hat\chi_{2}(1)$ & $\hat\chi'_2(1)$ & $\hat\chi'_3(1)$ & $\eta(1)$ & $\eta'(1)$   \\ \hline 
   th:$L_{w\geq 1}$ + SR & ~$ -0.06 \pm 0.02$~  &  ~$-0.00\pm 0.02$ & ~$0.04 \pm 0.02$~ & ~$0.31\pm 0.04$~ & ~$0.05\pm 0.10$~  \\ \hline  \hline
   
     \end{tabular}
\end{table}
 
 Following the notation in~\cite{Bernlochner:2017jka}, the form factors up to $O(e_{b,c})$ and $O(\alpha_s)$ can  be expressed by factoring out $\xi$ as: $ h_{i} = \hat h_{i}\, \xi$, where the $\hat h_i$ for the $\bar B \to D$ decay  are given as~\cite{Bernlochner:2017jka}: 
 \begin{subequations}
 \begin{align}
 \hat h_+ & = 1 + \hat\alpha_s \left[ C_{V_1} + \frac{w + 1}{2} (C_{V_1} + C_{V_3} )\right] + ( e_c + e_b) \hat L_1\,,  \\
 \hat h_{-} & = \hat\alpha_s \frac{w + 1}{2} (C_{V_2} - C_{V_3} ) + ( e_c - e_b) \hat L_4\,, \\
 \hat h_S & = 1 + \hat\alpha_s C_S + ( e_c + e_b) \left[ \hat L_1 - \hat L_4 \frac{w-1}{w+1} \right]\,, \\
 \hat h_T &= 1+ \hat\alpha_s (C_{T_1} - C_{T_2} + C_{T_3}) + (e_c + e_b) (\hat L_1 - \hat L_4)\,;
 \end{align}
 \end{subequations}
 for $\bar B \to D^*$, the associated $\hat h_{i}$ are shown as~\cite{Bernlochner:2017jka}:
  \begin{subequations}
  \begin{align}
  \hat h_V & = 1 + \alpha_s C_{V_1} + e_c (\hat L_2 -\hat L_5) + e_b (\hat L_1 -\hat L_4)\,, \\
   \hat h_{A_1} & = 1 + \hat \alpha_s C_{A_1} + e_c  \left(\hat L_2 - \hat L_5 \frac{w-1}{w+1} \right) + e_b \left(\hat L_1 - \hat L_4 \frac{w-1}{w+1} \right)\,, \\
  \hat h_{A_2} & = \hat \alpha_s C_{A_2} + e_c (\hat L_3 + \hat L_6)\,, \\
  \hat h_{A_3} &= 1+ \hat \alpha_s (C_{A_1}+C_{A_3}) + e_c (\hat L_2 - \hat L_3 + \hat L_6 - \hat L_5) + e_b (\hat L_1 - \hat L_4)\,, \\
  \hat h_P &= 1+ \hat \alpha_s C_P + e_c \left[ \hat L_2 + \hat L_3 (w-1) + \hat L_5 - \hat L_6 (w+1) \right] + e_b (\hat L_1 - \hat L_4)\,, \\
  \hat h_{T_1} &= 1 + \hat \alpha_s \left[ C_{T_1} + \frac{w-1}{2} (C_{T_2} - C_{T_3}) \right] + e_c \hat L_2 + e_b \hat L_1\,,\\
  \hat h_{T_2} &= \hat \alpha_s \frac{w+1}{2} (C_{T_2}+ C_{T_3}) + e_c \hat L_5 - e_b \hat L_4\,,\\
  \hat h_{T_3} &= \hat \alpha_s C_{T_2} + e_c (\hat L_6 - \hat L_3)\,.
  \end{align}  
  \end{subequations}
  The $w$-dependent functions $C_{\Gamma_i}$ can be found in~\cite{Neubert:1992qq,Bernlochner:2017jka}, and the sub-leading Isgur-Wise functions are~\cite{Falk:1992wt}:
   \begin{align}
   &\hat L_1 = -4(w-1) \hat\chi_2 + 12 \hat \chi_3\,, \ \hat L_2 = -4 \hat\chi_3\,, \ \hat L_3 = 4 \hat\chi_2\,,  \nonumber \\
   & \hat L_4 = 2 \eta -1\,, \ \hat L_5= -1\,, \  \hat L_6 = -2 \frac{1+\eta}{w+1}\,,
   \end{align}
   where the $w$-dependent functions $\hat\chi_i$ and $\eta$ can be approximated as:
    \begin{align}
    \hat\chi_2 (w) & \simeq \hat\chi_2(1) + \hat\chi'_2(1) (w-1)\,, \nonumber \\
    \hat\chi_3(w) & \simeq \hat\chi'_3 (1) (w-1) \,, \nonumber \\
    \eta(w) & \simeq \eta(1) + \eta'(1)(w-1)\,.
    \end{align}
 

%
%
 
\section*{Acknowledgments}

This work was partially supported by the Ministry of Science and Technology of Taiwan,  
under grants MOST-106-2112-M-006-010-MY2 (CHC). \\

\bibliographystyle{unsrt}

\end{document}